\documentclass{article}
\usepackage{graphicx,amssymb}
\usepackage{float}

\title{Global geometry of space-time  \\with the charged shell}
\author{V. A. Berezin\thanks{berezin@ms2.inr.ac.ru} \ и
        V. I. Dokuchaev\thanks{dokuchaev@inr.ac.ru}    \\
        {\small\em Institute for Nuclear research of the Russian Academy of Sciences, Moscow}}
\date{}

\begin{document}
\maketitle

\begin{abstract}
It is elaborated the complete classification of the possible types of the spherically symmetric global geometries for two types of electrically charged shells: (1) The charged shell as a single source of the gravitational field, when internal space-time is flat, and external space-time is the Reissner--Nordstr\"om metric; (2) The neutralizing shell with an electric charge opposite to the charge of the internal source with the Reissner--Nordstr\"om metric and with the Schwarzschild metric outside the shell.
\end{abstract}

\newpage
\tableofcontents

\newpage
\section*{List of notations}

{}\indent

$G$ --- The Newton constant

$e$ --- electric charge

$\rho(\tau)$ --- Shell radius as a function of the proper time $\tau$

$M$ --- ``bare'' mass of the shell

$m_{\rm in}>0$ --- mass of the internal metric

$m_{\rm out}>0$ --- mass of the external metric

$\Delta m=m_{\rm out}-m_{\rm in}$ --- see definition in (\ref{sigmaout2})

$m_{\rm run}(\rho)$ --- running mass (\ref{minrun}) and (\ref{moutrun})

$m_{\rm in,run}(\rho)$ --- mass of the internal metric (\ref{minrun})

$m_{\rm in,run}(\rho)$ --- mass of the external metric (\ref{moutrun})

$r_g=2Gm_{\rm out}>0$ --- gravitational radius of the external metric

$r_\pm=Gm_{\rm in}\pm\sqrt{G^2m_{\rm in}^2-Ge^2}$ --- radii of horizons of the internal metric

$R_\pm$--region --- space-time region with the signature $(+,-,-,-)$

$T_\pm$--region --- space-time region with the signature $(-,+,-,-)$

$\sigma_{\rm in}(\rho)=\pm1$ and $\sigma_{\rm out}(\rho)=\pm1$ --- sign-changing functions in (\ref{israel})

$\rho_0$ --- turning point (rejection) in (\ref{turning})

$\rho_{0,\rm min}$ --- minimal turning point

$\rho_{0,-}$ --- smaller turning point

$\rho_{0,+}$ --- larger turning point

$\rho_d$ --- turning point (\ref{mout1}) at the coincidence $\rho_{0,-}=\rho_{0,+}$

$\rho_{\sigma_{\rm in}}$ --- point of sign changing of $\sigma_{\rm in}$ in $T_\pm$--regions

$\rho_{\sigma_{\rm out}}$ --- point of sign changing of $\sigma_{\rm out}$ in $T_\pm$--regions (\ref{sigmaout})

$m_{\rm out,min}$ --- minimally possible mass $m_{\rm out}$ (\ref{fin43min})

$m_{\rm in,min}$ --- minimally possible mass $m_{\rm in}$  (\ref{fin43min})

$\Delta m_{\rm cr}$ --- critical value $\Delta m$  (\ref{Deltacr})

\newpage

\section{Introduction}

The structure of any spherically symmetric space-time is completely
determined by two invariant functions of two variables. Indeed,
locally, the general spherically symmetric metric can be written as
\begin{equation}
ds^2 = A^2 dt^2 + 2 H dt dq - B^2 dq^2 - R^2 d\sigma^2 \, ,
\end{equation}
where $A(t,q),\, H(t,q)$ and $B(t,q)$ are functions of the time
coordinate, $t$, and some radial coordinate, $q$, $d\sigma ^2$ is
the line element of a $2- dim $ unit sphere, and $R(t,q)$ is the
radius of this sphere in the sense that its area equals $4 \pi \,
R^2$. Therefore, we are, actually, dealing with the invariant
function $R(t,q)$ and the two-dimensional metric, which by suitable
coordinate transformation can always be put in the conformally flat
form
\begin{equation}
ds^2_2 = \gamma _{ik} dx^i dx^k = \omega^2 (t,q) (dt^2 - dq^2)
\, , \;\;\; i,k = 0,1 \, .
\end{equation}
This proves the above statement about two functions of two
variables.

The first invariant function is, of course, the radius $R(t,q)$. By
geometrical reasons, we choose for the second function the invariant
(notations are obvious)
\begin{equation}
\Delta = \gamma ^{ik} \frac{\partial R}{\partial x^i} \frac{\partial
R}{\partial x^k} = \frac{1}{\omega ^2} \left( \dot R^2 - R^{\prime
2}\right) \, .
\end{equation}
This is nothing more but the square of the normal vector to the
surfaces of constant radii, $R (t,q) = const$. The invariant
function $\Delta$ brings a very important geometrical information.
If $\Delta < 0$, the surfaces $R = const$ are time-like, such
regions are called the $R_{\pm}$-regions, the signs $"\pm"$ being
denote the sign of a spatial derivative of the radial function $R$.
If $\Delta > 0$, the regions are called the $T_{\pm}$-regions,
depending on the sign of the corresponding time derivative
(inevitable expansion or inevitable contraction), and the surfaces
$R = const$ are space-like. The $R_{\pm}-$ and $T_{\pm}-$ regions
are separated by the apparent horizons with $\Delta = 0$. It is the
set of these regions and horizons together with the boundaries
(infinities and that determines the global geometry. The boundaries
are to be chosen in such a way that the space-time becomes
geodesically complete, namely, all the time-like and null geodesics
should start and end either at infinities or at singularities.

The causal structure of geodesically complete spherically symmetric
space-times can be best seen on the conformal Carter--Penrose
diagrams where each point represents a sphere, and infinities are
brought to the final distances. Since every 2-dimensional space-time
is (locally) conformally flat, its Carter--Penrose diagram is the set
of that for the $2 - dim$ Minkowski manifold. To see how the latter
looks like, let us, first, transform the Minkowski  metric $ds^2 =
dt^2 - dx^2 $ to the double-null coordinates $u = t -x$ (retarded
time) and $v = t + x$ (advanced time), then $ds^2 = du dv$. We will
use the convention that on the diagram the time coordinate increases
from down to up, the spatial coordinate - from left to right, and
the null curves $(u = const , \; v = const$ are the straight lines
with the slope $\pm 45^{\circ}$. Making one more transformation
\begin{eqnarray}
u^{\prime} &=& \arctan {u} \, , \;\;\; - \frac{\pi}{2} \le u^{\prime}
\le \frac{\pi }{2} \nonumber \\
v^{\prime} &=& \arctan {v} \, , \;\;\; - \frac{\pi}{2} \le v^{\prime}
\le \frac{\pi }{2}
\end{eqnarray}
one gets
\begin{eqnarray}
ds^2 &=& \Omega ^2 ds^{\prime 2} \, , \;\;\; \Omega =
\frac{1}{\cos {u^{\prime}} \cos {v^{\prime}}} \nonumber \\
ds^{\prime 2} &=& du^{\prime} dv^{\prime} = dt^{\prime 2} -
dx^{\prime 2} \, .
\end{eqnarray}
Formally, the metric $ds^{\prime 2}$ looks exactly as the starting
one, but now coordinates $(u^{\prime}, \, v^{\prime})$ and
$(t^{\prime}, \, x^{\prime})$ run the finite intervals.

\begin{figure}[t]
\begin{center}
\includegraphics[angle=0,width=0.8\textwidth]{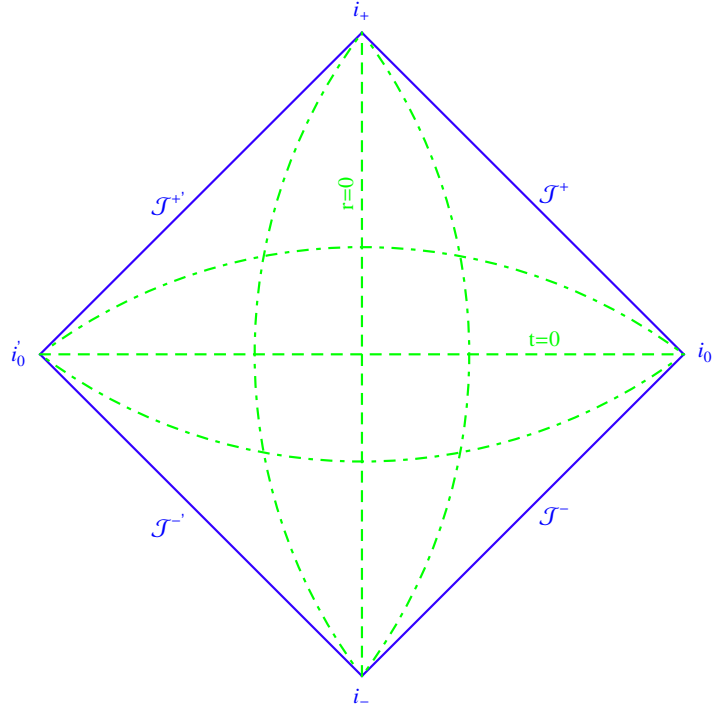}
\end{center}
\caption{The Carter--Penrose diagram for the complete $2 - dim$
Minkowski space-time $( - \infty < t < \infty, \; - \infty < x <
\infty)$. The horizontal dashed curves represent $t = const$
lines, while the vertical ones are for $x = const$.}
\label{2DimMinkowski}
\end{figure}

The Carter--Penrose diagram for the complete $2 - dim$ Minkowski
space-time $( - \infty < t < \infty, \; - \infty < x < \infty)$ is
shown in Fig.~\ref{2DimMinkowski}.
Here $J^{\pm} (J^{\prime \pm})$ are null future $(v^{\prime}
(u^{\prime}) = \frac{\pi}{2}, \, v (u) = \infty)$ and past
$(u^{\prime} (v^{\prime}) = - \frac{\pi}{2}, \, u (v) = - \infty)$
infinities, $i_{\pm}$ are future and past $(t^{\prime} = \pm
\frac{\pi}{2})$ temporal infinities, and $i_0 (i^{\prime}_0)$ are
spatial $(x^{\prime} = \pm \frac{\pi}{2}, \, x = \pm \infty)$
infinities. If the corres\-ponding conformally flat metric is not
complete in the sense that one of the coordinates starts from or
ends at the finite boundary value (like, for example, the zero
radius value in the case of spherical symmetry), then one should
cut the above square along the corresponding diagonal (in general,
along some time-like os space-like curve), and such part of the
complete Carter--Penrose diagram will be a triangle with the
vertical (left for $R_+$-regions and right for $R_-$-regions) or
horizontal (for $T_{\pm}$-regions) boundary.

\section{Conformal Carter--Penrose diagrams}

Both the Schwarzschild and Reissner--Nordstr\"om metrics look the same
in the so-called curvature coordinates:
\begin{equation}
ds^2 = F dt^2 - \frac{1}{F} dR^2 - R^2 (d \vartheta ^2 +
\sin^2{\vartheta} d\varphi^2) \, ,
\end{equation}
where $R$ - radius $(0 \le R < \infty), \, F = F(R)$, and
$\vartheta$ and $\varphi$ are spherical angles. The two-dimensional
part can easily be written in the conformally flat form by
introducing the "tortoise" coordinate $R^{\star}$:
\begin{eqnarray}
dR^{\star} &=& \frac{d R}{|F|} \, , \nonumber \\
ds^2_2 &=& F \left( d \xi^2 - dR^{\star 2}\right) \, .
\end{eqnarray}
In the $R_{\pm}$-regions $F = - \Delta > 0$ and $R^{\star}$ plays
the role of the spatial (radial) coordinate $q$, while $\xi$ is the
time coordinate $t$. In the $T_{\pm}$-regions, $R^{\star}$ plays the
role of the time coordinate $t$, while $\xi$ is the spatial
coordinate $q$.

\begin{figure}[t]
\begin{center}
\includegraphics[angle=0,width=0.8\textwidth]{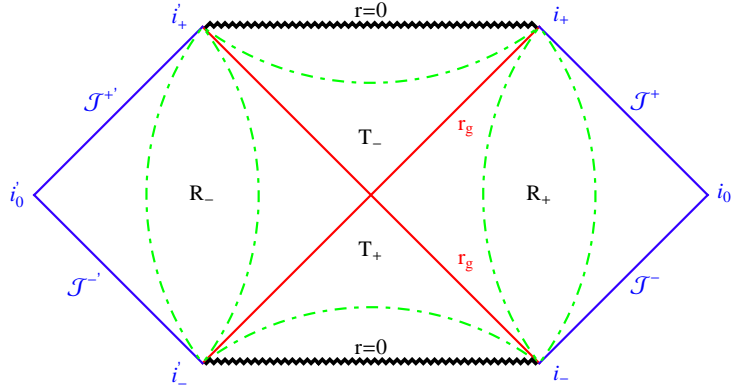}
\end{center}
\caption{The complete Carter--Penrose diagram of the Schwarzschild
metric.} \label{SchwTotal}
\end{figure}

Consider, first, the  Schwarzschild metric. In this case
\begin{equation}
F = 1 - \frac{2 \, G \, m}{R} \, ,
\end{equation}
where $G$ is the Newton's gravitational constant $m$ is the total
mass of the gravitating system measured by distant observers (at
infinity), and we put the speed of light $c = 1$. For $R > r_g = 2
\, G \, m$ we have the the $R$-region, and for $R < r_g$ - the
$T$-region. The event horizon coincides with the apparent horizon
at $R = r_g$ (gravitational, or Schwarzschild, radius). At $R = 0$
we encounter the (space-like) curvature singularity. The complete
Carter--Penrose diagram looks as follows in Fig.~\ref{SchwTotal}.
There are two isometric $R_{\pm}$-regions bounded by two apparent
(past and future) horizons at $R = r_g$ and two asymptotically flat
regions with corresponding future and past temporal $(i_{\pm}, \,
i^{\prime}_{\pm})$, future and past null $(J_{\pm}, \,
J^{\prime}_{\pm})$ and spatial $(i_0, \, i^{\prime}_0)$ infinities.
Also we have two $T$-regions ($T_+$ and $T_-$) bounded by the
apparent horizons at $R = r_g$ and future and past space-like
singularities at $R = 0$. This is called the eternal Schwarzschild
black hole. The gravitational source is concentrated on these two
space-like singularities, i.e., it exists only for one moment in the
past and reappears again for one moment in the future.

The causal structure of the Reissner--Nordstr\"om space-time is much
more complex. The function $F$ equals now
\begin{equation}
F = 1 - \frac{2 \, G \, m}{R} + \frac{G \, e^2}{R^2},
\end{equation}
$e$ is the electric charge. There are three different cases

\begin{figure}
\begin{center}
\includegraphics[angle=0,width=0.7\textwidth]{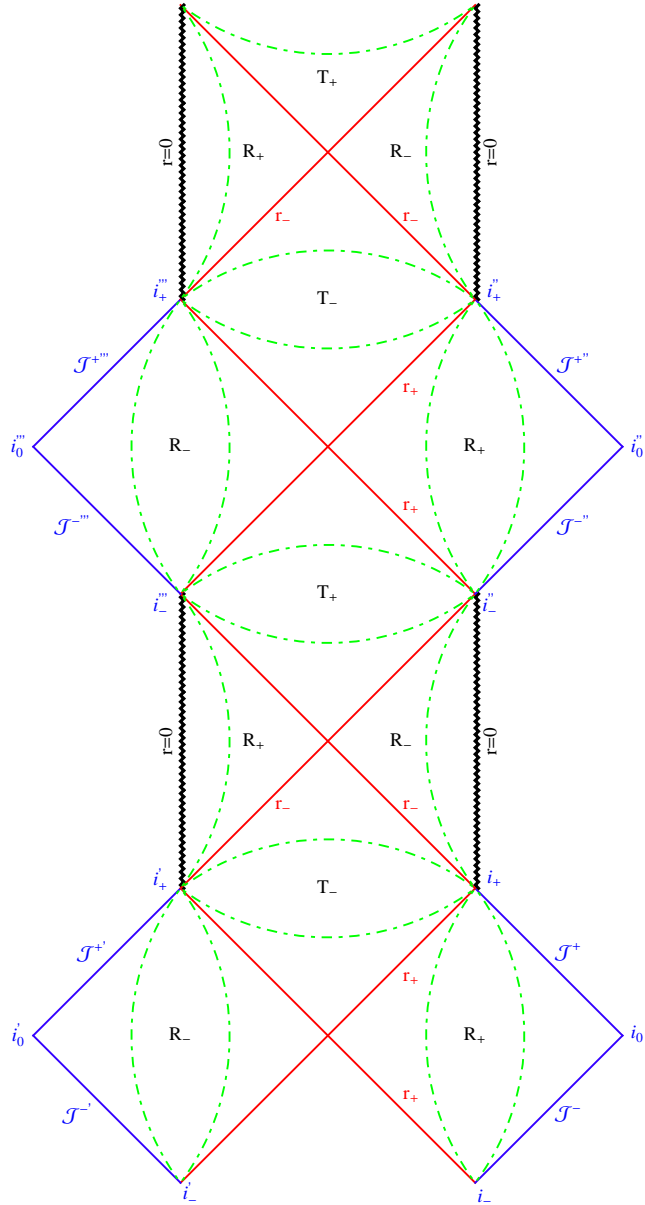}
\end{center}
\caption{The complete Carter--Penrose diagram of the Reissner--Nordstr\"om black hole, $G \, m^2 < e^2$.} \label{RN}
\end{figure}

(1) $G \, m^2 > e^2$ - Reissner--Nordstr\"om black hole, equation $F =
0$ has two nonequal real roots $r_{\pm}$,
\begin{equation}
r_{\pm} = G \, m \pm \sqrt{G^2 \, m^2 - G \, e^2} \, .
\end{equation}
According to the signs of $F$, we have the $R$-regions for $r_+ <
R < \infty$ and $0 \le R < r_-$, $T$-regions in-between, $r_- < R
< r_+$, and two apparent horizons at $R = r_{\pm}$, the external
one, $r_+$, playing the role of the event horizon, and the inner,
$r_-$, - the Cauchy horizon. The geodesically complete
Carter--Penrose diagram is the ladder extended infinitely to the
past and to the future as shown in
Fig.~\ref{RN}.
In the complete (eternal) Reissner--Nordstr\"om black hole space-time
both the the gravitational source and the electric charge(s) are
concentrated on two (for each part of the ladder) time-like
singularities $R = 0$ (left and right on the diagram), the signs
of the electric charges on them being opposite..

\begin{figure}
\begin{center}
\includegraphics[angle=0,width=0.5\textwidth]{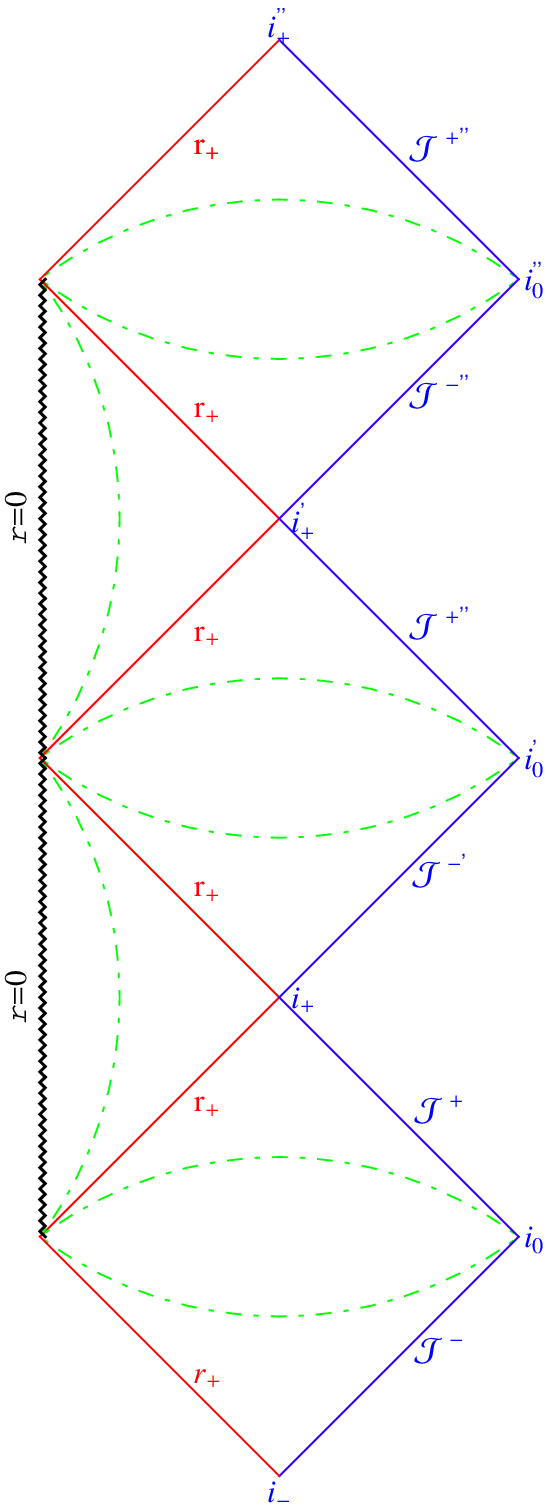}
\end{center}
\caption{Extreme Reissner--Nordstr\"om black
hole, $G \, m^2 = e^2$.} \label{RNextremal}
\end{figure}

(2) $G \, m^2 = e^2$ --- extremal Reissner--Nordstr\"om black hole.
Equation $F = 0 $ has the double root $r_+ = r_- = G \, m =
\sqrt{G} |e|$.  We have $R$-regions everywhere except the apparent
(event) horizon at $R = r_+ = r_-$, as shown in ~\ref{RNextremal}.

\begin{figure}[ht]
\begin{center}
\includegraphics[angle=0,width=0.5\textwidth]{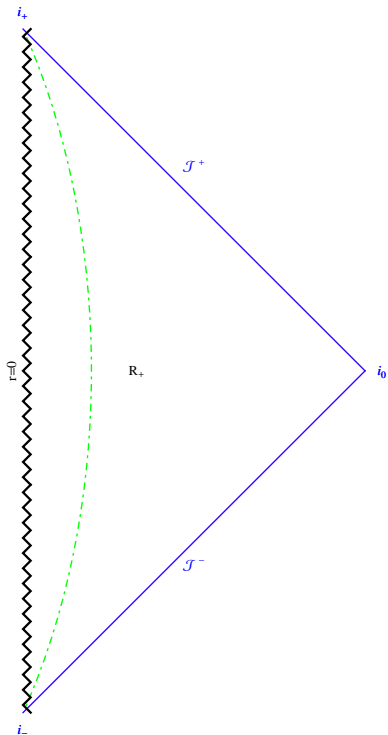}
\end{center}
\caption{The complete Carter--Penrose diagram of the
Reissner--Nordstr\"om naked singularity, $G \, m^2 < e^2$.}
\label{NS}
\end{figure}

 (3) $G \, m^2 < e^2$ - no black hole, naked singularity at
$R - 0$. The Carter--Penrose diagram is very simple (see
Fig.~\ref{NS}.

\section{Formalism of the thin shells}

The thin shell is a hyper-surface in the space-time on which the
energy-momen\-tum tensor is singular. If such a hyper-surface is time-
or space-like, one can introduce in its vicinity the so-called
Gaussian normal coordinates, and the line element can be written as
\begin{equation}
ds^2 = \varepsilon dn^2 + \gamma_{ij}(n,x) dx^i dx^j \, ,
\end{equation}
$n$ is the coordinate in the normal direction to the shell, and
$x^i$ - coordinates on the shell, $\varepsilon = + 1$ in the
space-like case and $\varepsilon = - 1$ in the time-like case. The
surface is supposed to be located at $n = 0$. The energy-momentum
tensor $T^{\mu}_{\nu}$ is proportional to $\delta$-function,
\begin{equation}
T^{\mu}_{\nu} = S^{\mu}_{\nu} \, \delta (n) \, ,
\end{equation}
$S^{\mu}_{\nu}$ is called the surface energy-momentum tensor. The
dynamics of the thin shell is governed by the Israel equations \cite{Israel} (see also \cite{BerKuzTkachPRD87} -- \cite{BerKuzTkachJETP}),
obtained by integrating the Einstein equations across the shell.
First of all, one gets $S^n_n = S^i_n = 0$, this can be considered
as the definition of the thin shell. The Israel equations are
\begin{equation}
\varepsilon \left( \left[K_{ij} \right] - \gamma_{ij} \left[ K\right] \right)
 = 8 \pi \, G S_{ij} \, ,
\end{equation}
supplemented by the Bianchi identity for the shell
\begin{equation}
S^j_{i|j} + \left[ T^n_i\right] = 0\, .
\end{equation}
Here $K_{ij} =-(1/2) \partial \gamma_{ij}/\partial
n$ is the extrinsic curvature tensor, $K$ is its trace, brackets
$[\;] = (out) - (in)$ is the jump across the shell, the vertical
line denotes the covariant derivative with respect to the metric
$\gamma_{ij}$. In what follows we will be dealing with the time-like
shells only, so, $\varepsilon = - 1$.

In the case of spherical symmetry everything is simplified
drastically. The metric becomes
\begin{equation}
ds^2 = - dn^2 + \gamma_{00} (n, \tau) d \tau^2 - \rho^2 (n,\tau) d \sigma^2,
\end{equation}
$\rho (0,\tau)$ is the shell radius as a function of the proper time
of the observer sitting on this shell, $n < 0$ inside and $n > 0$
outside. The mixed components of the surface energy momentum tensors
are $S^0_0$ (surface energy density) and $S^2_2 = S^3_3$ (surface
tension), and the Israel equations reduced to one constraint and one
dynamical equations, namely,
\begin{eqnarray}
\left[ K^2_2 \right] &=& 4 \pi \, G \, S^0_0 \nonumber \\
\left[ K^0_0 \right] + \left[ K^2_2 \right] &=& 8 \pi \, G \, S^2_2 \, .
\end{eqnarray}
The supplement equation is now
\begin{equation}
\dot S^0_0 + \frac{2 \,\dot \rho}{\rho} \left( S^0_0 - S^2_2 \right) +
\left[ T^n_0\right] = 0 \, .
\end{equation}
We are interested in the situation when both inside and outside the
shell the space-time is (electro)-vacuum one, hence, $T^n_0 = 0$.
For the sake of simplicity we will consider the dust shell, for
which $S^2_2 = 0$. Then,
\begin{equation}
S^0_0 = \frac{M}{4 \pi \, \rho^2} \, ,
\end{equation}
where $M = const$ is the ``bare'' mass of the shell (without the
gravitational mass defect). Thus, we need only the first,
constraint, equation. In order to go further we have to calculate
\begin{equation}
K^2_2 = - \frac{1}{\rho^2} K_{22} = - \frac{1}{2 \rho^2}
\frac{\partial (\rho^2)}{\partial n} = - \frac{\rho_{,n}}{\rho} \, .
\end{equation}
But, from definition of the invariant $\Delta$ it follows
\begin{eqnarray}
\Delta &=& \dot \rho^2 - \rho^2_{,n} \nonumber\\
\rho_{,n} &=& \sigma \sqrt{\dot \rho^2 - \Delta} \nonumber \\
K^2_2 &=& - \frac{\sigma}{\rho} \sqrt{\dot \rho^2 - \Delta} \, .
\end{eqnarray}
Here $\sigma = \pm 1$ depending on whether radii increasee $(\sigma
= + 1)$ in the normal outward direction or decrease $(\sigma = -
1)$. Thus, the sign of $\sigma$ coincides with that of the
$R$-region, and it can change only in the $T$-regions. Finally, the
only equation we will need in our analysis is
\begin{equation}
\sigma_{in} \sqrt{\dot \rho^2 - \Delta_{in}} -
\sigma_{out} \sqrt{\dot \rho^2 - \Delta_{out}} = \frac{G \, M }{\rho} \, .
\end{equation}
Since in our case $\Delta = - F$, we have
\begin{equation}
\sigma_{in} \sqrt{\dot \rho^2 + 1 - \frac{2 \, G \, m_{in}}{\rho} +
\frac{G \, e^2_{in}}{\rho^2}} - \sigma_{out} \sqrt{\dot \rho^2 + 1 -
\frac{2 \, G \, m_{out}}{\rho} + \frac{G \, e^2_{out}}{\rho^2}} = \frac{G \, M }{\rho} \, .
\label{shell}
\end{equation}
We will not consider exotic matter shells, so $M > 0$. From the
above constraint equation (that is nothing more but the energy
conservation law) it follows that for the qualitative analysis one
needs to investigate the behavior of the function $\rho (\tau)$ only
at several special points: $\rho \to \infty, \; \dot \rho = 0, \;
\rho = 0$ and $\rho = \rho_{\sigma}$ where $\sigma_{out} (\sigma
_{in})$ changes its sign. Examples for using the thin shell formalism see, e.\,g., in \cite{BerKuzTkachJETP} -- \cite{BerKuzTkach3}.

\section{Charged shell with the Minkowski space inside}

We start by considering the case of a thin charged shell with the
Minkowski space-time (containing the world-line $r = 0$) inside and
Reissner--Nordstr\"om one outside . This means that $m_{\rm in}=e_{\rm
in}=0$, $F_{\rm in}=1$, $\sigma_{\rm in}=+1$ and the "naked" mass of
the shell $M=const$. The equation for shell dynamics (\ref{shell})
in this particular case takes the form
\begin{equation}
 \sqrt{\dot{\rho}^2+1}-\sigma_{\rm out}\sqrt{\dot{\rho}^2
 -\frac{2\, G \, m}{\rho}+\frac{Ge^2}{\rho^2}}=\frac{GM}{\rho},
 \label{shellM}
\end{equation}
where it is written $m=m_{\rm out}$ for brevity. By squaring
equation (\ref{shellM}) we get
\begin{equation}
 m=M\sqrt{\dot{\rho}^2+1}-\frac{2GM-e^2}{2\rho}.
 \label{shellM2}
\end{equation}
and
\begin{equation}
 \sigma_{\rm out}\sqrt{\dot{\rho}^2+1-\frac{2\, G \, m}{\rho}+\frac{Ge^2}{\rho^2}}
 =\frac{m}{M}-\frac{GM^2+e^2}{2M\rho}.
 \label{shellM3}
\end{equation}
It easily seen from  (\ref{shellM3}) that $m > M$ for infinite
(unbound) motions and $m < M$ for finite (bound) ones. From
 (\ref{shellM3}) we obtain the relations which define the sign of
$\sigma_{\rm out}$:
\begin{equation}
 \sigma_{\rm out}=\mbox{sign}\left(\frac{m}{M}-\frac{GM^2+e^2}{2M\rho}\right).
 \label{sigmaout}
\end{equation}
and determine the radius $\rho_\sigma$, where $\sigma_{\rm out}$
changes its sign:
\begin{equation}
 \rho_\sigma=\frac{GM^2+e^2}{2m}.
 \label{rhosigma}
\end{equation}
It happens when $\dot{\rho}^2+F_{\rm out}=0$ where $F_{\rm out}<0$,
so, it may take place only in $T$-region, i.e., only in the case
$e^2 < G m^2$, the Reissner--Nordstr\"om black hole metric. Formally,
the value for $\rho_{\sigma}$ can always be calculated. But it lies
on the shell trajectory, only if
\begin{equation}
r_-<\rho_\sigma<r_+,
 \label{r-rhosigmar+}
\end{equation}
This is equivalent (as can be easily verified) to
\begin{equation}
r_-<GM<r_+.
 \label{r-GMr+}
\end{equation}
For infinite motions $G M < G m < r_+$, so, the value of $\sigma$
remains unchanged during the shell evolution, if $G M < r_- = G m -
\sqrt{G^2 m^2 - G e^2}$, and this is equivalent to $\frac{m}{M} <
\frac{1}{2} + \frac{e^2}{2 G M^2}$. In the case of finite motions
$r_- < G m < G M$, so, the value of $\sigma$ remains unchanged if $G
M > r_+ = G m + \sqrt{G^2 m^2 - G e^2}$, and we obtain the same
relation for the total mass $m$ as before. Thus, we have the
universal relations for the shell motion in the case when the
external solutions are the Reissner--Nordstr\"om black holes with $G
m^2 > e^2$: $\sigma$ changed its sign on the shell trajectory if
\begin{equation}
\label{sc}
\frac{m}{M} >
\frac{1}{2} + \frac{e^2}{2 G M^2}
\end{equation}
and it remains unchanged if
\begin{equation}
\label{suc}
\frac{m}{M} < \frac{1}{2} + \frac{e^2}{2 G M^2}
\end{equation}
Now, let us define the relation for the turning point $\rho=\rho_0$,
where $\dot\rho=0$. The turning points can be situated only in the
$R_\pm$-regions. By putting $\dot\rho=0$ in  (\ref{shellM2}) we
find that it may be no turning points at all or only one turning
point:
\begin{equation}
 \rho_0=\frac{GM^2-e^2}{2(M-m)}.
 \label{rho0}
\end{equation}
Evidently, the turning point exists if $sign (M - m) = sign (G M^2 -
e^2)$, i.e., for infinite motions $(m > M)$ it happens for $e^2 > G
M^2$, while for finite motions $(m < M)$ the turning point exists by
definitions, so the finite motion itself is possible, only if $e^2 <
G M^2$.

\begin{figure}
\begin{center}
\includegraphics[angle=0,width=0.7\textwidth]{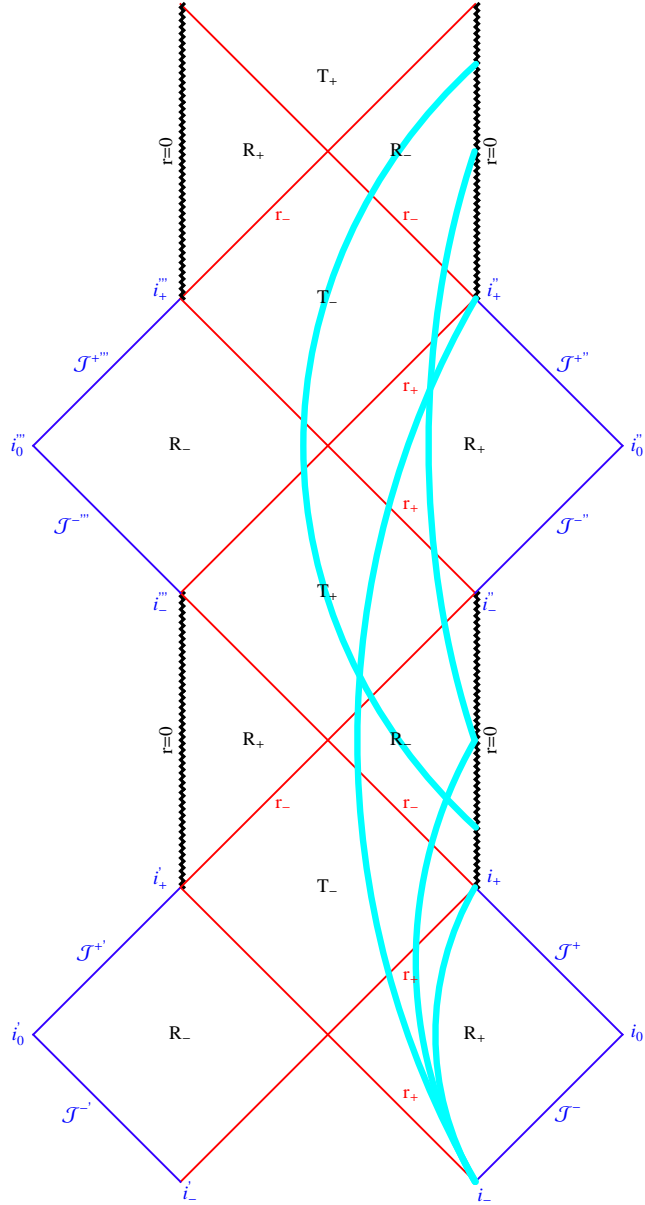}
\end{center}
\caption{The Carter--Penrose diagram for Reissner--Nordstr\"om metric (trajectories of the shell with turning points).} \label{TurningPoints}
\end{figure}
It is clear from (\ref{shellM}) that the shell evolution is
completely determined by the values of two parameters, bare mass $M$
and electric charge $e$ that characterizes the shell itself, and the
total mass of the system, $m$, depending on the initial conditions
which includes the initial value of $\sigma_{out}$. For infinite
motions it is natural to put the initial conditions at infinity,
$\rho \to \infty$. Since we adopt the physically acceptable signs
for the masses, $M > 0,\, m > 0$, we find that $\sigma_{out}(\infty)
= + 1$, and the value of the total mass $m$ is determined by the
initial velocity of the shell there. For finite motions at $\rho \to
0, \;\; \sigma (0) = - 1$, and the value of the total mass is
determined by the value of the turning point $\rho_0$ and the value
of $\sigma$ there. It should be noted here that, given the shell
parameters and initial values for $\rho(0)$ and $\dot\rho(0)$, we
are not able to continue unambiguously the solutions beyond the
Cauchy horizons present in the complete Reissner--Nordstr\"om black
hole space-times. To avoid any inconvenience we decided to do this
using equations of motion for our shells and demanding $\rho(\tau)$
and $\dot\rho(\tau)$ to be continuous fanctions.

We start our classification by specifying the relation between the
shell parameters, $M$ and $e$, and then determine the trajectories
of the shells and corresponding global geometries depending on the
values of the total mass $m$. Evidently, we should distinguish two
types of the shells, that ones with $G M^2 < e^2$ and with $e^2 < G
M^2$.

\begin{figure}
\begin{center}
\includegraphics[angle=0,width=0.8\textwidth]{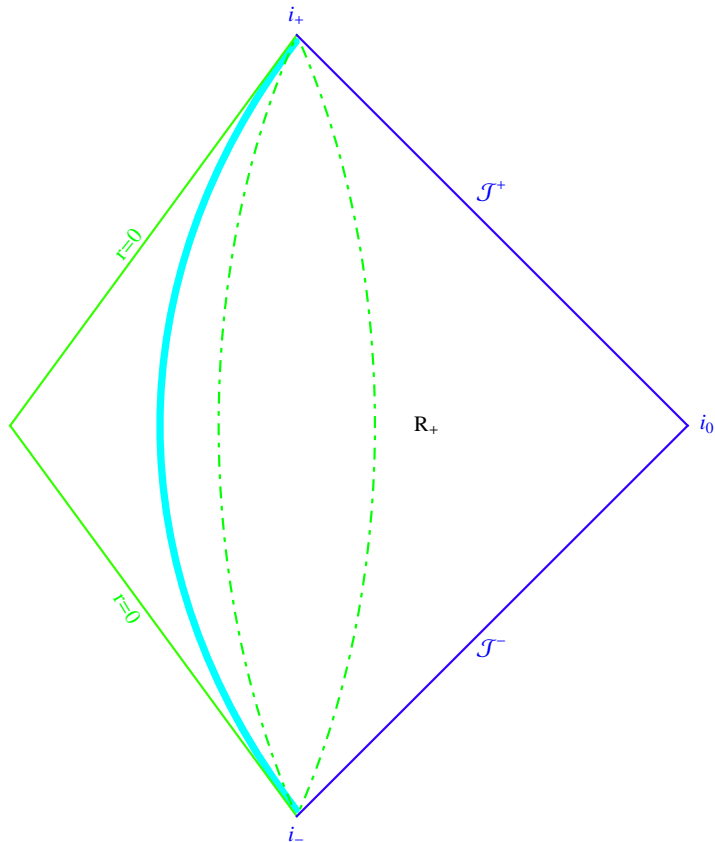}
\end{center}
\caption{Case I-2: Infinite motion with a turning point, $G M^2 < G m^2 < e^2$.}
 \label{Minkowskinfinite1}
\end{figure}
\subsection{Type I: $\;\;G M^2 < e^2$}

There are three different cases, depending on where we insert the
total mass $m$ into the above inequality: to the left, in the
middle, or to the right.
\begin{figure}
\begin{center}
\includegraphics[angle=0,width=0.55\textwidth]{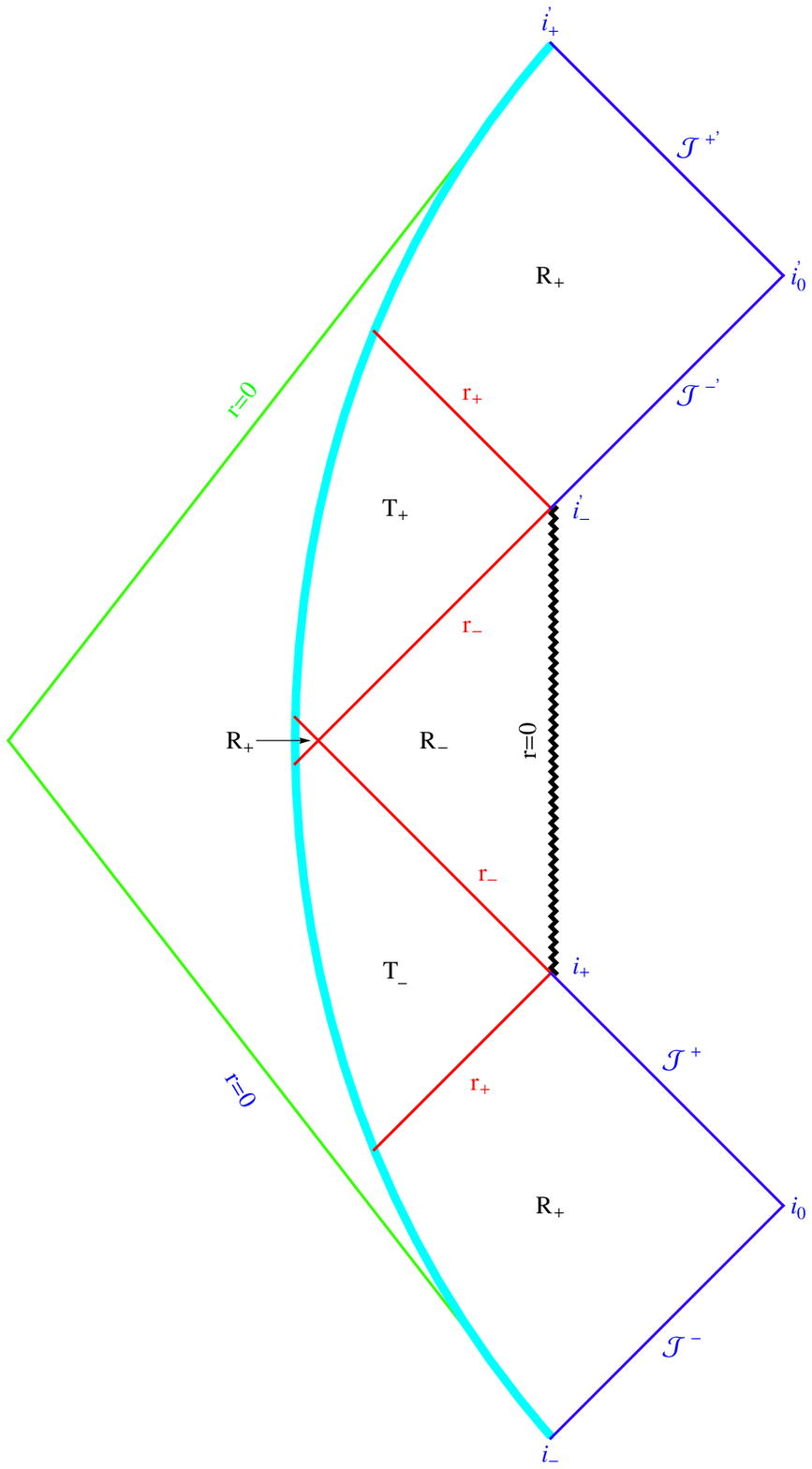}
\end{center}
\caption{Subcase I-3(a): On the shell trajectory $\sigma_{out}$
does not change its sign.} \label{MinkowskinfiniteFig7}
\end{figure}

\subsubsection {I-1: $\;\; G m^2 < G M^2 < e^2$}

Finite motion with no turning point, what is impossible.

\subsubsection {I-2: $\;\; G M^2 < G m^2 < e^2$}

Infinite motion with a turning point. Outside we have the
Reissner--Nordstr\"om metric with naked singularity. But, since
$\rho_0 < r < \infty$, this singularity is "hidden" inside the
shell where, instead, the space-time is flat. The Carter--Penrose
diagram looks as follows in Fig.~\ref{Minkowskinfinite1}.
\begin{figure}[t]
\begin{center}
\includegraphics[angle=0,width=0.4\textwidth]{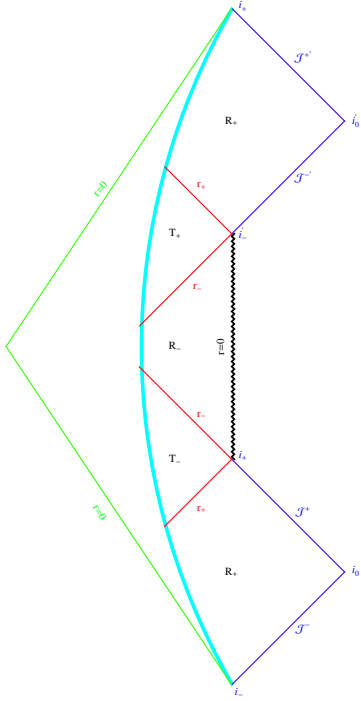}
\end{center}
\caption{Subcase I - 3(b): $\sigma_{out}$ changes its sign, and
the turning point lies in $R_-$-region, $0 < \rho_0 < r_-$.}
\label{Minkowskinfinite5}
\end{figure}
\begin{figure}
\begin{center}
\includegraphics[angle=0,width=0.5\textwidth]{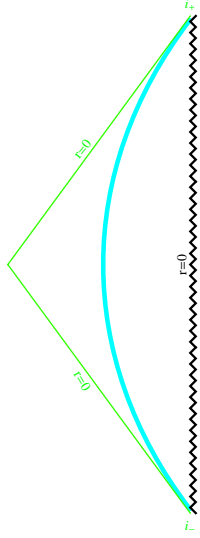}
\end{center}
\caption{Subcase II-1: Finite motion with turning point, $G m^2 < e^2 < G M^2$.}
 \label{MinkowskifiniteFig9}
\end{figure}

\subsubsection {I-3: $\;\; G M^2 < e^2 < G m^2$}

Infinite motion with turning point. The exterior metric is the
Reissner--Nordstr\"om black hole with two horizons at $r_{\pm} = G m
\pm \sqrt{G^2 m^2 - G e^2}$. A turning point should lie in
$R$-region, but it could happen outside the event horizon, $\rho_0 >
r_+$, as well as inside the inner horizon, $\rho_0 < r_-$. It is not
difficult to show that the first of the inequalities is equivalent
to $(r_+ - G M)^2 < 0$ what is impossible. Since for $0 < r < r_-$
there exist two $R$-regions, $R_+$ and $R_-$, and at infinity
$\sigma_{out} = + 1$, we have, accordingly, two subcases.

Subcase I-3(a):

On the shell trajectory $\sigma_{out}$ does not change its sign,
consequently,
\begin{equation}
\label{suc1}
\frac{m}{M} < \frac{1}{2} + \frac{e^2}{2 G M^2} \, ,
\end{equation}
the turning point lies in $R_+$-region, $0 < \rho_0 < r_-$, and
the Carter--Penrose diagram looks as follows in
Fig.~\ref{MinkowskinfiniteFig7}.

Subcase I-3(b):

$\sigma_{out}$ changes its sign, and the turning point lies in
$R_-$-region, $0 < \rho_0 < r_-$,
\begin{equation}
\label{sc1}
\frac{m}{M} > \frac{1}{2} + \frac{e^2}{2 G M^2} \, ,
\end{equation}
The conformal diagram is shown in
Fig.~\ref{Minkowskinfinite5}.
\begin{figure}[t]
\begin{center}
\includegraphics[angle=0,width=0.45\textwidth]{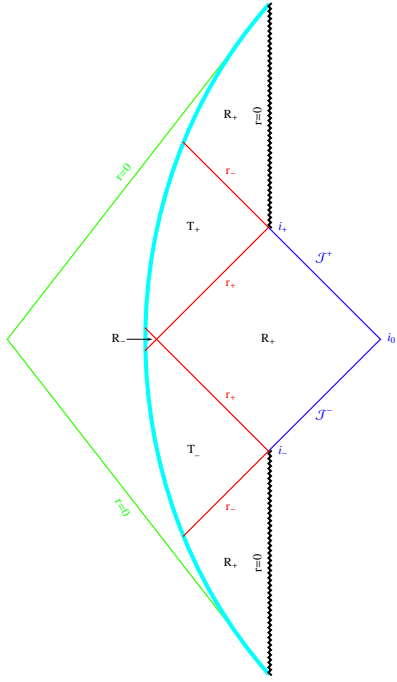}
\end{center}
\caption{Subcase II-2(a): The turning point lies in the
asymptotically flat $R_-$-region on the other side of the
Einstein-Rosen bridge.} \label{MinkowskiFinite3}
\end{figure}

\subsection{Type II: $\;\; e^2 < G M^2$}

Again, we have three different cases depending on the values of
total mass $m$.

\subsubsection {II-1: $\;\; G m^2 < e^2 < G M^2$}

Finite motion with turning point. Outside is the
Reissner--Nordstr\"om geometry with naked singularity, and
$\sigma_{out} = -1$ everywhere. The conformal diagram is very
simple and shown in Fig.~\ref{MinkowskifiniteFig9}.
This diagram is rather curious. The difference between the left
and right parts is only in the nature of the world lines $r = 0$,
the first one is nonsingular, while the other does, and it is
there the electric charge with the opposite sign to that of the
shell's is concentrated. By dashed curves we show the curves of
constant radii, $0 < r < \rho_0$.
\begin{figure}[h]
\begin{center}
\includegraphics[angle=0,width=0.48\textwidth]{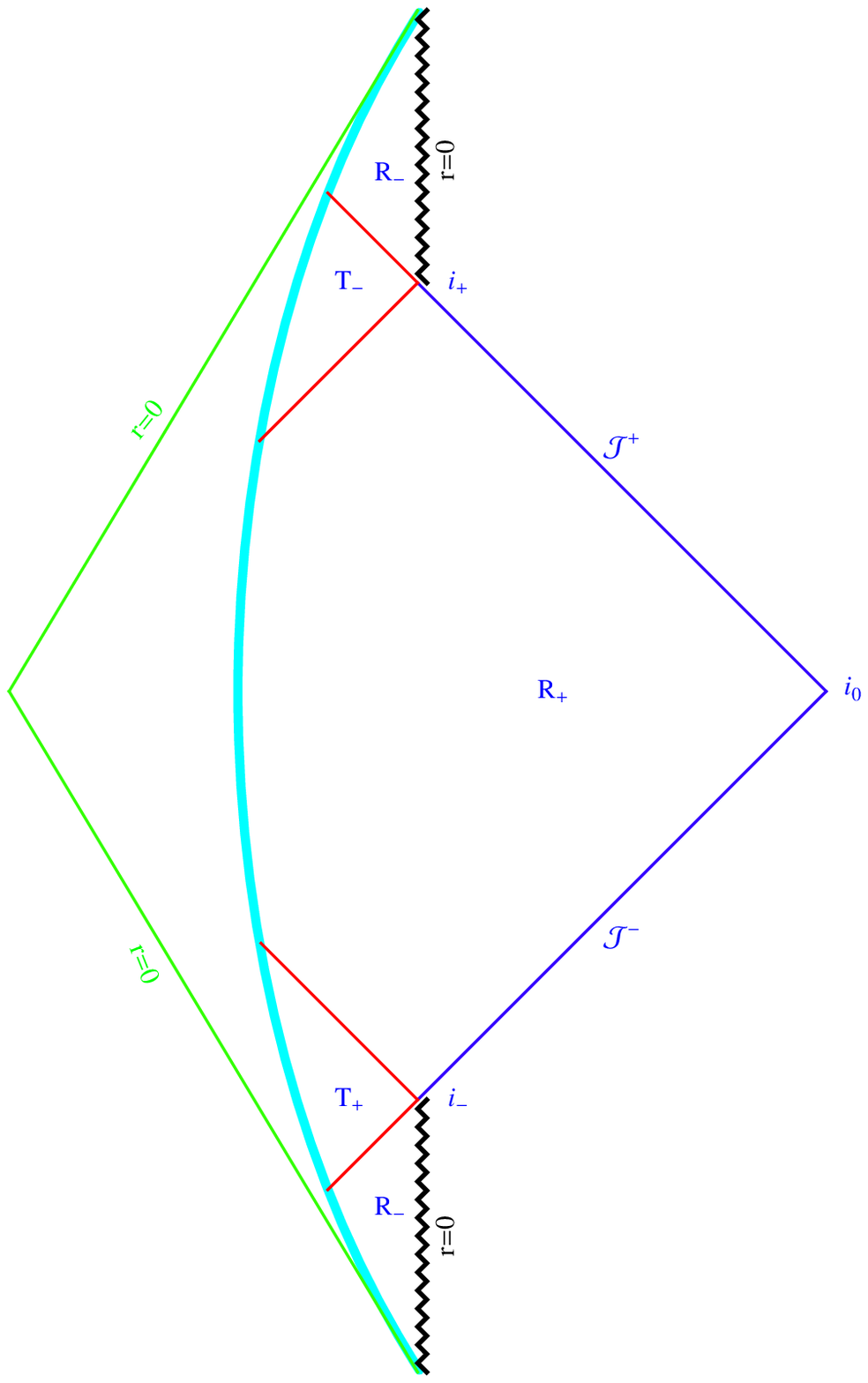}
\end{center}
\caption{Subcase II-2(b): $\sigma_{out}$ changes its sign during
the shell evolution, and the turning point lies in the
asymptotically flat $R_+$-region on "our" side of the
Einstein-Rosen bridge.} \label{MinkowskiFinite4}
\end{figure}
\begin{figure}[h]
\begin{center}
\includegraphics[angle=0,width=0.6\textwidth]{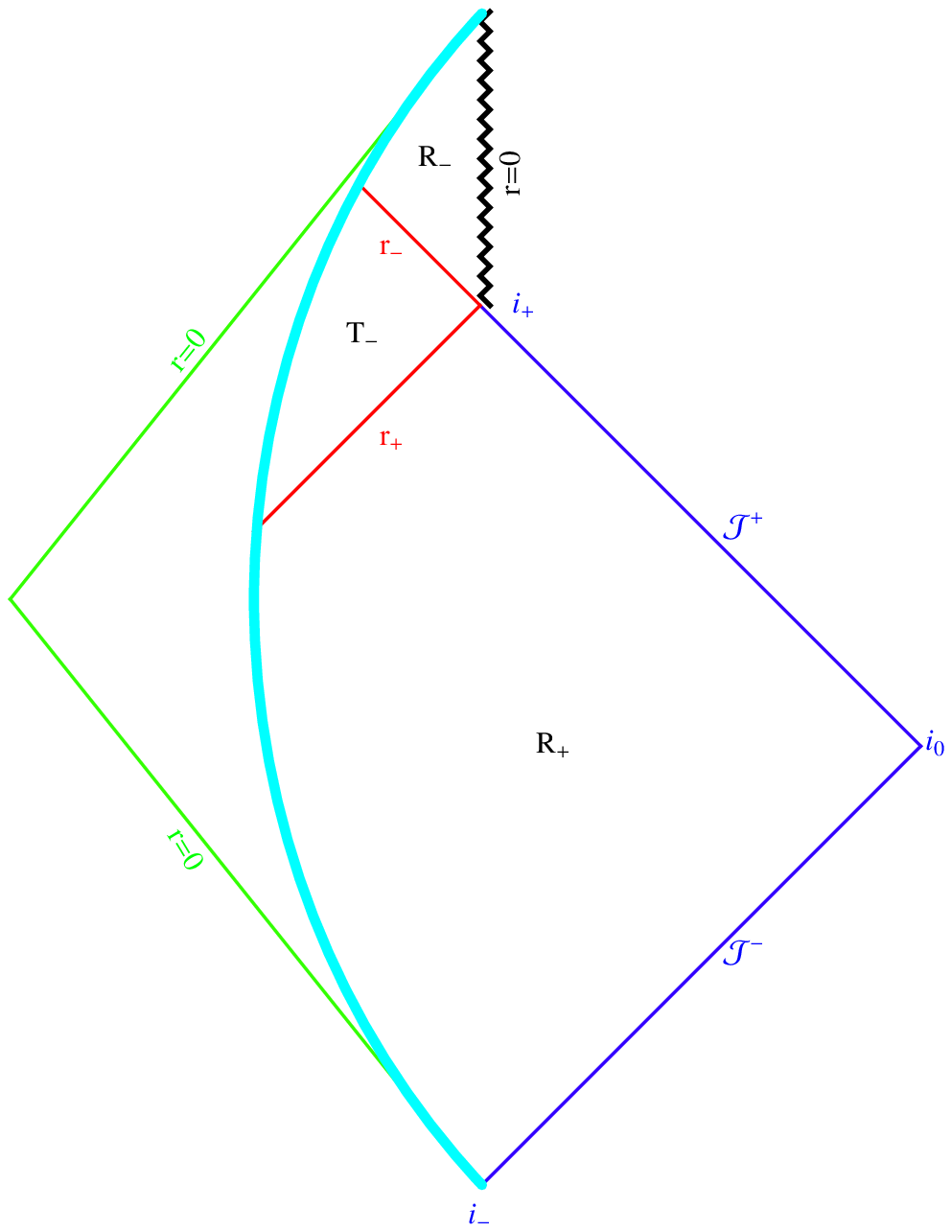}
\end{center}
\caption{Subcase II-3: Infinite motion with no turning point, $e^2 < G M^2 < G m^2$.} \label{Minkowskinfinite2}
\end{figure}

\subsubsection {II-2: $\;\; e^2 < G m^2 < G M^2$}

Finite motion, therefore, $0 \le \rho \le \rho_0$. Again, we have
two possibilities depending on whether $\sigma_{out}$ changes its
sign on the trajectory, or not. But now trajectories start at $\rho
= 0$ where $\sigma_{out} = - 1$, and, as can be shown, the turning
point $\rho_0$ lies outside the event horizon, $\rho_0 > r_+$. Thus,

Subcase II-2(a):
\begin{equation}
\label{suc2}
\frac{m}{M} < \frac{1}{2} + \frac{e^2}{2 G M^2} \, .
\end{equation}
Everywhere on the trajectory $\sigma_{out} = - 1$, so, the turning
point lies in the asymptotically flat $R_-$-region on the other
side of the Einstein-Rosen bridge. The Carter--Penrose diagram
looks as follows in Fig.~\ref{MinkowskiFinite3}.

Subcase II-2(b):
\begin{equation}
\label{sc2}
\frac{m}{M} > \frac{1}{2} + \frac{e^2}{2 G M^2} \, ,
\end{equation}
$\sigma_{out}$ changes its sign during the shell evolution, and
the turning point lies in the asymptotically flat $R_+$-region on
"our" side of the Einstein-Rosen bridge. The conformal diagram is
shown in Fig.~\ref{MinkowskiFinite4}.

\subsubsection {II-3: $\;\; e^2 < G M^2 < G m^2$}

Infinite motion with no turning point. It starts at infinity in
$R_+$-region with $\sigma_{out} = + 1$ and ends at $\rho = 0$ with
$\sigma_{out} = - 1$. Of course, there exists also the reverse
motion. Both of them are shown on Carter--Penrose diagram below in
Fig.~\ref{Minkowskinfinite2}.

\section{Neutralizing shell}

Now we consider the charged shell with the electric charge
opposite in sign to the charge of the internal
Reissner--Nordstr\"om metric with the mass $m_{\rm in}$ and
electric charge $e$. The shell has a bare mass $M$ and charge $-e
$, which neutralizes the charge of the internal source. As a
result, the external metric is the Schwarzschild one with a mass
$m_{\rm tot}=m$.

Now equation for shell dynamics (\ref{shell}) can be written as
\begin{equation}
 \sigma_{\rm in}\sqrt{\dot{\rho}^2+1-\frac{2Gm_{\rm in}}{\rho}+\frac{Ge^2}{\rho^2}}
 -\sigma_{\rm out}\sqrt{\dot{\rho}^2+1-\frac{2Gm}{\rho}}
 =\frac{GM}{\rho},
 \label{shellM3b}
\end{equation}
where the total mass $m_{\rm tot}=m_{\rm out}=m$.

Now both $\sigma_{\rm in}$ and $\sigma_{\rm out}$ can change sign
during shell motion.

\subsubsection{Changing sign in $\sigma_{\rm out}$}

By squaring  (\ref{shellM3b}) to exclude $\sigma_{\rm in}$ we
obtain
\begin{equation}
 m-m_{\rm in}=\Delta m=
 M\sigma_{\rm out}\sqrt{\dot{\rho}^2+1-\frac{2Gm}{\rho}}
 +\frac{GM^2-e^2}{2\rho},
 \label{sigmaout2}
\end{equation}
and so
\begin{equation}
 \sigma_{\rm out}=\mbox{sign}\left(\Delta m-\frac{GM^2-e^2}{2\rho}\right).
 \label{sigmaout2b}
\end{equation}\
In the limiting cases we have $\sigma_{\rm out}(\infty)=\Delta m$
and $\sigma_{\rm out}(0)=-\mbox{sign}(GM^2-e^2)$. From
(\ref{sigmaout2b}) we find now the radius of changing sign for
$\sigma_{\rm out}$:
\begin{equation}
 \rho_\sigma=\frac{GM^2-e^2}{2\Delta m}.
 \label{rhosigma2}
\end{equation}
The radius $\rho_\sigma$ can change sign only in the $T$-regions
of the Schwarzschild metric, i.\,e.
\begin{equation}
 0<\rho_\sigma<2Gm.
 \label{rhosigmaT}
\end{equation}
From the left inequality in (\ref{rhosigmaT}) it follows that \
$\mbox{sign}[(GM^2-e^2)/\Delta m]=+1$ only if \
$\mbox{sign}(GM^2-e^2)=\mbox{sign}(\Delta m)$, and so $\sigma_{\rm
out}(0)=-\sigma_{\rm out}(\infty)$. From the right inequality in
(\ref{rhosigmaT}) it follows that
\begin{equation}
\frac{GM^2-e^2}{\Delta m}<4Gm.
 \label{right}
\end{equation}
If $\sigma_{\rm out}(\infty)+1$, then we have relations $\Delta
m>0$ and $GM^2-e^2>0$. Inequality (\ref{rhosigmaT}) now can be
written as
\begin{equation}
GM^2-e^2<4Gm\Delta m=4Gm(\Delta m)^2+4 Gm_{\rm in}\Delta m.
 \label{right2}
\end{equation}
or
\begin{equation}
4Gm(\Delta m)^2+4 Gm_{\rm in}\Delta m-(GM^2-e^2)>0.
 \label{right3}
\end{equation}
For the r.h.s. there are one negative and one positive root. For a
positive root we obtain:
\begin{equation}
 \Delta m_\sigma
 =\frac{m_{\rm in}}{2}\left(\sqrt{1+\frac{GM^2-e^2}{Gm_{\rm in}^2}}-1\right).
 \label{Deltamsigma}
\end{equation}
It can be seen that $\Delta m>\Delta m_\sigma$.

In the opposite case $\sigma_{\rm out}(\infty)=+1$ we have
relations $\Delta m<0$ and so $GM^2-e^2<0$. Inequality
(\ref{rhosigmaT}) now can be written as $GM^2-e^2>4Gm\Delta m$ or
\begin{equation}
4Gm(\Delta m)^2+4 Gm_{\rm in}\Delta m+(e^2-GM^2)<0.
 \label{right4}
\end{equation}
Now for the r.h.s. there are two negative roots or the roots are
absent at all:
\begin{equation}
 \Delta m_{\rm cr}
 =\frac{m_{\rm in}}{2}\left(-1\pm\sqrt{1-\frac{e^2-GM^2}{Gm_{\rm in}^2}}\right).
 \label{Deltamsigma2}
\end{equation}
From (\ref{right4}) we see that condition for the existence of two
roots is
\begin{equation}
Gm_{\rm in}^2>e^2-GM^2.
 \label{right5}
\end{equation}
Now we consider in details the behavior of $\sigma_{\rm in}$

\subsubsection{Changing sign in $\sigma_{\rm in}$}

Analogously, by squaring  (\ref{shellM3b}) to exclude $\sigma_{\rm
out}$ we obtain
\begin{equation}
 M\sigma_{\rm in}\sqrt{\dot{\rho}^2+1-\frac{2Gm_{\rm in}}{\rho}+\frac{Ge^2}{\rho^2}}
 =\Delta m+\frac{GM^2+e^2}{2\rho},
 \label{sigmain2}
\end{equation}
and so
\begin{equation}
 \sigma_{\rm in}=\mbox{sign}\left(\Delta m+\frac{GM^2+e^2}{2\rho}\right).
 \label{sigmain2b}
\end{equation}
In the limiting cases we have $\sigma_{\rm
in}(\infty)=\mbox{sign}(\Delta m)=\sigma_{\rm out}(\infty)$ and
$\sigma_{\rm in}(0)=+1$. From (\ref{sigmain2b}) we find now the
radius of changing sign for $\sigma_{\rm in}$:
\begin{equation}
 \rho_\sigma({\rm in})=-\frac{GM^2+e^2}{2\Delta m}.
 \label{rhosigmain}
\end{equation}
Now there are possible the both cases, $\Delta m\gtrless0$.

If $\Delta m>0$, then $\sigma_{\rm in}=+1$ everywhere. In
particular, $\sigma_{\rm in}(0)=+1$ at $\rho\to0$.

The change of sign in $\sigma_{\rm in}=+1$ is possible only if
$\Delta m<0$. This point must be in the $T$-region of the
Reissner--Nordstr\"om metric, and so
\begin{equation}
 r_-<\rho_\sigma({\rm in})<r_+,
 \label{rhosigmain2}
\end{equation}
where $r_\pm=Gm_{\rm in}\pm\sqrt{G^2m_{\rm in}^2-Ge^2}$.
Inequality (\ref{sigmain2b}) may be written also as
\begin{equation}
 r_-<\rho_\sigma({\rm in})<r_+,
 \label{rhosigmain2a}
\end{equation}
These relations may rewritten as *** see page 5 in draft ??? ***
\begin{equation}
 r_-<\frac{e^2(\Delta m)}{GM^2m_{\rm in}}<r_+,
 \label{rhosigmain2b}
\end{equation}

\subsection{Turning points $\dot\rho^2=0$}

From (\ref{sigmaout2}) we have an equation for the turning point $\rho=\rho_0$:
\begin{equation}
\left(\Delta m-\frac{GM^2-e^2}{2\rho_0}\right)^2=
 M\left(1-\frac{2Gm}{\rho_0}\right)
 \label{turning}
\end{equation}
By introducing a new variable $x=(2M\rho_0)^{-1}$, we write equation
$\dot{\rho}^2=0$ as
\begin{equation}
 (GM^2-e^2)^2x^2+2x\left[\frac{\Delta m}{M}e^2+GM(m+m_{\rm in})\right]
 +\frac{(\Delta m)^2}{M^2}-1=0
 \label{turningx}
\end{equation}
with roots
\begin{equation}
 x_\pm=\frac{1}{(GM^2-e^2)^2}
 \left\{-B\pm
 \sqrt{B^2-(GM^2-e^2)^2\left[\frac{(\Delta m)^2}{M^2}-1\right]}\right\},
 \label{roots}
\end{equation}
where
\begin{equation}
 B=\left[\frac{\Delta m}{M}e^2+GM(m+m_{\rm in})\right].
 \label{B}
\end{equation}
Finite motion --- there are always two real root. Infinite motion ---
there are no real roots at all or there are two negative roots.

\subsection{Infinite motion with $\rho\to\infty$}

\begin{figure}
\begin{center}
\includegraphics[angle=0,width=0.6\textwidth]{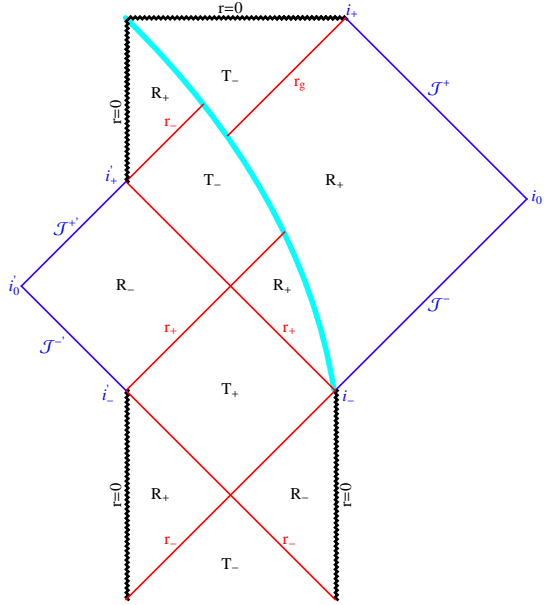}
\end{center}
\caption{Infinite motion $\rho\to\infty$: black hole case} \label{InfiniteI}
\end{figure}

We start our consideration with the case of infinite motion, when
$\rho\to\infty$. In infinite motion $\sigma_{\rm
out}(\infty)=\pm1$ ($R_+$-region). From $\Delta
m=M\sqrt{\dot\rho^2+1}$ at $\rho\to\infty$ it follows that $\Delta
m>M>0$. From (\ref{rhosigma2}) it follows that $\rho_\sigma(in)<0$
and does not exist. Therefore, $\sigma_{\rm in}=+1$ for infinite
motion. See Fig.~\ref{InfiniteI}.

\begin{figure}
\begin{center}
\includegraphics[angle=0,width=0.4\textwidth]{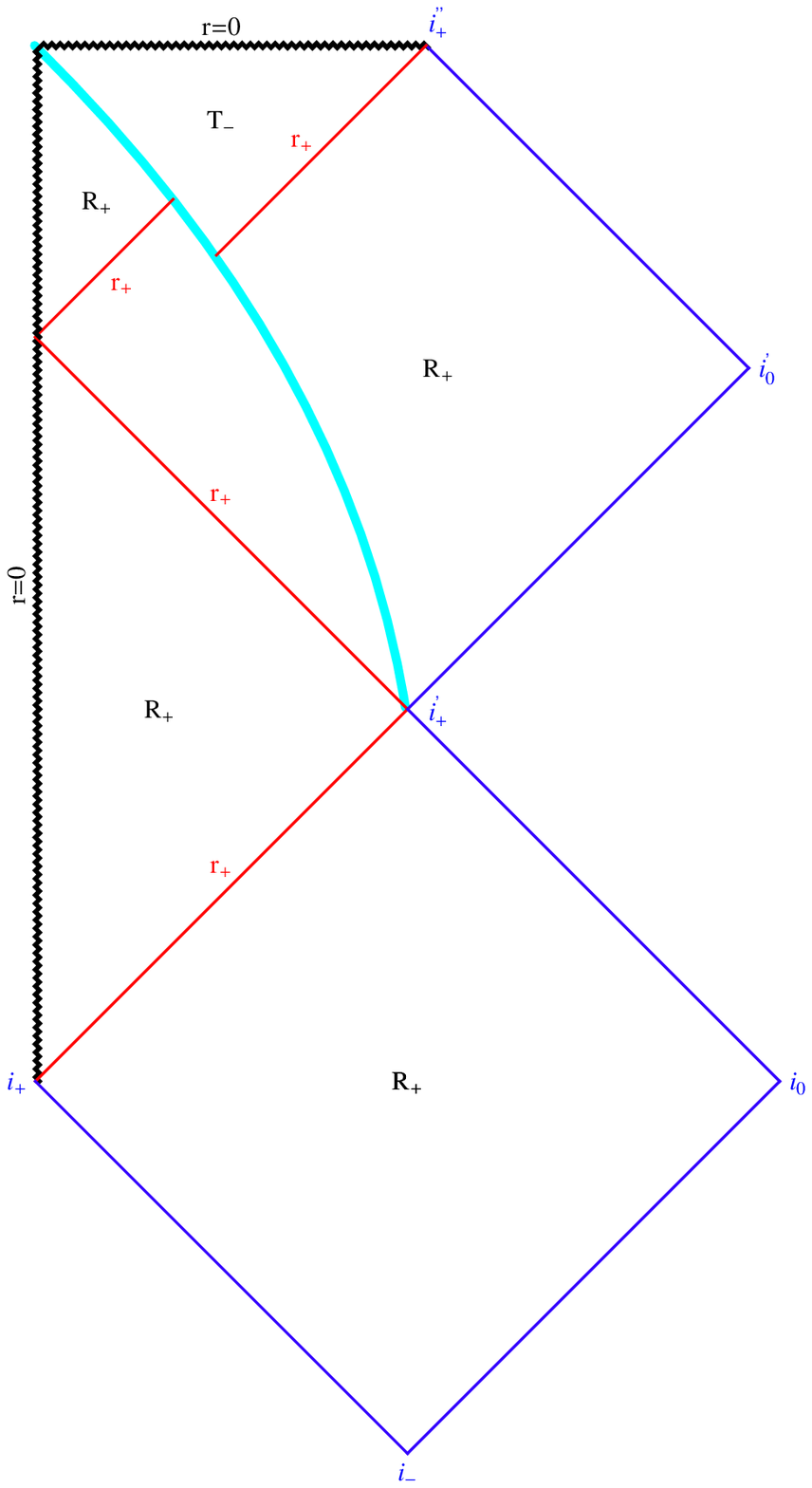}
\hfill
\includegraphics[angle=0,width=0.4\textwidth]{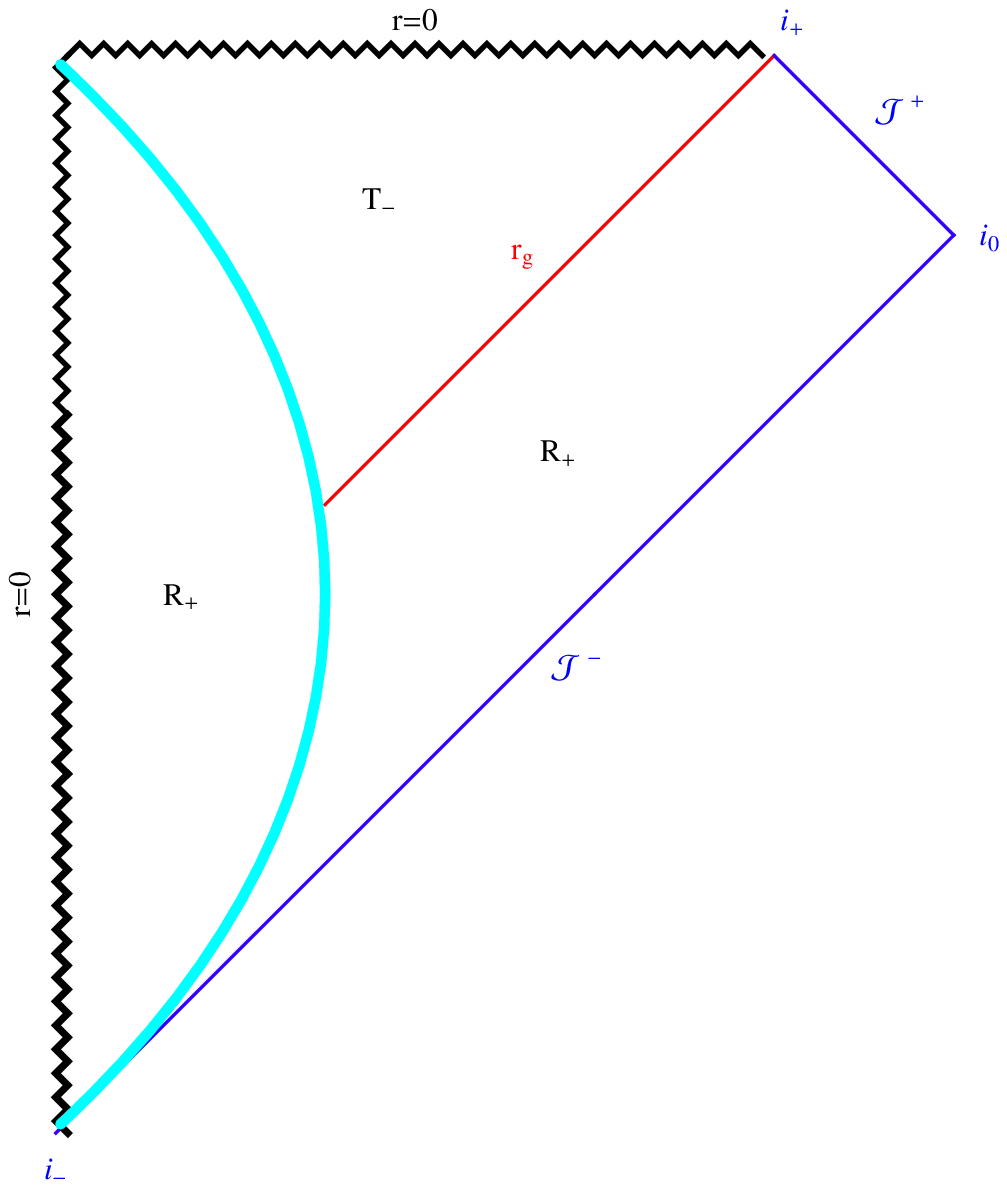}
\end{center}
\caption{Infinite motion $\rho\to\infty$: cases of extreme black hole and naked singularity.}
\label{InfiniteNS}
\end{figure}

\subsubsection{The case $\sigma_{\rm out}(\infty)={\rm sign}(\Delta m)=+1$}
\label{infiniteII} 
\begin{figure}
\begin{center}
\includegraphics[angle=0,width=0.48\textwidth]{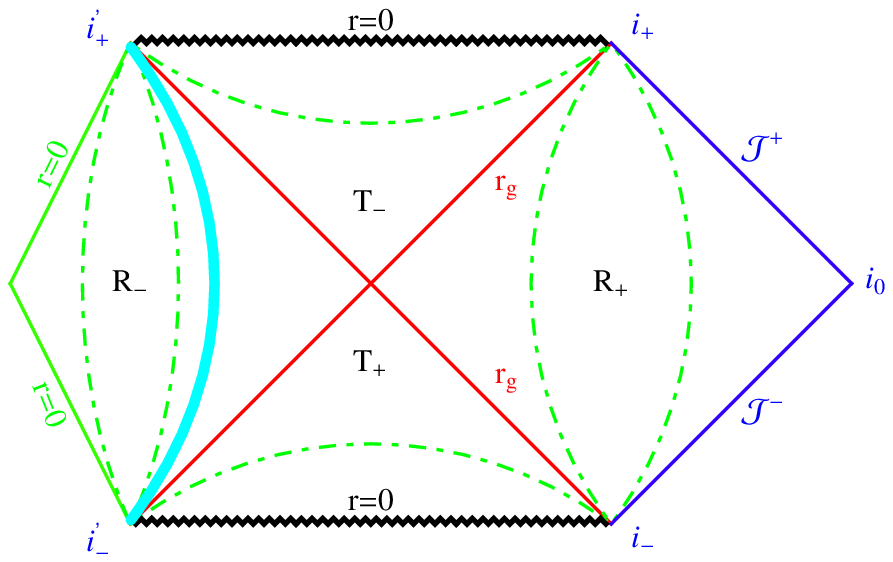}
\hfill
\includegraphics[angle=0,width=0.48\textwidth]{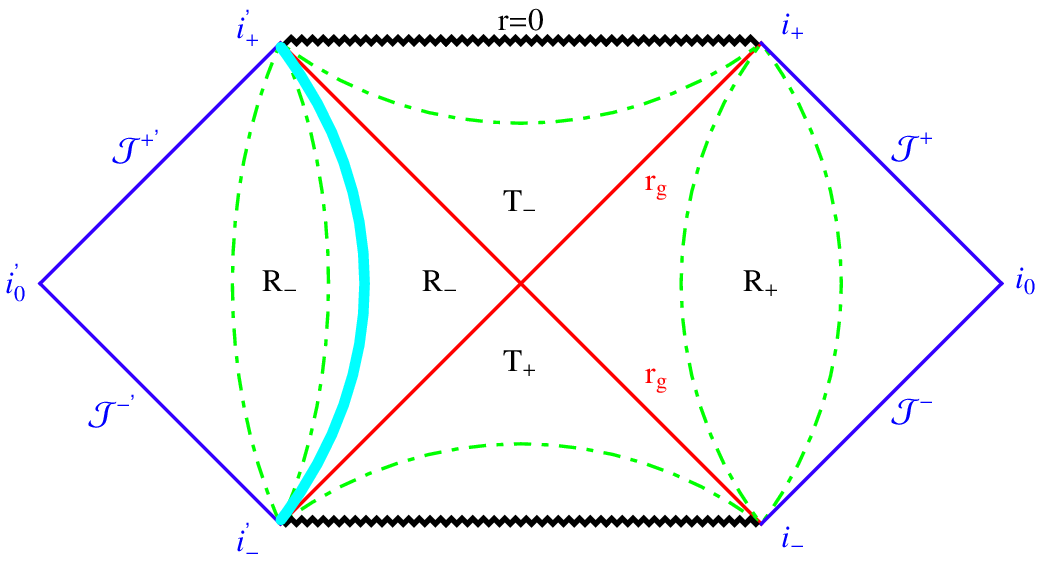}
\end{center}
\caption{Infinite motion (Case \ref{infiniteII}).}
\label{InfiniteIIa}
\end{figure}

\begin{figure}
\begin{center}
\includegraphics[angle=0,width=0.6\textwidth]{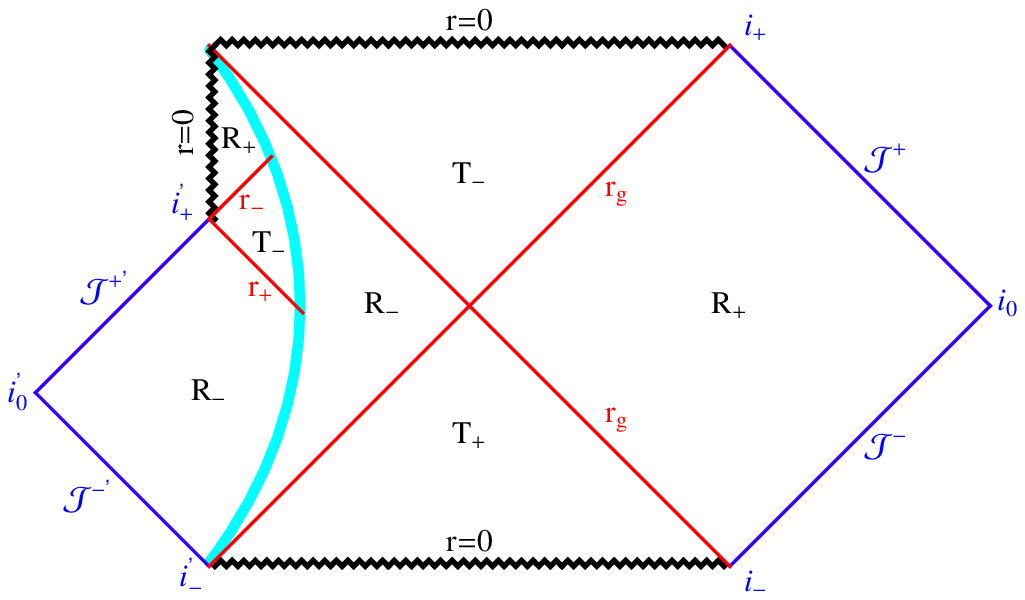}
\end{center}
\caption{Infinite motion (Case \ref{infiniteII}).}
\label{InfiniteIIb}
\end{figure}

In this case $\Delta m>0$ and so $\sigma_{\rm
in}=1$ everywhere, $\Delta m>M$ and $\sigma_{\rm out}$ can change
its sign if $GM^2>e^2$. From the equation for turning points it
follows that both roots are negative, i\,e. $0\leq\rho<\infty$.
This result is a rather evident because in the considered case
both gravitation and Coulomb force are attractive.  See
Figs.~\ref{InfiniteIIa} and \ref{InfiniteIIb}.

\subsubsection{The case $\sigma_{\rm out}(\infty)=-1$ with turning point}
\label{sigmaoutminus1}

In this case $\Delta m<0$, $(\Delta m)^2>M^2$, $\sigma_{\rm
in}(\infty)=-1$, $\sigma_{\rm in}(0)=+1$. If the shell is falling
into the $T$-region in the Schwarzschild metric (``out''), then
the infall to the central singularity is inevitable. This means
that the point of changing sign $\rho_\sigma(in)$ is always exists
and is on the shell trajectory.

\begin{figure}
\begin{center}
\includegraphics[angle=0,width=0.48\textwidth]{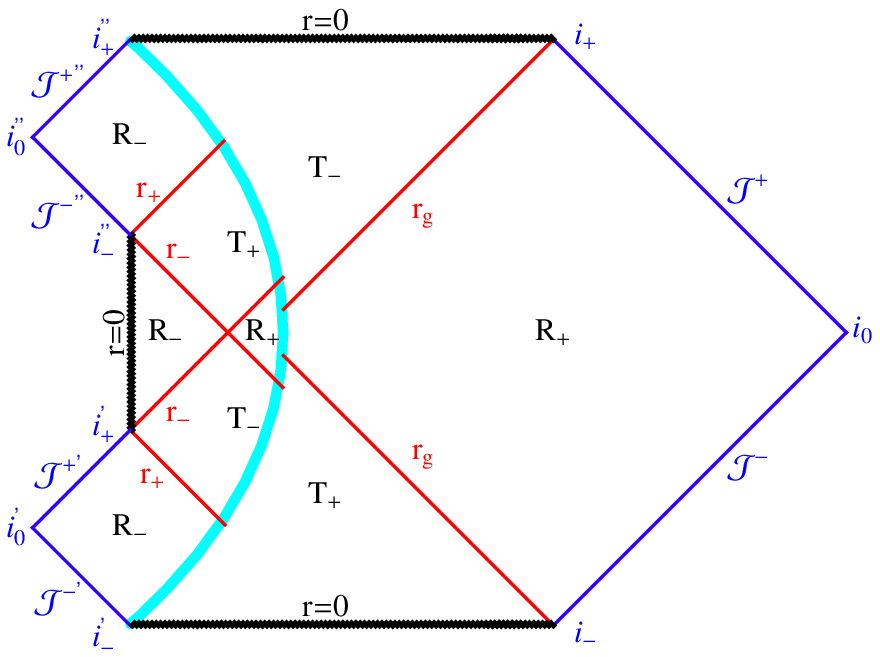}
\hfill
\includegraphics[angle=0,width=0.48\textwidth]{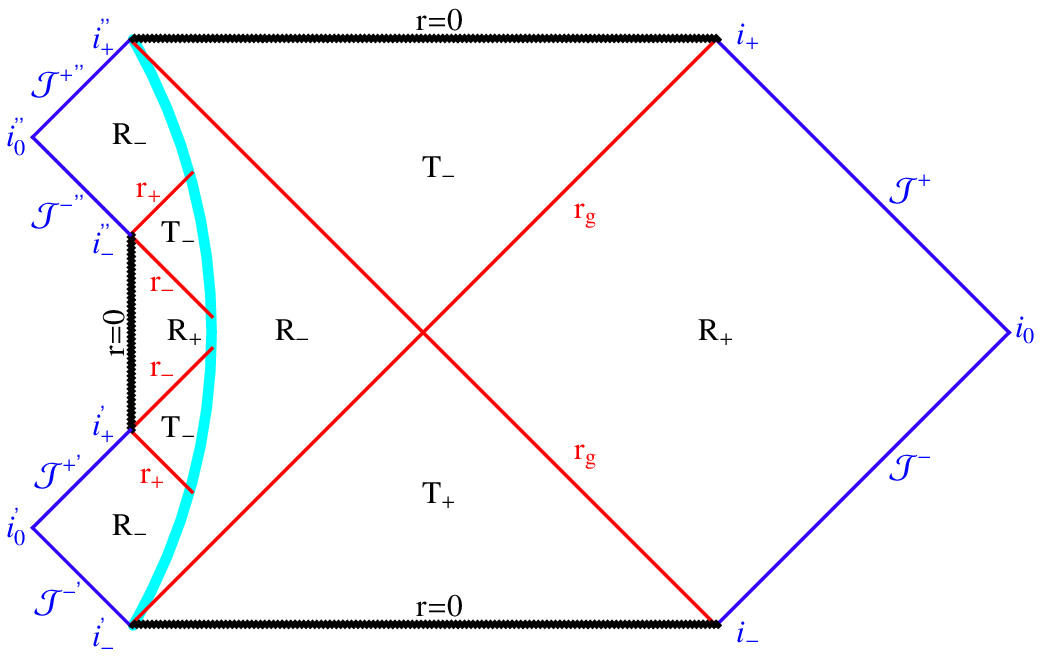}
\end{center}
\caption{Infinite motion with turning point (Case
\ref{sigmaoutminus1}).} \label{InfinRplusminus}
\end{figure}

There is also a possibility that a turning point $\rho_0$ exists
and is in the $R$-region of both internal and external metrics,
and so $\rho_0>r_g>r_+$. This means that a turning point is in the
$R_-$-regions in the external (``out'') and internal (``in'')
metrics. Respectively in this case the both
 $\sigma_{\rm in}$ and  $\sigma_{\rm out}$ do not change signs.
See Figs.~\ref{InfinRplusminus}

\subsubsection{The case $\sigma_{\rm out}(\infty)=-1$ without turning point}
\label{noturning}

\begin{figure}
\begin{center}
\includegraphics[angle=0,width=0.48\textwidth]{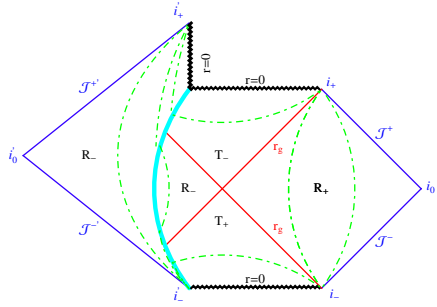}
\end{center}
\caption{Case \ref{noturning} with no turning point.}
\label{Page11RN}
\end{figure}

This corresponds to $B=0$ in (\ref{B}) or
\begin{equation}
\frac{\Delta m}{M}+GM(m+m_{\rm in})>0
 \label{turning2}
\end{equation}
and
\begin{equation}
\frac{\Delta m}{M}(e^2+GM^2)+2GMm_{\rm in}>0.
 \label{turning3}
\end{equation}
From these relations  it follows
\begin{equation}
-\Delta m<2m_{\rm in}\left(1+\frac{e^2}{GM^2}\right)^{-1}.
 \label{turning4}
\end{equation}
In this case there are two negative roots for turning point:
\begin{equation}
 \left[\frac{\Delta m}{M}(e^2+GM^2)+2GMm_{\rm in}\right]^2<
 (GM-e^2)^2\left[\frac{(\Delta m)^2}{M^2}-1\right].
 \label{turning5}
\end{equation}
That is there are no turning point. The last inequality may be
expressed as
\begin{equation}
 \frac{(\Delta m)^2}{M^2}+\frac{\Delta m}{M}\frac{m_{\rm in}}{M}\frac{e^2+GM^2}{e^2}
 +\frac{Gm_{\rm in}^2}{M^2e^2}+\frac{(GM^2-e^2)^2}{4GM^2e^2}<.
 \label{turning6}
\end{equation}
$e^2<Gm_{\rm in}^2$. See Fig.~\ref{Page11RN}.

\subsubsection{The case $\sigma_{\rm out}(\infty)=-1$ with turning point}

Now we have
\begin{equation}
\frac{\Delta m}{M}+GM(m+m_{\rm in})<0
 \label{turning2b}
\end{equation}
and
\begin{equation}
\frac{\Delta m}{M}(e^2+GM^2)+2GMm_{\rm in}<0.
 \label{turning3b}
\end{equation}
From these relations  it follows
\begin{equation}
-\Delta m>2m_{\rm in}\left(1+\frac{e^2}{GM^2}\right)^{-1}.
 \label{turning4a}
\end{equation}
There are two roots for turning point
\begin{equation}
 \left[\frac{\Delta m}{M}(e^2+GM^2)+2GMm_{\rm in}\right]^2
 >(GM-e^2)^2\left[\frac{(\Delta m)^2}{M^2}-1\right].
 \label{turning5a}
\end{equation}
The roots with
\begin{equation}
 \Delta m-\frac{GM^2-e^2}{2\rho_0}\gtrless0.
 \label{root1}
\end{equation}
correspond respectively to $\sigma_{\rm in}(\rho_0)=\pm1$.

\subsection{Infinite motion starting in $R_-$-region}

It must be mentioned also the specific case of infinite motion
starting in the $R_-$-region, where $\sigma_{\rm out}(\infty)=-1$,
with a turning point in the $R_+(in)$-region, where $\sigma_{\rm
in}(\rho_0)=+1$. This type of motion is realized if
$GM^2<e^2<Gm_{\rm in}^2$ and so $M<m_{\rm in}$.

\subsection{Finite motion with $\rho<\infty$}

$|\Delta m/M|<1$.

\subsubsection{The case $\sigma_{\rm out}(\rho_0)=+1$}
\label{FiniteRN1}

In this case $\Delta m>0$.

If $GM^2>e^2$, then $\sigma_{\rm out}(0)=-1$, From existence of
$\rho_\sigma$ it follows $\Delta m>0$.

\begin{figure}
\begin{center}
\includegraphics[angle=0,width=0.48\textwidth]{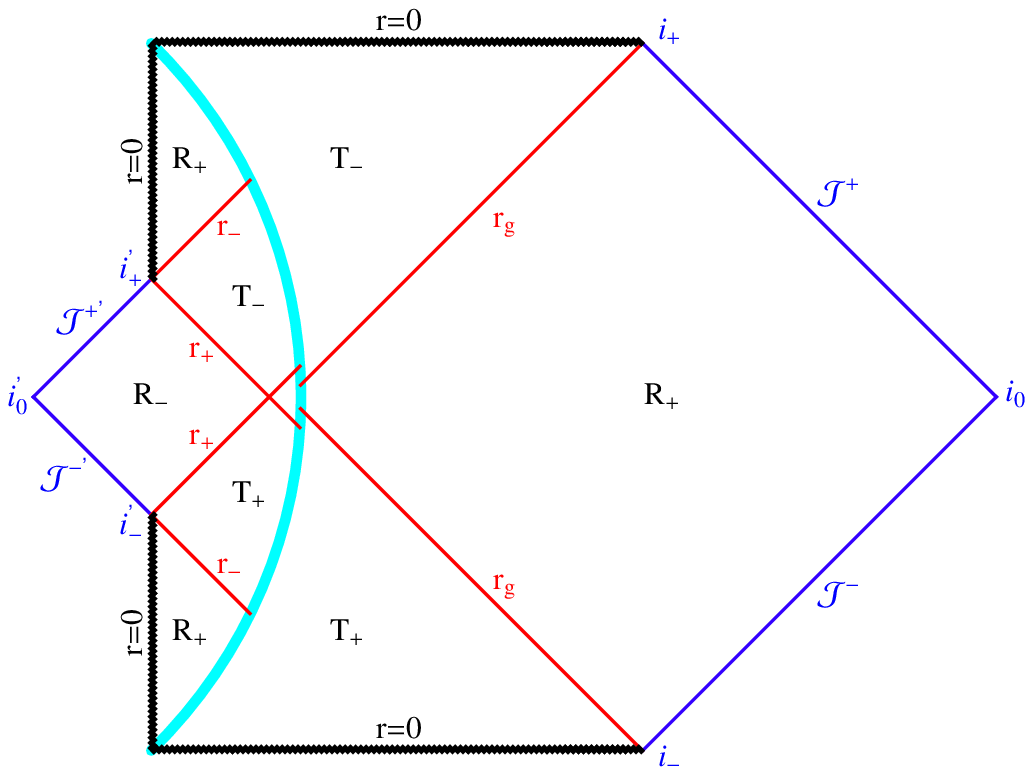}
\hfill
\includegraphics[angle=0,width=0.48\textwidth]{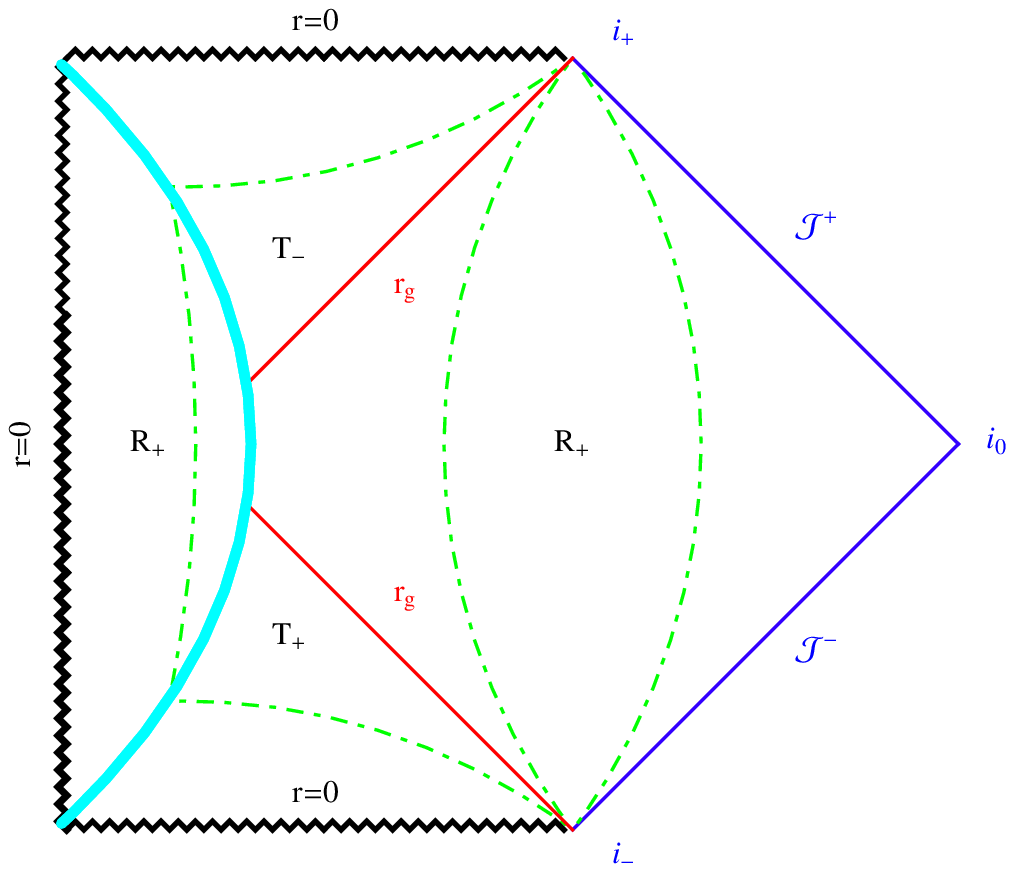}
\end{center}
\caption{The Case \ref{FiniteRN1} with a finite motion and $\Delta
m>0$.} \label{FiniteRNa}
\end{figure}

If $GM^2<e^2$, then $\sigma_{\rm out}(0)=+1$, $\rho_\sigma$ does
not exist and $\Delta m>0$.
\begin{equation}
 \Delta m-\frac{GM^2-e^2}{2\rho_0}>0.
 \label{root1c}
\end{equation}
$\sigma_{\rm in}=+1$. See Fig.~\ref{FiniteRNa}

\subsubsection{The case $\sigma_{\rm out}(\rho_0)=-1$}
\label{FiniteRN2}

\begin{equation}
 \Delta m-\frac{GM^2-e^2}{2\rho_0}<0.
 \label{root1d}
\end{equation}
The case $\sigma_{\rm out}(0)=-1$ corresponds to $GM^2>e^2$. There
are possible the both cases $\Delta m\gtrless0$. In this case
$\rho_\sigma>\rho_0$ and there is one root.

\begin{figure}
\begin{center}
\includegraphics[angle=0,width=0.48\textwidth]{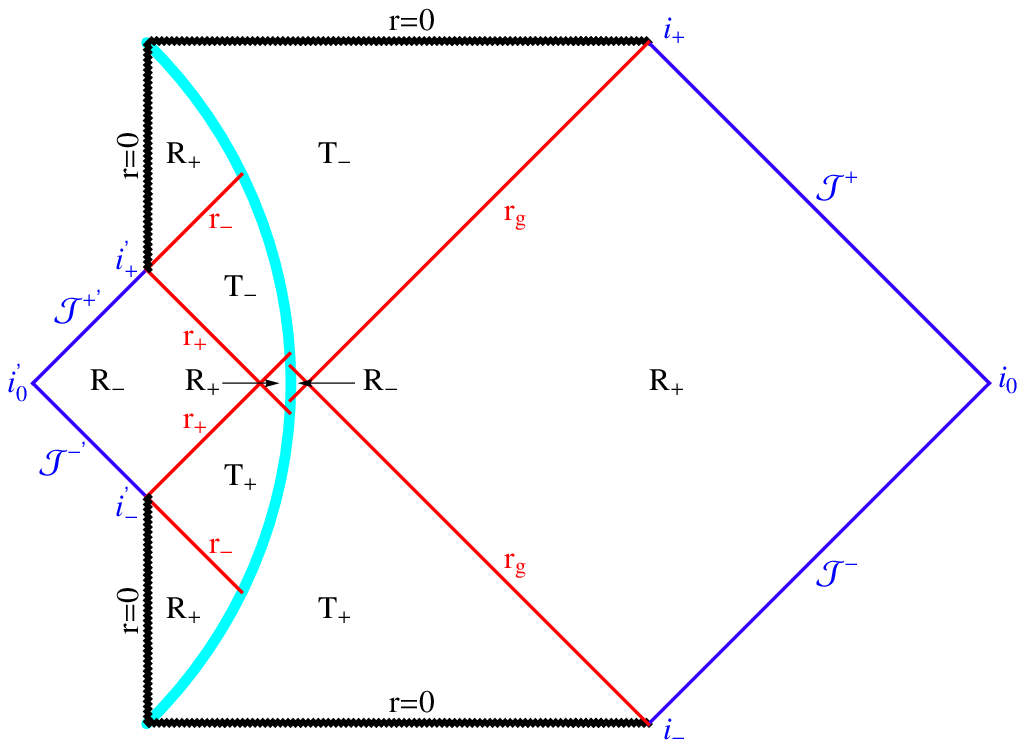}
\hfill
\includegraphics[angle=0,width=0.48\textwidth]{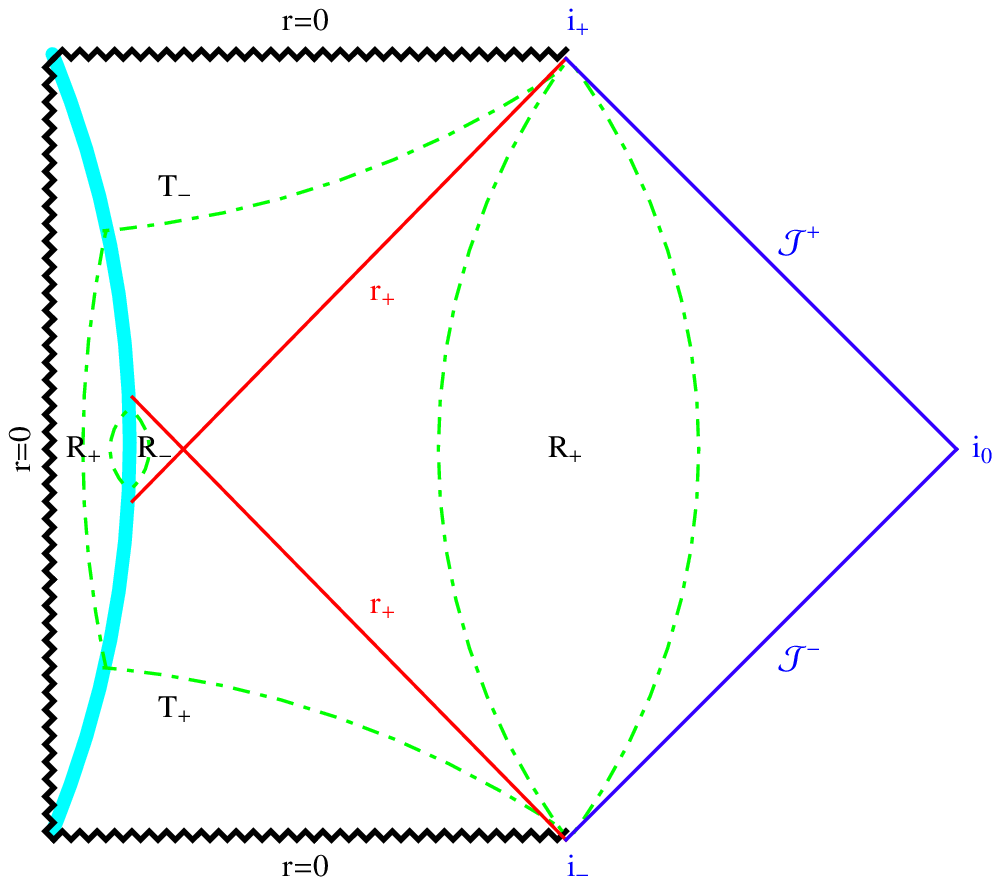}
\end{center}
\caption{The Case \ref{FiniteRN2} with a finite motion and $\Delta
m>0$.} \label{FiniteRNb}
\end{figure}

The case $\sigma_{\rm out}(0)=+1$ corresponds to $GM^2<e^2$ and so
$\Delta m<0$. See Fig.~\ref{FiniteRNb}
\begin{figure}
\begin{center}
\includegraphics[angle=0,width=0.48\textwidth]{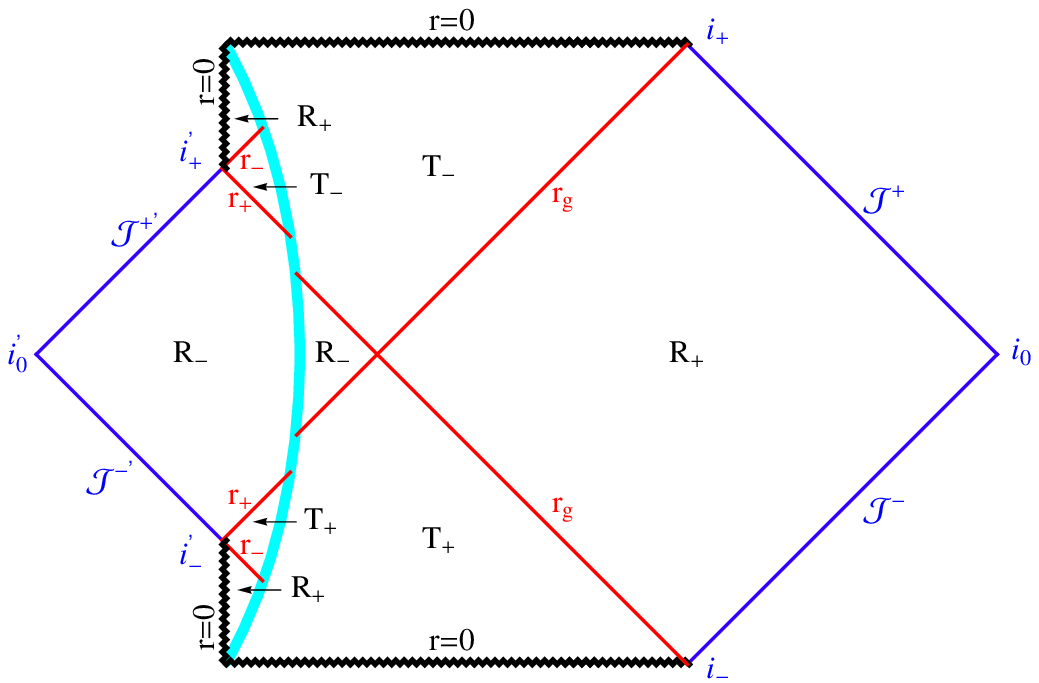}
\hfill
\includegraphics[angle=0,width=0.48\textwidth]{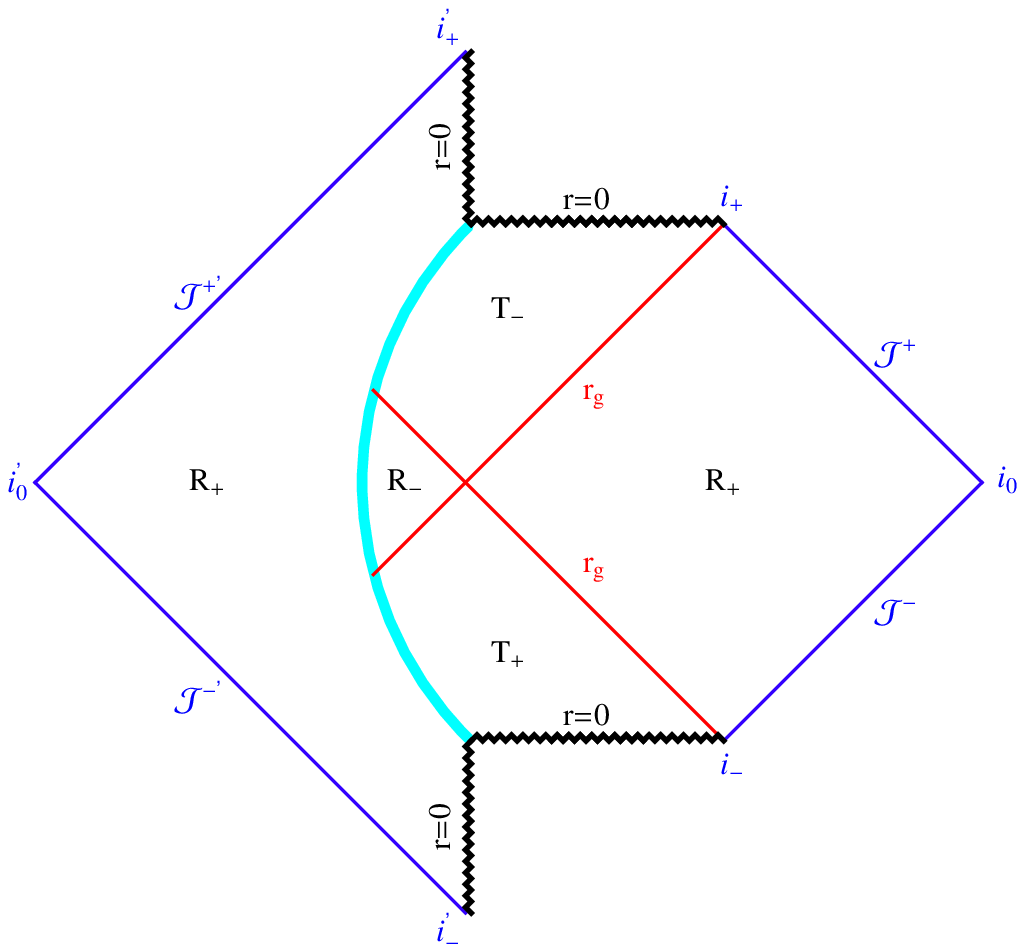}
\end{center}
\caption{The Case of a naked singularity and $\Delta m>0$.}
\label{NakedSingularity}
\end{figure}

\section{Neutralizing shell --- capacitor}

We describe here the possible types of the spherically symmetric global geo\-metri\-es for the moving shell with the electric charge, which is equal and opposite to the corresponding charge of the internal Reissner--Nordstr\"om metric with mass $m_{\rm in}$ and electric charge $e$. This shell has a "naked" mass $M$ and charge $-e$, which is neutralizing the charge of the internal source. As a result, the external metric is the Schwarzschild metric with the mass $m_{\rm tot}=m$.

The complete space-time consists of three parts. The first part --- the internal (in) one is a piece of the Reissner--Nordstr\"om metric, defined by two parameters, the mass $m_{\rm in}>0$ and charge $e$. The source of this metric may be the charged shell, described in the Part~I. Here for simplicity we suppose that the sources of mass and charge are confined in the central singularity at $r=0$.

The second part --- the external (out) one is a piece of the Schwarzschild metric with the mass $m_{\rm tot}>0$. This two parts are separated off each other by the third part --- the thin shell with the a "naked" mass $M$ and a compensating charge $-e$. The Carter--Penrose diagram for the total Schwarzschild space-time is shown in Fig.~\ref{SchwTotal}. The corresponding conformal diagrams for the Reissner--Nordstr\"om metric (see Figs.~\ref{RN}, \ref{RNextremal} and \ref{NS}) depends on the relation of parameters $e^2\gtreqless Gm_{\rm in}^2$. The horizon radii are $r_\pm=Gm_{\rm in}\pm\sqrt{G^2m_{\rm in}^2-Ge^2}$. In the case of the extreme black hole $r_-=r_+=Gm_{\rm in}$.
In the Carter--Penrose diagrams, the part of the Reissner--Nordstr\"om space-time manyfold would be at left from the shell (in), and the corresponding part of the Schwarzschild manyfold would be at right from the shell (out).

\subsection{Equations}

The corresponding W. Israel equation for the dynamics of the neutralizing shell
\begin{equation}
\sigma_{\rm in} \sqrt{\dot \rho^2 + 1 - \frac{2 G  m_{\rm in}}{\rho}
+ \frac{G  e^2}{\rho^2}} - \sigma_{\rm out} \sqrt{\dot \rho^2 + 1 -
\frac{2 G m_{\rm out}}{\rho}} = \frac{G  M }{\rho} \, .
 \label{israel}
\end{equation}
where the total mass $m_{\rm tot}=m_{\rm out}=m$, $\rho(\tau)$ --- a shell radius as a function of the proper time $\tau$, $\sigma=+1$ --- radii are growing in the direction of the external normal to the shell, $\sigma=-1$ --- the corresponding radii are diminishing. At the same time, $\sigma=+1$ at the $R_+$-regions, and, respectively, $\sigma=-1$ at $R_-$-regions. The sign of $\sigma$ may be changed only in the $T_\pm$-regions, when a corresponding subradical expression in (\ref{israel}) is equal to zero.
In this way $\sigma_{\rm out}$ is changing sign at the point $\rho_{\sigma_{\rm out}}$, for which are satisfied the following equations:
\begin{equation}
\dot \rho^2 + 1=\frac{2 G m_{\rm out}}{\rho_{\sigma_{\rm out}}},
\quad \sigma_{\rm in} \sqrt{\frac{2 G \Delta m}{\rho_{\sigma_{\rm
out}}} + \frac{G  e^2}{\rho_{\sigma_{\rm out}}^2}} = \frac{G M
}{\rho_{\sigma_{\rm out}}},
 \label{sigmaout7}
\end{equation}
where
\begin{equation}
\Delta m=m_{\rm out}-m_{\rm in}.
\label{Deltam}
\end{equation}
Certainly, this point may be absent on the specific shell trajectory. We see, that $\sigma_{\rm in}(\rho_{\sigma_{\rm out}})=1$ and, furthermore,
\begin{eqnarray}
\Delta m &=& \frac{G M^2-e^2}{2\rho_{\sigma_{\rm out}}}
 >-\frac{e^2}{2\rho_{\sigma_{\rm out}}},  \\
\rho_{\sigma_{\rm out}} &=& \frac{G M^2-e^2}{2\Delta m},
 \quad({\rm sign}[\Delta m]={\rm sign}[G M^2-e^2]).
\end{eqnarray}
At the same time, $\sigma_{\rm in}$ changes the sign at the point
$\rho_{\sigma_{\rm in}}$, which is a solution of equations
\begin{equation}
\dot \rho^2 + 1=\frac{2 G m_{\rm in}}{\rho_{\sigma_{\rm in}}}
 -\frac{G e^2}{\rho_{\sigma_{\rm in}}^2}, \quad
-\sigma_{\rm out} \sqrt{-\frac{2 G \Delta m}{\rho_{\sigma_{\rm in}}}
- \frac{G e^2}{\rho_{\sigma_{\rm in}}^2}} = \frac{G M
}{\rho_{\sigma_{\rm in}}}.
\end{equation}
We see, that $\sigma_{\rm out}(\rho_{\sigma_{\rm in}})=-1$ and, furthermore,
\begin{eqnarray}
\Delta m &=& -\frac{G M^2+e^2}{2\rho_{\sigma_{\rm in}}}
<- \frac{e^2}{2\rho_{\sigma_{\rm in}}} < 0 \,,  \\
\rho_{\sigma_{\rm in}} &=& \frac{G M^2+e^2}{2(-\Delta m)}.
\end{eqnarray}
At $\Delta m<0$
\begin{equation}
\rho_{\sigma_{\rm out}} = \frac{e^2-G M^2}{2(-\Delta m)}
\end{equation}
and, therefore, $\rho_{\sigma_{\rm out}}<\rho_{\sigma_{\rm in}}$.
From the condition of ``non exoticism'' it follws, that at $\sigma_{\rm in}=-1$
it is necessary must be $\sigma_{\rm out}=-1$.

For investigation of the shell dynamics, it is needed to square the Israel equation (\ref{israel}):
\begin{equation}
\Delta m=M \sigma_{\rm out} \sqrt{\dot \rho^2 + 1 - \frac{2 G m_{\rm
out}}{\rho}}+ \frac{G  M^2-e^2 }{2\rho} \, .
 \label{deltamout}
\end{equation}
From here it follows, that
\begin{equation}
\sigma_{\rm out}={\rm sign}\left[\Delta m-\frac{G  M^2-e^2 }{2\rho}\right] \, .
\end{equation}
With a help of (\ref{israel}) and (\ref{deltamout}) we find
\begin{equation}
\Delta m=M \sigma_{\rm in} \sqrt{\dot \rho^2 + 1 - \frac{2 G m_{\rm
in}}{\rho} + \frac{Ge^2}{\rho^2}} - \frac{G  M^2+e^2 }{2\rho} \, .
\end{equation}
For physical interpretation of this equation it is useful to introduce the ``running'' mass
$m_{\rm run}(\rho)$ as an effective total mass (energy) inside the sphere of radius $\rho$. In our case this is a total energy with the deduction of energy, distributed beyond the sphere of radius $\rho$. For the inside metric this is
\begin{equation}
m_{\rm in,run}(\rho)=m_{\rm in}-\frac{e^2}{2\rho},
 \label{minrun}
\end{equation}
i.\,e., the total mass at infinity (without the shell) with the deduction of electrosta\-tic energy beyond the sphere. For the external metric this is
\begin{equation}
m_{\rm out,run}(\rho)=m_{\rm out},
 \label{moutrun}
\end{equation}
since outside the shell the electric field is absent. The difference of these running masses
\begin{equation}
\Delta m_{\rm run}=m_{\rm out,run}-m_{\rm in,run}= \Delta
m+\frac{e^2}{2\rho}
\end{equation}
is a running mass of the shell . By substituting from here $\Delta m$ to the squared
equation, we have
\begin{equation}
\Delta m_{\rm run}= \sigma_{\rm in} M\sqrt{\dot \rho^2 + 1 - \frac{2
G m_{\rm run}}{\rho}} - \frac{G  M^2}{2\rho} \, .
 \label{deltamrun}
\end{equation}
This equation in this form is viewed similar to the self-gravitating neutral shell with the only difference, that now $\Delta m_{\rm run}$ is already non constant. This is because the work of the Coulomb forces is not taken into account. The internal mass is changing due to the changing of the electrostatic energy, which influences the attraction inside the shell. Equation (\ref{deltamrun}) at $\sigma_{\rm in}=+1$ has the sense of the energy conservation law. The term with the square root is an effective kinetic energy with addition of potential energy inside. The second term is an negative energy of the shell self-action. The kinetic energy formally changes sign if $\sigma_{\rm in}=-1$.

Up to now, we tacitly suppose the integration over the radius for defining the mass (energy) beyond the definite radius. The corresponding bare mass by definition is defined by integration along the direction of external normal of the shell, i.\,e. from left to right (subject to agreement) on the Carter--Penrose diagram. However, at $\sigma_{\rm in}=-1$ these directions are opposite. Therefore, though $M$ is always positive, $\Delta m$ may have any sign. Additionally, in the case $\sigma_{\rm in}=-1$ the center of the sphere (which is sometimes only imaginary) is placed beyond the sphere or at least not the inside. For this reason it is requested to change the sins in the definitions of as $m_{\rm run}$,  and
$\Delta m_{\rm run}$. In result, the shell is gravitationally attracted from inside to the outside (from left to right on the conformal diagram), however, as a matter of fact the shell is gravitationally attracted toward the (possible) center from the point of the internal observer. This is a quite realistic because the source is outside the shell. It must be taken into account for a qualitative understanding and physical interpretation of the described global geometries.

It is possible also the different interpretation. From the initial Israel equation it follows, that it is remained the same under the simultaneous changing of signs of the both $\sigma$ with the additional changing the sign of the bare mass $M\to -M$. This transformation is equivalent to the exchange between (in) and (out), when sign $M$ is changed automatically due to the changing of radial direction of integration.

The Israel equation is the Einstein constraint equation, integrated along the normal to the shell the Gauss coordinate. For fixed parameters of the internal metric $m_{\rm in}$ and $e$, and the shell parameters $M_{\rm in}$ and $-e$, the corresponding solution is defined by the initial conditions $\rho=\rho(0)$ and $\dot \rho=\dot \rho(0)$, $\sigma_{\rm in,0}=\sigma_{\rm in}(\rho(0))$, $\sigma_{\rm out,0}=\sigma_{\rm out}(\rho(0))$. At the same time, a total mass of the system, $m_{\rm out}$, is calculated from the constraint equation. For infinite motion it is natural to define  $\rho(0)=\infty$. With this determination $\sigma_{\rm in}(\infty)$, $\sigma_{\rm out}(\infty)$ and $m_{\rm out}$ define the shell velocity at infinity. For finite motion it useful to use the turning point $\rho_0$, which is fixing the other initial parameters: $\dot \rho_0=0$, $\sigma_{\rm in,0}=\sigma_{\rm in}(\rho_0)$, $\sigma_{\rm out,0}=\sigma_{\rm out}(\rho_0)$ and $m_{\rm out}$.
By putting $\rho_0=\infty$ in the squared equations, we have
\begin{equation}
\sigma_{\rm in}(\infty)=\sigma_{\rm out}(\infty)={\rm sign}[\Delta m].
\end{equation}
It is clear, that the cases $\Delta m>0$ and $\Delta m<0$ must be considered separately. For the qualitative description of the dynamical shell trajectory we need to know $\sigma(0)$, i.\,e., the value  $\sigma(\rho)$ at $\rho=0$. We have
\begin{equation}
\sigma_{\rm out}(0)={\rm sign}\left[{e^2-GM^2}\right], \quad
 \sigma_{\rm in}(0)={\rm sign}\left[{e^2+GM^2}\right],
\end{equation}
i.\,e., $\rho_{\sigma_{\rm out}}$ at $\Delta m>0$ exists, if only $e^2>GM^2$, and, respectively, at $\Delta m<0$, if only $e^2<GM^2$. Accordingly, $\rho_{\sigma_{\rm in}}$ exists only at $\Delta m<0$.

The infinite motion is realized at $(\Delta m/M)^2>1$, while the finite one at $(\Delta m/M)^2<1$.
The turning points are defined from the twice squared Israel equation (\ref{israel}):
\begin{equation}
 \frac{(GM^2-e^2)^2}{4\rho_0^2M^2} +\frac{1}{\rho_0M}\left[
 \frac{\Delta m}{M}(e^2+GM^2)+2GMm_{\rm in}\right]+
\left(\frac{\Delta m}{M}\right)^2-1=0.
\label{turning1}
\end{equation}
The roots of this equation are
\begin{equation}
\frac{1}{2\rho_0M}=\frac{-B\pm\sqrt{D}}{(GM^2-e^2)^2},
 \label{bounce}
\end{equation}
where
\begin{eqnarray}
 \label{B2}
 B &=& \frac{\Delta m}{M}(e^2\!-\!GM^2)\!+\!2GMm_{\rm out}
 =\frac{\Delta m}{M}(e^2\!+\!GM^2)\!+\!2GMm_{\rm in}, \\
 D &=& B^2- (GM^2-e^2)^2\left[\left(\frac{\Delta m}{M}\right)^2-1\right].
 \label{BD}
\end{eqnarray}

\subsection{Neutralizing shell at $\Delta m>0$}

We consider all possible types of shell trajectories at $\Delta m>0$.
In this case always $B>0$, and roots of the quadratic equation are complex or negative at the infinite motion. Therefore, the turning points are absent. Under the finite motion --- one of the roots is positive and other is negative, i.\,e., there is only turning point.

\subsubsection{$\Delta m>0$: infinite motion}

In the infinite motion at $\Delta m>0$ from the relation
$\Delta m>M>0$ follows that $M<m_{\rm out}$. Additionally, we have
$\sigma_{\rm in}=+1$ --- everywhere, $\sigma_{\rm out}(\infty)=+1$, $\sigma_{\rm out}(0)={\rm
sign}[e^2-GM^2]$ and equation for the point, where $\sigma_{\rm out}$ changes the sign:
\begin{equation}
\rho_{\sigma_{\rm out}}=\frac{GM^2-e^2}{2\Delta m}.
 \label{rhosigmaout}
\end{equation}
The point  $\rho_{\sigma_{\rm out}}$ is absent, if $e^2>GM^2$ (the self-repulsive shell). It corresponds to $\sigma(0)=+1$. On the contrary, for self-attractive shell, i.\,e., at $e^2<GM^2$, the trajectory is inevitably has the point for changing the sign of $\sigma$, and $\sigma_{\rm out}(0)=-1$. Therefore, the self-repulsive shell is collapsing without obstruction.

\begin{figure}[H]
\begin{center}
\includegraphics[angle=0,width=0.45\textwidth]{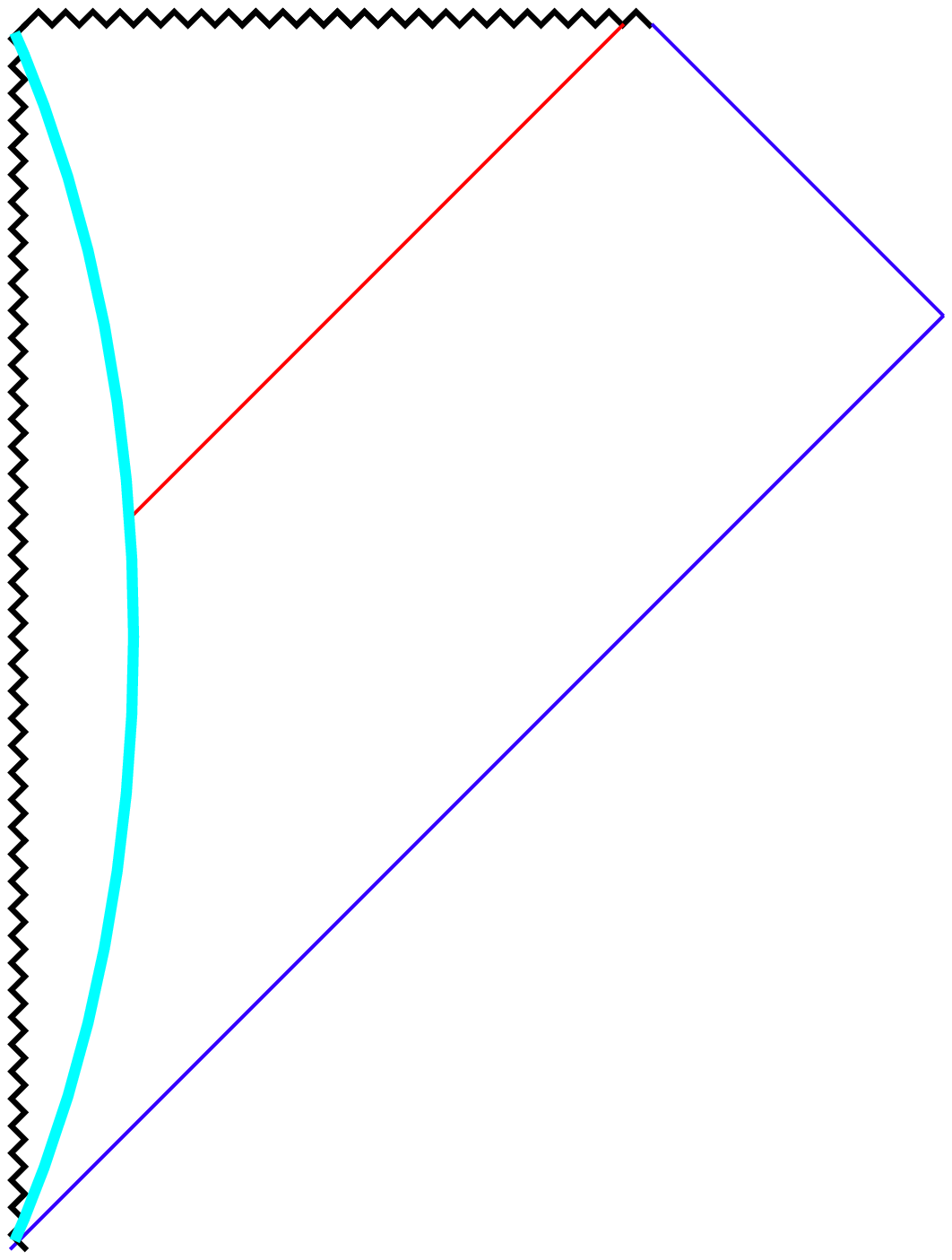}
\hfill
\includegraphics[angle=0,width=0.44\textwidth]{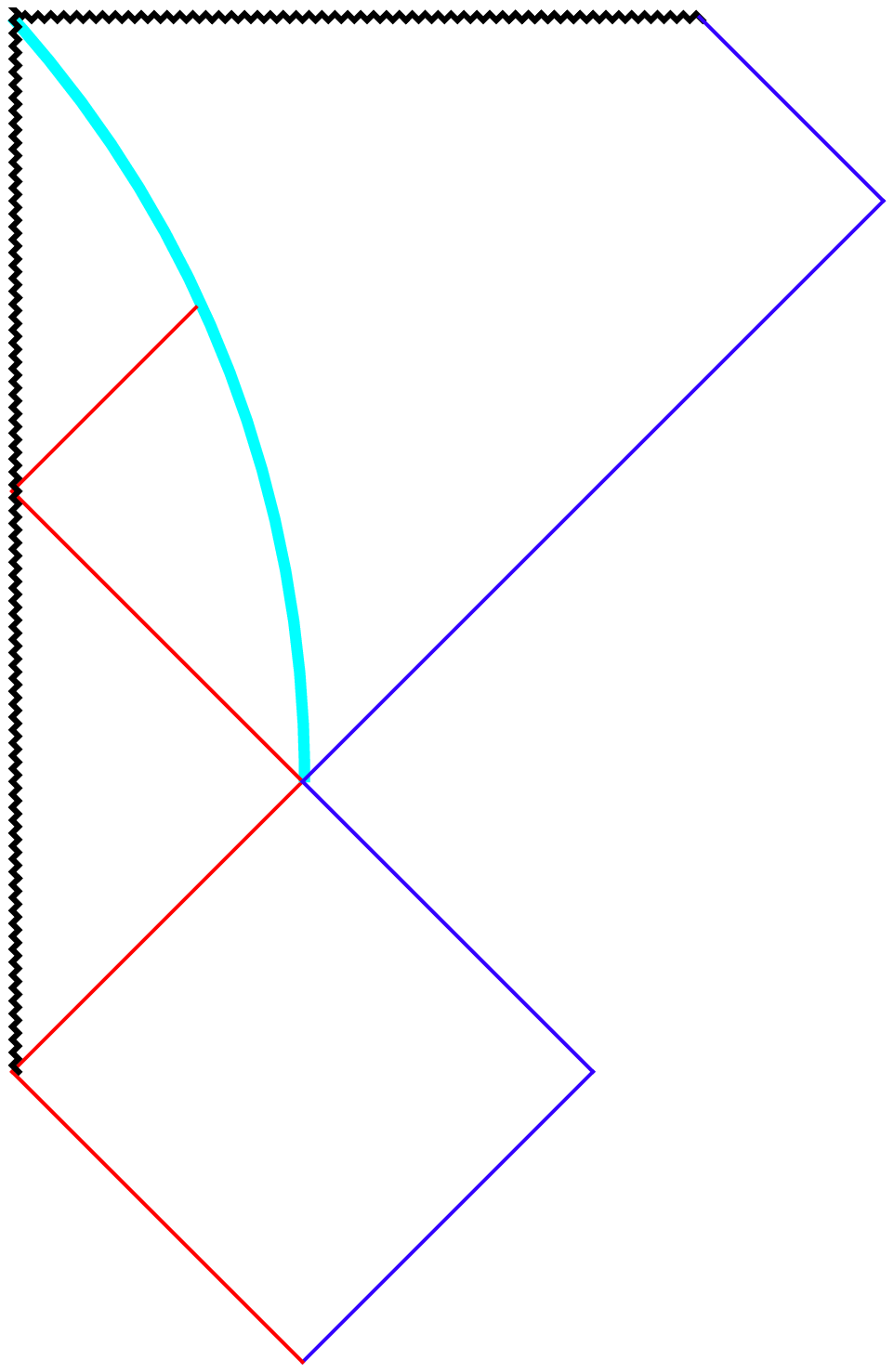}
\end{center}
\caption{Infinite motion of the shell at $\Delta m>M$ and $e^2>Gm^2_{in}$
(at the left panel) and $e^2=Gm^2_{in}$ (at the right panel).} \label{NS26a}
\end{figure}

\begin{figure}[H]
\begin{center}
\includegraphics[angle=0,width=0.49\textwidth]{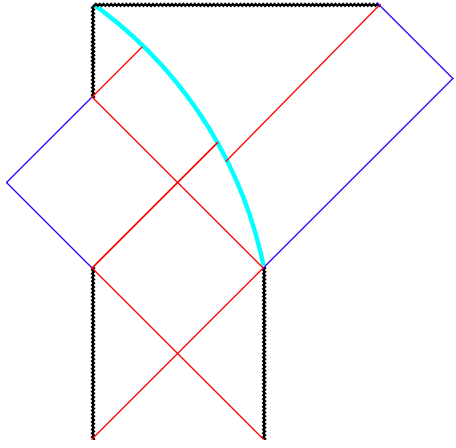}
\end{center}
\caption{Infinite motion of the shell at $\Delta m>M$ and $e^2<Gm^2_{in}$.}
\label{NS26c}
\end{figure}

In the second case it is possible the restriction due to the inequality
$\rho_{\sigma_{\rm out}}<2Gm_{\rm out}$. Let us verify this inequality:
\begin{eqnarray}
 \label{rhosigmaoutless}
\rho_{\sigma_{\rm out}} &=& \frac{GM^2-e^2}{2\Delta m}<2Gm_{\rm out}
\\
GM^2 &<& G\Delta m^2<4G\Delta m^2+4G\Delta m \,m_{\rm in}+e^2.
\end{eqnarray}
It is clear that inequality is held automatically. The corresponding Carter--Penrose diagrams are shown in Figs.~\ref{NS26a}--\ref{NS26c}.

\subsubsection{$\Delta m>0$: infinite motion}

\begin{figure}[H]
\begin{center}
\includegraphics[angle=0,width=0.9\textwidth]{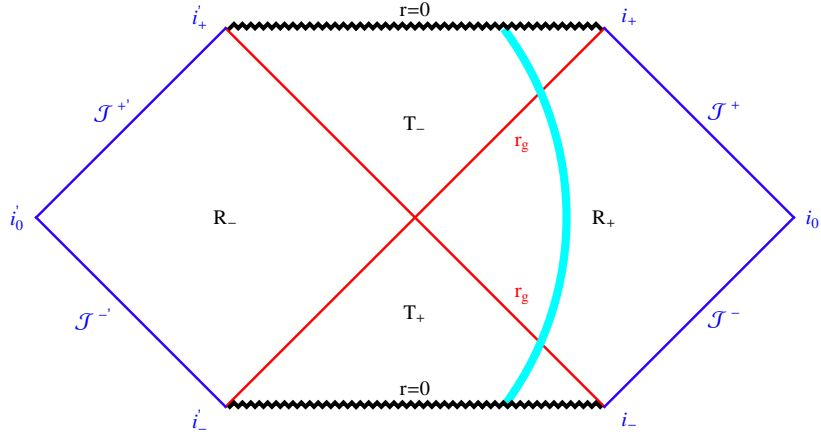}
\end{center}
\caption{Finite motion of the shell in the external metric at $e^2>GM^2$ and $0<\Delta m<M$.} \label{SchwFinP22}
\end{figure}

Now consider the infinite motion at $\Delta m>0$. It is a more complicated case, because of the possible existence of the turning point in $R_\pm$-regions, as in the internal and in the external metrics. Besides of this, $\sigma$ may have the different signs at the point $r=0$. In fact, we have
\begin{equation}
0<\Delta m<M, \quad 0\leq\rho\leq\rho_0.
\end{equation}
The turning point now is only one:
\begin{eqnarray}
\frac{1}{2\rho_0M} &=&\frac{\sqrt{D}-B}{(GM^2-e^2)^2}, \\
 B &=& \frac{\Delta m}{M}(e^2\!-\!GM^2)\!+\!2GMm_{\rm out} \\
 &=& \frac{\Delta m}{M}(e^2\!+\!GM^2)\!+\!2GMm_{\rm in}, \\
 D &=& B^2+(GM^2-e^2)^2\left[1-\left(\frac{\Delta m}{M}\right)^2\right]\geq B^2.
\end{eqnarray}
In the limiting case $e^2=GM^2$ we have
\begin{eqnarray}
\frac{1}{2\rho_0M} &=& \frac{1-\left(\frac{\Delta
m}{M}\right)^2}{2B}=\frac{1-\left(\frac{\Delta
m}{M}\right)^2}{4GMm_{\rm out}}, \\
\rho_0 &=& \frac{2Gm_{\rm out}}{1-\left(\frac{\Delta
m}{M}\right)^2}>2Gm_{\rm out}=r_g.
\end{eqnarray}
Let us verify this inequality in the general case, i.\,e., find any limitation to the parameters:
\begin{equation}
\label{rho01}
 \frac{1}{2\rho_0M} =\frac{\sqrt{D}-B}{(GM^2-e^2)^2}
 <\frac{1}{4GMm_{\rm out}} \quad \Rightarrow \quad
 (\sqrt{D}-2GMm_{\rm out})^2 \geq 0.
\end{equation}
Let us define condition, when turning point is placed directly at the horizon of the external metric:
\begin{equation}
\sqrt{D}=2GMm_{\rm out} \quad \Rightarrow \quad
(GM^2-e^2)(GM^2-e^2+4G\Delta m\, m_{\rm out})=0.
\end{equation}
The possibility $GM^2-e^2=0$ is realized only, if additionally
$\Delta m=0$. However
\begin{equation}
\Delta m=\frac{e^2-GM^2}{4Gm_{\rm out}},
\end{equation}
Condition $\Delta m>0$ is realized only in the case of the self-repulsive shell with $e^2>GM^2$,
\begin{equation}
 \Delta m<M \: \Rightarrow \:
 0<\!e^2\!-\!GM^2\!<4GM m_{\rm out}, \: GM^2\!<\!e^2\!<GM(M\!+\! 4m_{\rm out}).
\end{equation}
Now define $\rho_{\sigma_{\rm in}}$ and $\rho_{\sigma_{\rm
out}}$, which both determine the global geometry of the total space-time. From condition $\Delta m>0$ it follows, that
\begin{equation}
\rho_{\sigma_{\rm out}}=\frac{GM^2-e^2}{2\Delta m}
\end{equation}
exists only for the self-attractive shell with $GM^2>e^2$. In this case  $\sigma_{\rm out}(0)={\rm sign}[e^2-GM^2]$. In result, for the self-repulsive shell we have $\sigma_{\rm out}(0)=+1$, and, therefore, the point $\rho_{\sigma}$ is absent. This means, that turning point is inevitably placed in the $R_+$-region of the external metric. The conformal diagram for the Schwarzschild manifold at $e^2>GM^2$ and $0<\Delta m<M$ is shown in Fig.~\ref{SchwFinP22}.

For the self-attractive shell $e^2>GM^2$ and $\rho_{\sigma}(0)=-1$. The sign $\sigma(\rho_0)$ depends on the existence of the  point $\rho_{\sigma_{\rm out}}$ in the $T$-region of the external metric:
\begin{eqnarray}
\rho_{\sigma_{\rm out}} &<& 2Gm_{\rm out} \quad \Rightarrow \quad
\sigma(\rho_0)=+1, \\
\rho_{\sigma_{\rm out}} &>& 2Gm_{\rm out} \quad \Rightarrow \quad
\sigma(\rho_0)=-1.
\end{eqnarray}
From equations (\ref{rhosigmaout}) and (\ref{rhosigmaoutless}) we have
\begin{equation}
\rho_{\sigma_{\rm out}} = \frac{GM^2-e^2}{2\Delta m}<2GM_{\rm out}
\end{equation}
At $GM^2>e^2$ this condition transforms to the inequality
\begin{equation}
\Delta m>\frac{m_{\rm in}}{2}\left(\sqrt{\frac{M^2}{m_{\rm
in}^2}-\frac{e^2}{Gm_{\rm in}^2}-1}\right).
\end{equation}
Under this condition and, additionally, at $\Delta m<M$ we would have $\rho_{\sigma}(0)=+1$.

\begin{figure}
\begin{center}
\includegraphics[angle=0,width=0.49\textwidth]{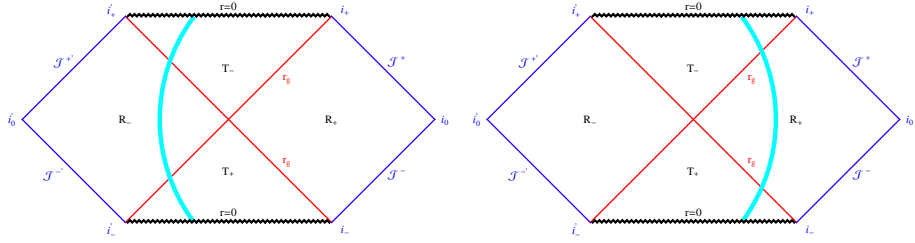}
\hfill
\includegraphics[angle=0,width=0.49\textwidth]{SchwFinP22.eps}
\end{center}
\caption{Finite motion of the shell in the external metric at
$GM^2>e^2$ and $\Delta m>(m_{\rm in}/2)[\sqrt{(M/m_{\rm in})^2-(e^2/Gm_{\rm
in}^2)+1}-1]$ \ at the right panel, and $\Delta m<(m_{\rm in}/2)
[\sqrt{(M/m_{\rm in})^2-(e^2/Gm_{\rm in}^2)+1}-1]$ at the left panel.} \label{SchwFinP25}
\end{figure}

\begin{figure}
\begin{center}
\includegraphics[angle=0,width=0.4\textwidth]{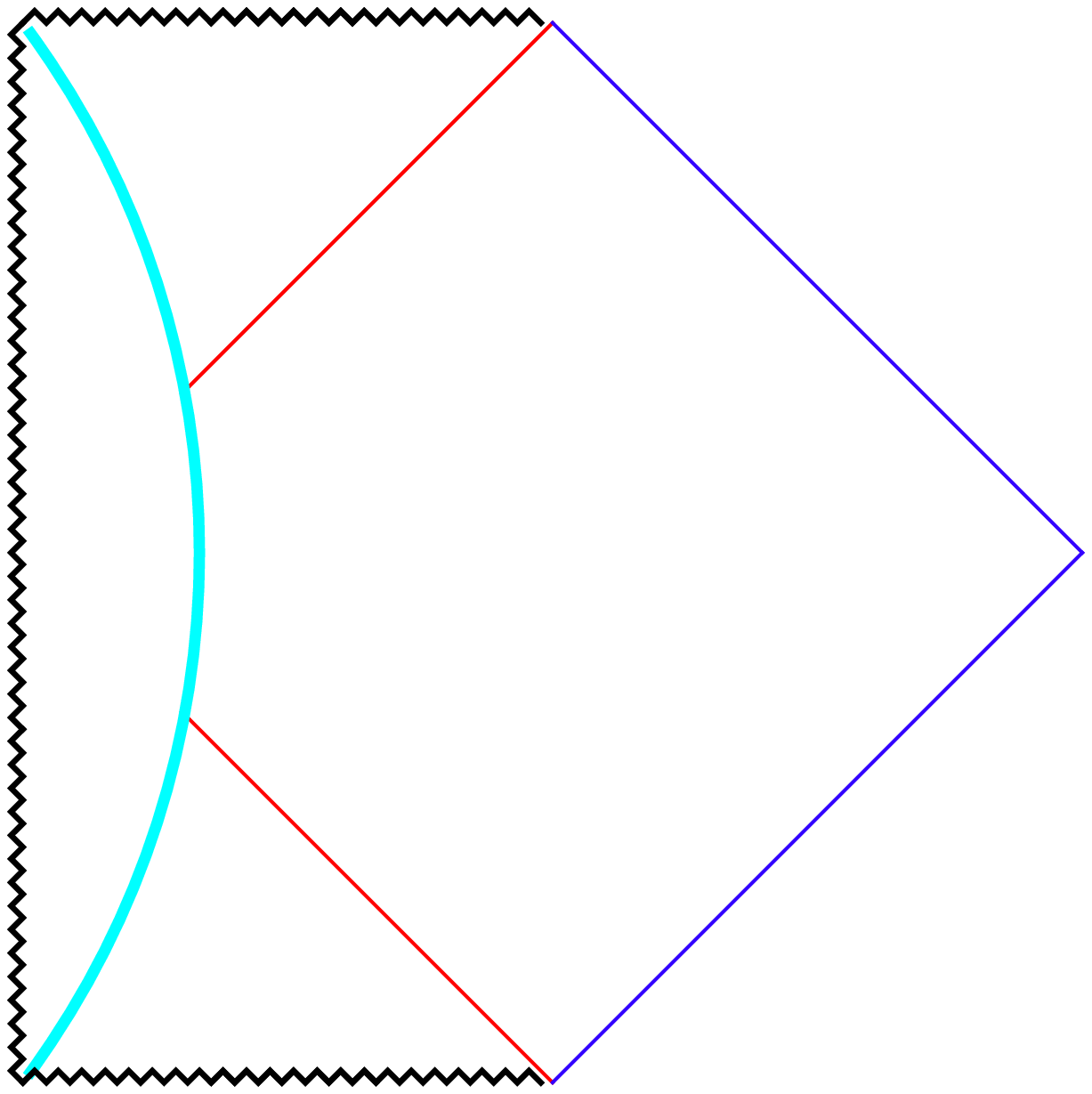}
\hskip0.5cm
\hfill
\includegraphics[angle=0,width=0.52\textwidth]{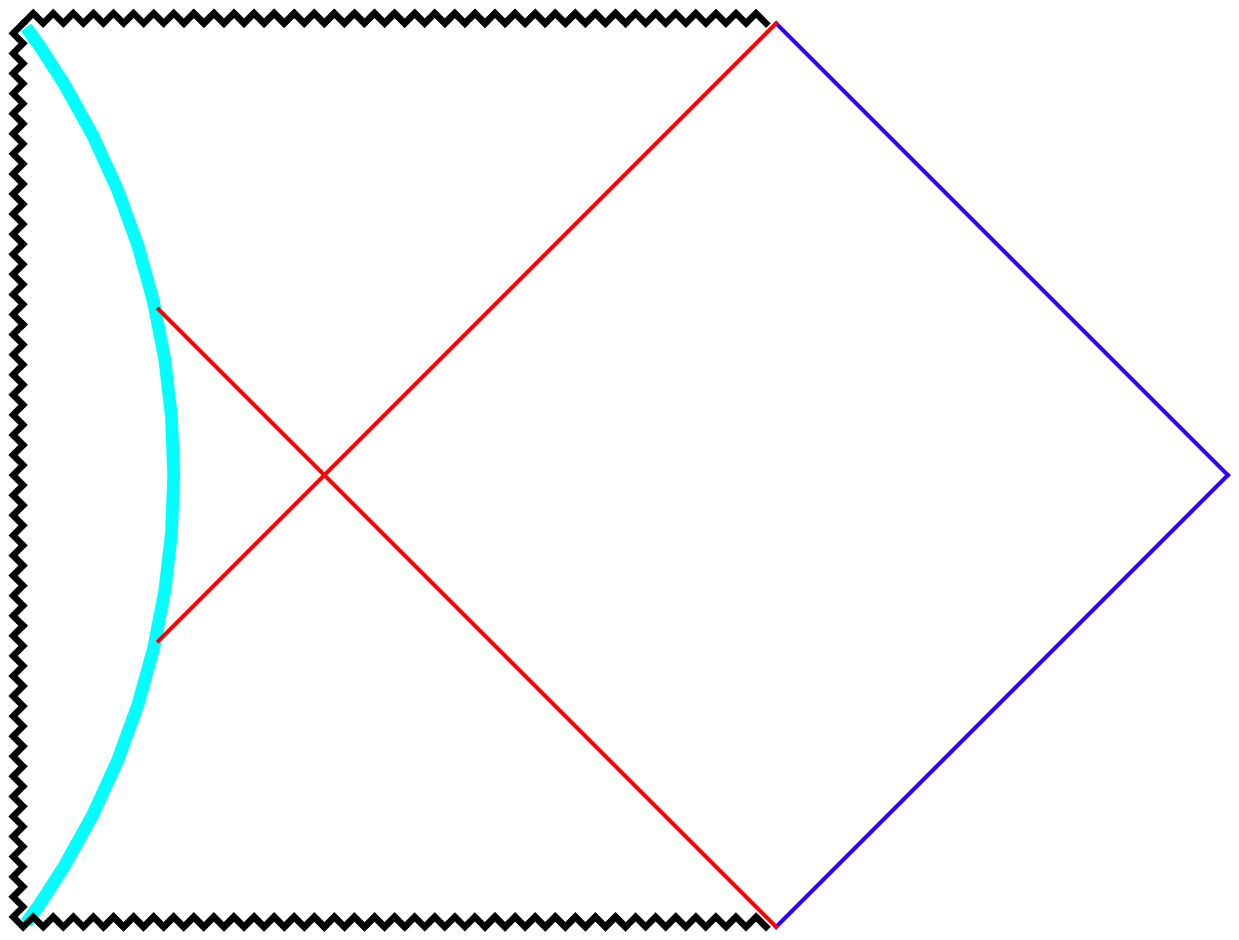}
\end{center}
\caption{Finite motion of the shell. At the left panel: at $e^2>GM^2$ and
$0<\Delta m/M<1$ or at $e^2>GM^2$ and $\Delta m > (m_{\rm in}/2)[\sqrt{(M/m_{\rm in})^2-(e^2/Gm_{\rm in}^2)+1}-1]$. At the right panel: at $e^2<GM^2$, $0<\Delta m/M<1$ or at $e^2>GM^2$ and $\Delta m<(m_{\rm in}/2)[\sqrt{(M/m_{\rm in})^2-(e^2/Gm_{\rm in}^2)+1}-1]$.} \label{NS27}
\end{figure}

\begin{figure}
\begin{center}
\includegraphics[angle=0,width=0.4\textwidth]{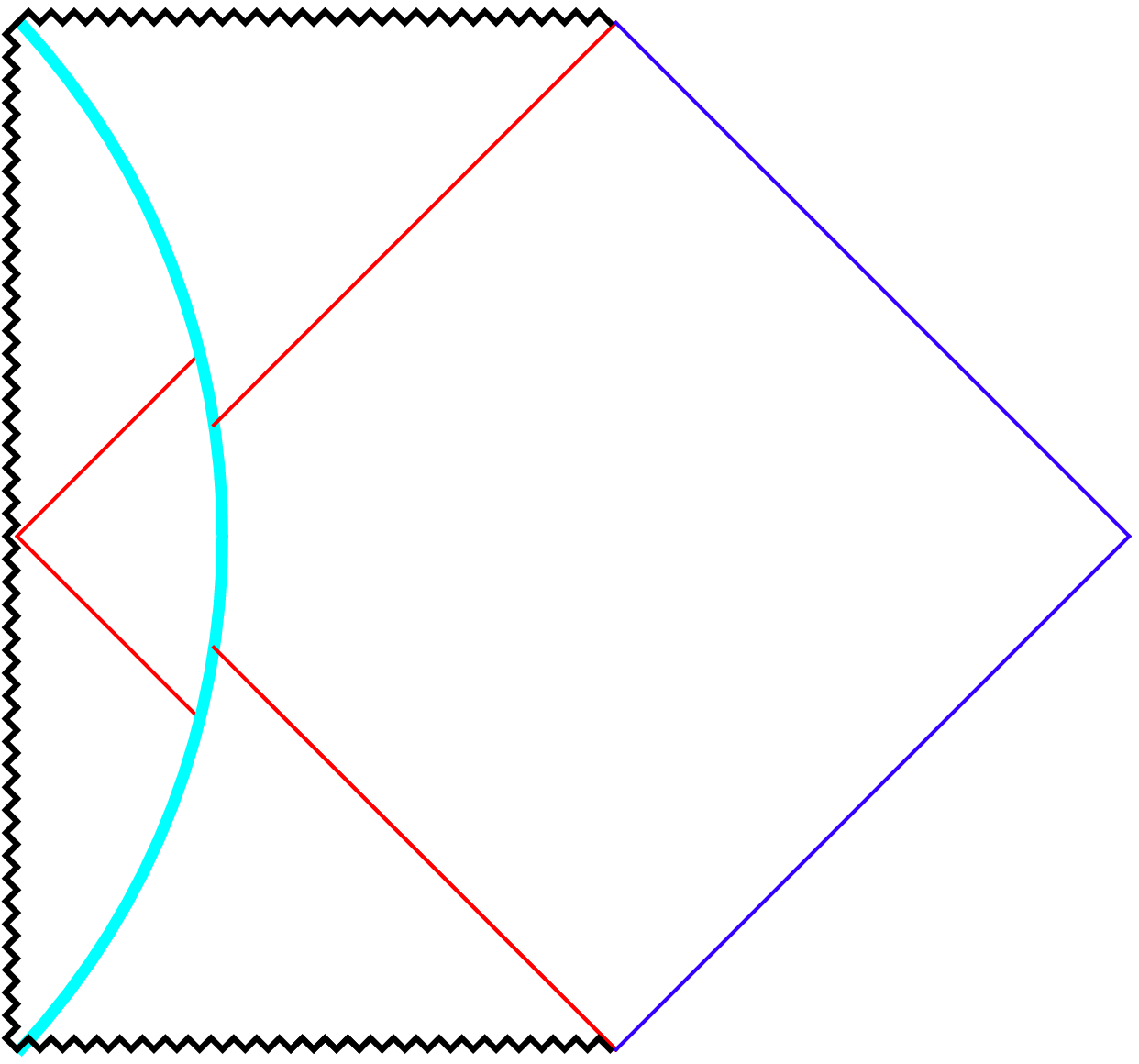}
\hfill
\includegraphics[angle=0,width=0.45\textwidth]{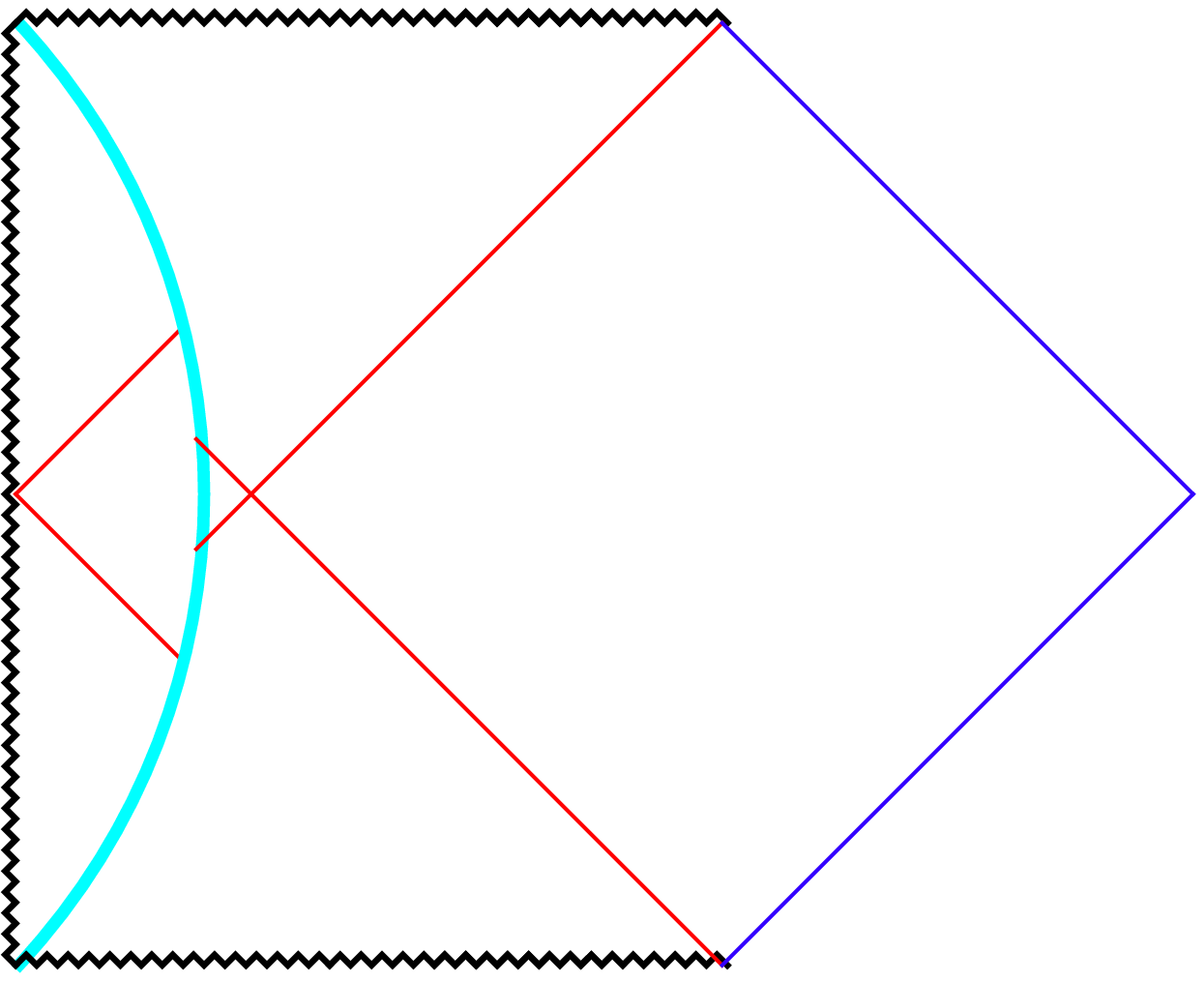}
\end{center}
\caption{Finite motion of the shell. At the left panel: at $e^2>GM^2$ and
$0<\Delta m/M<1$ or at $e^2<GM^2$ and $\Delta m>(M-m_{\rm in})/2$.
At the right panel: at $e^2<GM^2$, $0<\Delta m/M<1$ or at $e^2>GM^2$ and
$\Delta m>(M-m_{\rm in})/2$.} \label{NS27cd}
\end{figure}

\begin{figure}
\begin{center}
\includegraphics[angle=0,width=0.48\textwidth]{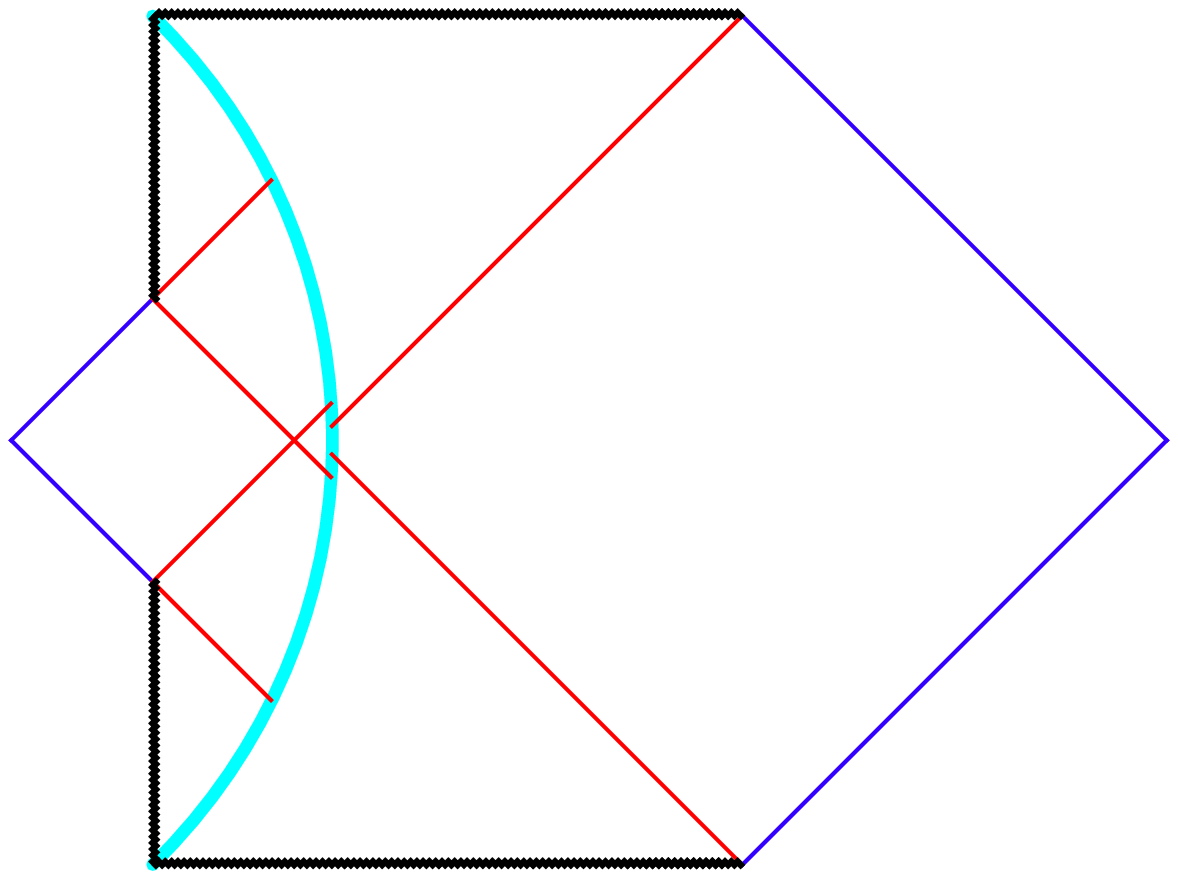}
\hfill
\includegraphics[angle=0,width=0.49\textwidth]{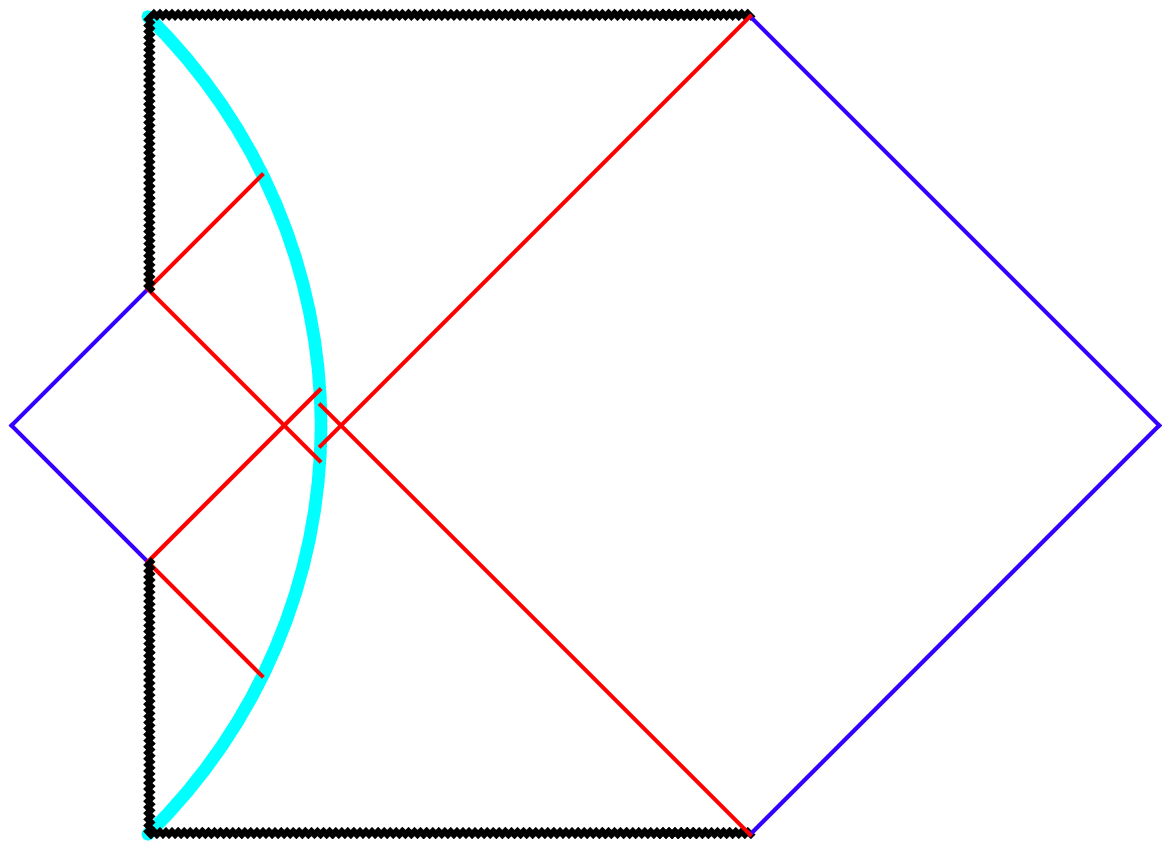}
\end{center}
\caption{The case of the finite motion of the shell at $\Delta m>0$.}
\label{NSp28}
\end{figure}

In the opposite case the point $\rho_{\sigma_{\rm out}}$ is placed in the $R$-region.
In principle, the shell must come out into the $R_-$-region under realization of the conditions $\rho_0<\rho_{\sigma_{\rm out}}$ and $\rho_{\sigma_{\rm out}}>2Gm_{\rm out}$. These conditions are equivalent to the inequality
\begin{equation}
GM^2>4G\Delta m \, m_{\rm out}+e^2,
\end{equation}
which by-turn corresponds to the condition $\rho_0>r_g=2Gm_{\rm
out}$. In Fig.~\ref{SchwFinP25} are shown two conformal diagrams for the cases $GM^2>e^2$ and
\begin{equation}
\Delta m\lessgtr\frac{m_{\rm
in}}{2}\left(\sqrt{\frac{M^2}{m_{\rm in}^2}-\frac{e^2}{Gm_{\rm
in}^2}+1}-1\right).
\end{equation}
We see that this classification does not depend on the properties of the internal metric at $\Delta m>0$, when $\sigma_{\rm in}=+1$ everywhere along the shell trajectory. The corresponding complete Carter--Penrose diagrams are shown in Figs.~\ref{NS27}--\ref{NSp28}.

\subsubsection{$\Delta m=0$: finite motion}

Before the description the more complicated and multivariate case $\Delta m<0$, we consider the separately the specific interjacent case $\Delta m=0$, and,conjointly, verify the previous formulas, as the limiting cases.

\begin{figure}
\begin{center}
\includegraphics[angle=0,width=0.48\textwidth]{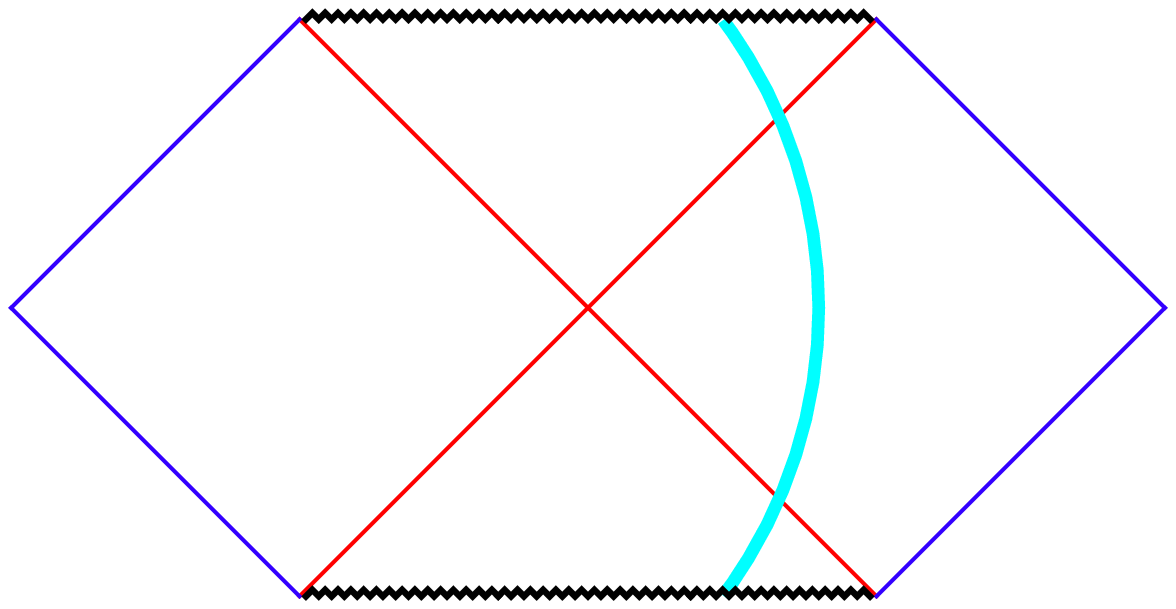}
\hfill
\includegraphics[angle=0,width=0.49\textwidth]{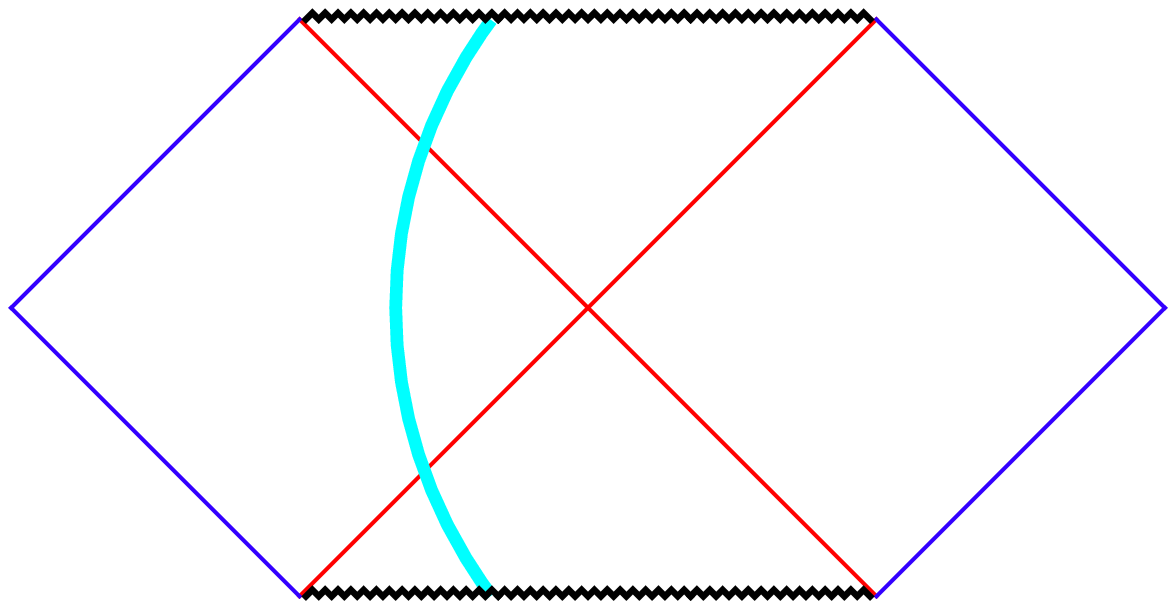}
\includegraphics[angle=0,width=0.49\textwidth]{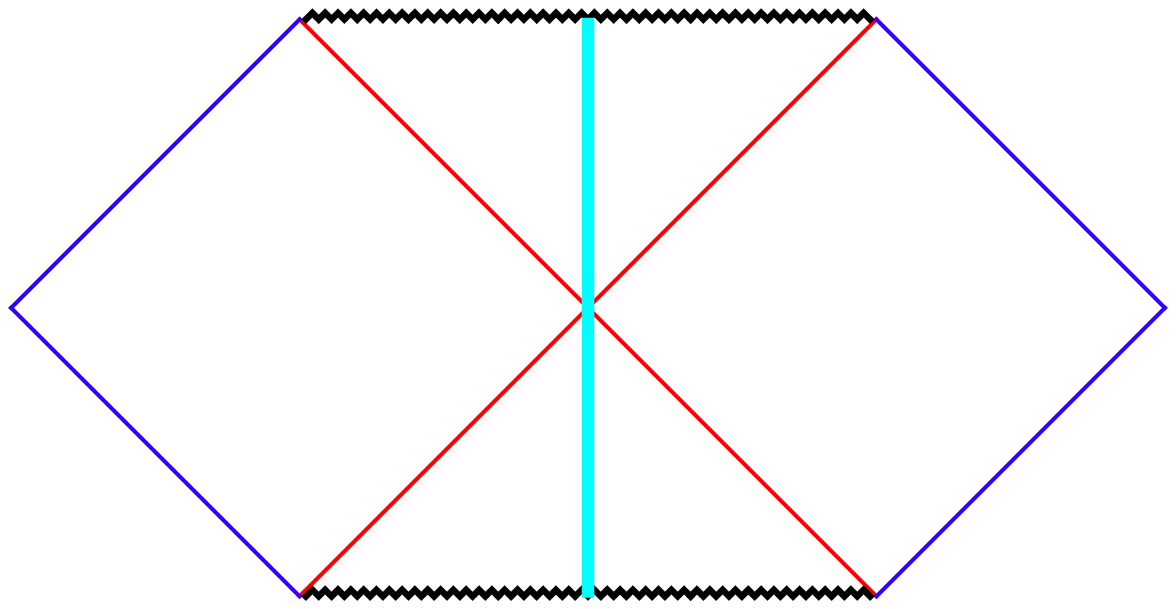}
\end{center}
\caption{The case of finite motion at $\Delta m>0$. The finite motion of the shell at $e^2>GM^2$ (the left upper panel), $e^2GM^2$ (the right upper panel) and $e^2=GM^2$ (the central lower panel).} \label{SchwP29abc}
\end{figure}

Forasmuch then, at the finite motion we have $\sigma_{\rm in}=+1$ and
\begin{eqnarray}
\Delta m &=& 0 \quad \Rightarrow \quad \sigma_{\rm out}={\rm
sign}[e^2-GM^2].
\end{eqnarray}
At the same tine
\begin{eqnarray}
 B &=& 2GMm_{\rm out}=2GMm_{\rm in} \\
 D &=& 4GM^2m_{\rm in}^2+(GM^2-e^2)^2, \\
 \frac{1}{2\rho_0M} &=&\frac{\sqrt{4GM^2m_{\rm in}^2+(GM^2-e^2)^2}
 -2GMm_{\rm out}}{(GM^2-e^2)^2}, \\
 GM^2 &=& e^2 \quad \Rightarrow \quad \rho_0=2GMm_{\rm out}=2GMm_{\rm in}.
\end{eqnarray}
Consequently, the point $\rho_0$ is placed at the external horizon. The corresponding Carter--Penrose diagrams for the cases $e^2\gtreqless GM^2$ are shown in Fig.~\ref{SchwP29abc}.

\subsubsection{Extreme black hole at $e^2=GM^2$: indifferent shell}

The one more  interjacent case: $e^2=GM^2$ --- indifferent shell, which corresponds to conditions:
\begin{eqnarray}
 \sigma_{\rm out}&=& {\rm sign}[\Delta m], \quad
 \sigma_{\rm in} = {\rm sign}\left [\Delta m+\frac{GM^2}{\rho}\right ], \\
 B &=& D=2GMm_{\rm out}>0, \\
 \rho_0 &=&\frac{2G\,m_{\rm in}}{1-\frac{\Delta m^2}{M^2}}>r_g.
\end{eqnarray}
Let us $\Delta m>0$. In this case  $\sigma_{\rm out}=\sigma_{\rm
in}=+1$. The turning point is placed in the $R_+$-region of as external and the internal metrics. The form of the corresponding conformal diagrams is evident.

The case $\Delta m <0$ is somewhat complicated because $\sigma_{\rm
out}=-1$, and the sign $\sigma_{\rm in}$ may be changed. For the internal metric we have
\begin{eqnarray}
 \sigma_{\rm in}(\infty)&=& -1, \quad \sigma_{\rm in}(0) = +1, \\
 \rho_{\sigma_{\rm in}} &=&-\frac{GM^2}{\Delta m}.
\end{eqnarray}
The problem is, first of all, for the infinite motion. On the one hand, the turning point is absent, and on the second hand, the sign $\sigma$ must be changed. It is possible only in the $T$-region, which is absent, if $e^2\geq Gm_{\rm in}^2$. We come to contradiction:
\begin{eqnarray}
e^2 &=& GM^2 \quad \Rightarrow \quad M\geq m_{\rm in} \\
\Delta m&<&0 \quad \Rightarrow \quad m_{\rm out}<m_{\rm in}, \quad
M\geq m_{\rm in}>m_{\rm out}, \\
\Delta m&<&-M, \quad 0<m_{\rm out}<m_{\rm in}-M \quad \Rightarrow
\quad m_{\rm in}>M.
\end{eqnarray}
Consequently, the infinite motion at $\Delta m<0$, $e^2=GM^2\geq
G\,m_{\rm in}^2$ is forbidden by condition $m_{\rm out}>0$.

For the possibility of infinite motion at $e^2=GM^2<G\,m_{\rm in}^2$
it is required that the turning point $\rho_{\sigma_{\rm in}}$ must be between the two horizons, $r_-<\rho_{\sigma_{\rm in}}<r_+$. Condition $M>0$ at $\sigma_{\rm in}=\sigma_{\rm out}=+1$ (i.\,e. in the  $R_-$-region) demands that
\begin{equation}
\Delta m<-\frac{e^2}{\rho}=-\frac{GM^2}{\rho}.
\end{equation}
At $|\Delta m<M|$ it is followed, that $\rho<GM$. This means that in this case the infinite motion is impossible.

Consider now the finite motion at $\Delta m<0$ and $\Delta m/M>-1$. For the turning point we have
\begin{equation}
\sigma_{\rm in}(\rho_0) =
 {\rm sign}\left[\Delta m+\frac{GM^2}{\rho_0}\right] =
 {\rm sign}\left[\Delta m
 +\frac{M^2\left(1-\frac{\Delta m^2}{M^2}\right)}{2m_{\rm
 out}}\right].
\end{equation}
\begin{equation}
\sigma(\rho_0) =
 {\rm sign}[M^2-\Delta m^2 +2 m_{\rm out}\Delta m].
\end{equation}
Roots of equation
\begin{equation}
M^2-\Delta m^2 +2 m_{\rm out}\Delta m=0
\end{equation}
are
\begin{equation}
 \left(\frac{\Delta m}{M}\right)_\pm=-\frac{m_{\rm in}}{M}\pm
 \sqrt{\frac{m_{\rm in}^2}{M^2}-1}.
\end{equation}
\begin{equation}
e^2=GM^2<GM^2m_{\rm in} \quad \Rightarrow \quad M<m_{\rm in}.
\end{equation}
It turns out, that $\sigma_{\rm in}(\rho_0)=+1$, if
\begin{equation}
 0>\frac{\Delta m}{M}>
 \sqrt{\frac{m_{\rm in}^2}{M^2}-1}-\frac{m_{\rm in}}{M},
\end{equation}
and $\sigma_{\rm in}(\rho_0)=-1$, if
\begin{equation}
 -1<\frac{\Delta m}{M}<
 \sqrt{\frac{m_{\rm in}^2}{M^2}-1}-\frac{m_{\rm in}}{M}.
\end{equation}
Now it is requested to examine where are placed $\rho_{\sigma_{\rm in}}$. $\rho_{\sigma_{\rm in}}<\rho_0$:
\begin{equation}
-\frac{GM^2}{\Delta m}<\frac{2Gm_{\rm out}2}{1-\frac{\Delta
m^2}{M^2}}  \quad \Rightarrow \quad M^2+\Delta m^2-2\Delta m
\,m_{\rm in}<0.
\end{equation}
This inequality is held for $\sigma_{\rm in}(\rho_0)=-1$. Consequently, the sign $\sigma_{\rm in}$ is changing on the trajectory. In the second case the sign $\sigma_{\rm in}$ is not changed on the trajectory. If $e^2=GM^2>G m_{\rm in}^2$, then $M>m_{\rm in}$, and in this case  $\sigma_{\rm in}(\rho_0)=+1$ without any limitations.
\begin{figure}[H]
\begin{center}
\includegraphics[angle=0,width=0.46\textwidth]{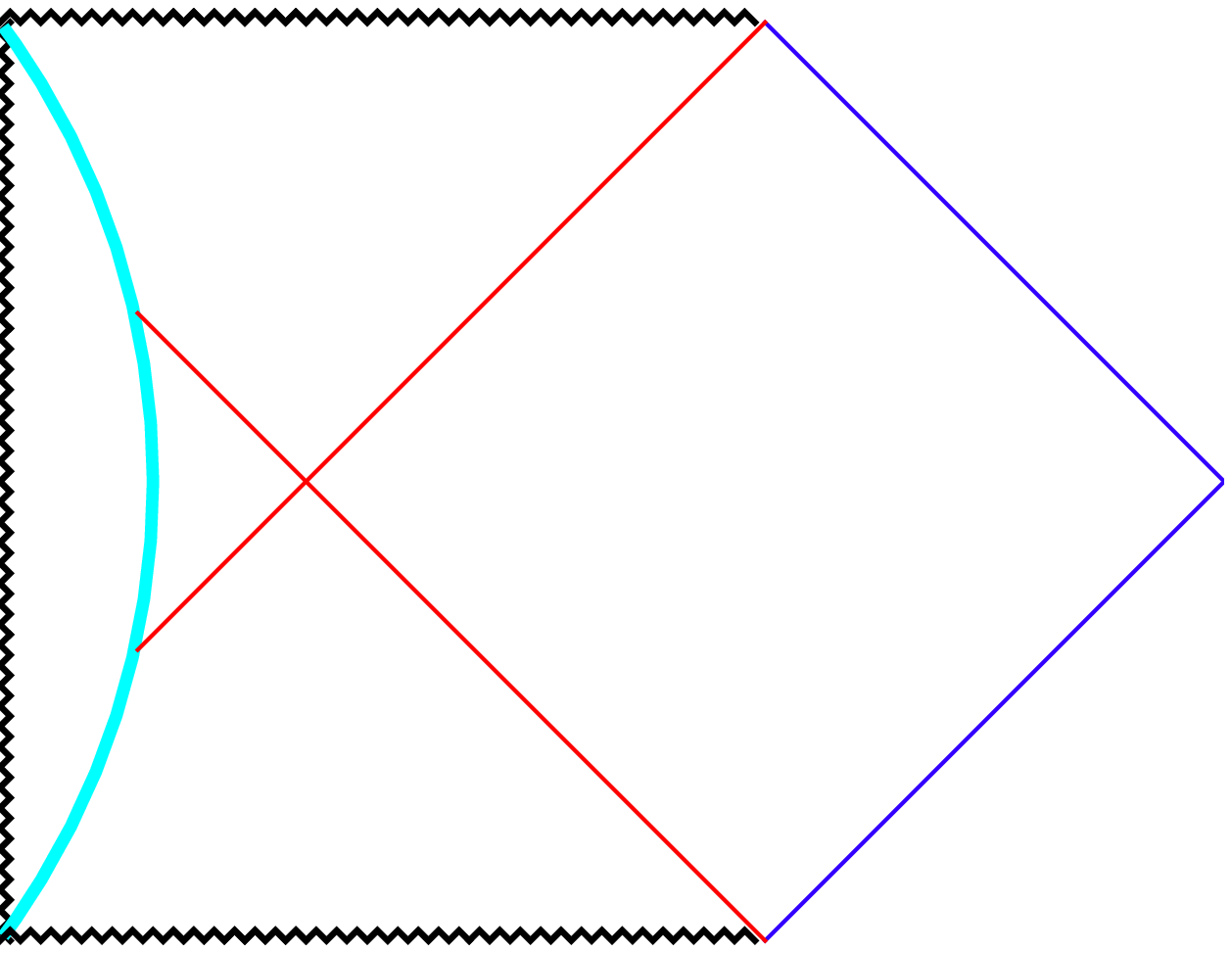}
\hskip0.5cm
\hfill
\includegraphics[angle=0,width=0.46\textwidth]{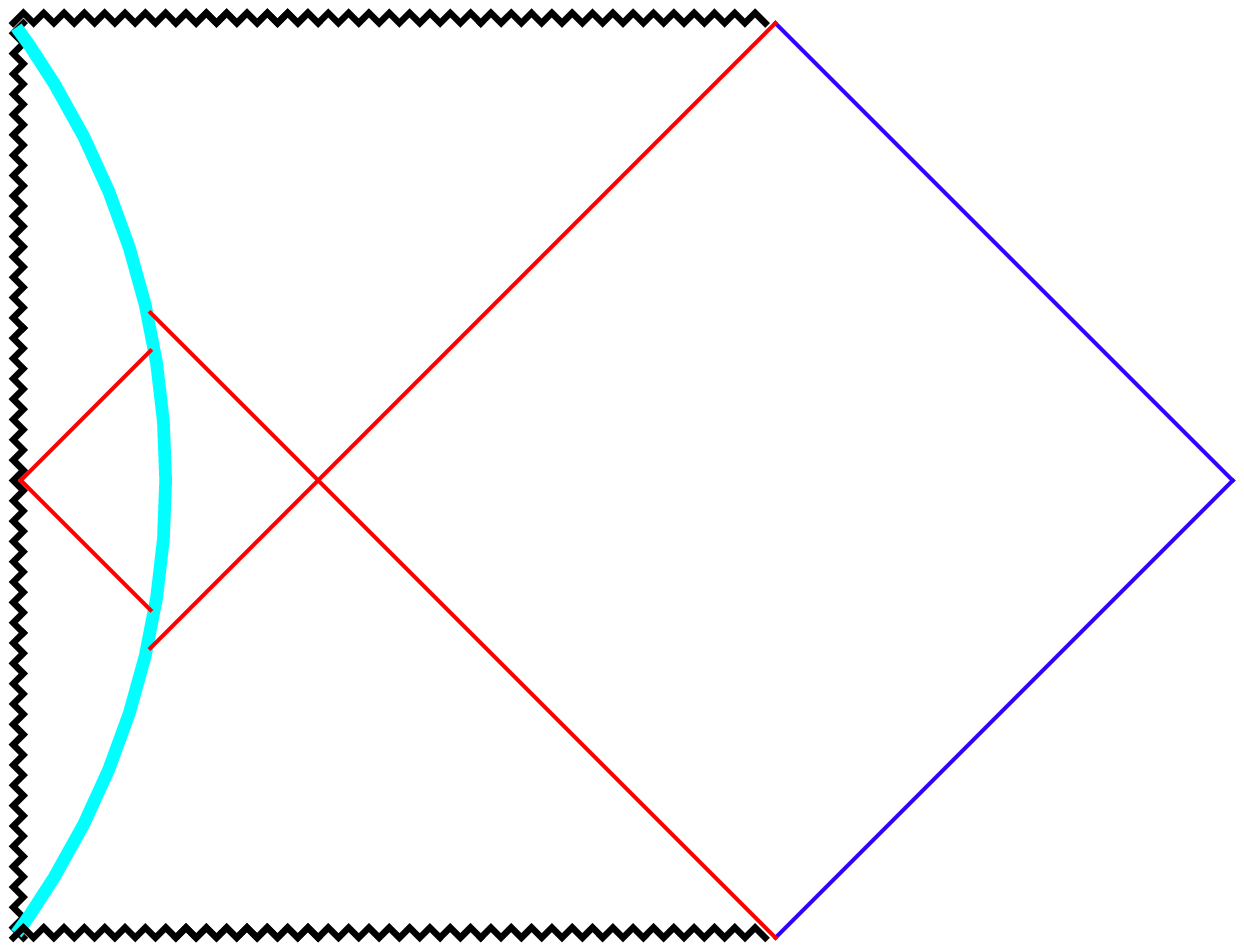}
\end{center}
\caption{Finite motion of the shell at $e^2=GM^2$. The left panel: at
$e^2=GM^2>Gm_{\rm in}^2$. The right panel: at $e^2=GM^2=Gm_{\rm in}^2$.}
\label{NS37}
\end{figure}
\begin{figure}[H]
\begin{center}
\includegraphics[angle=0,width=0.48\textwidth]{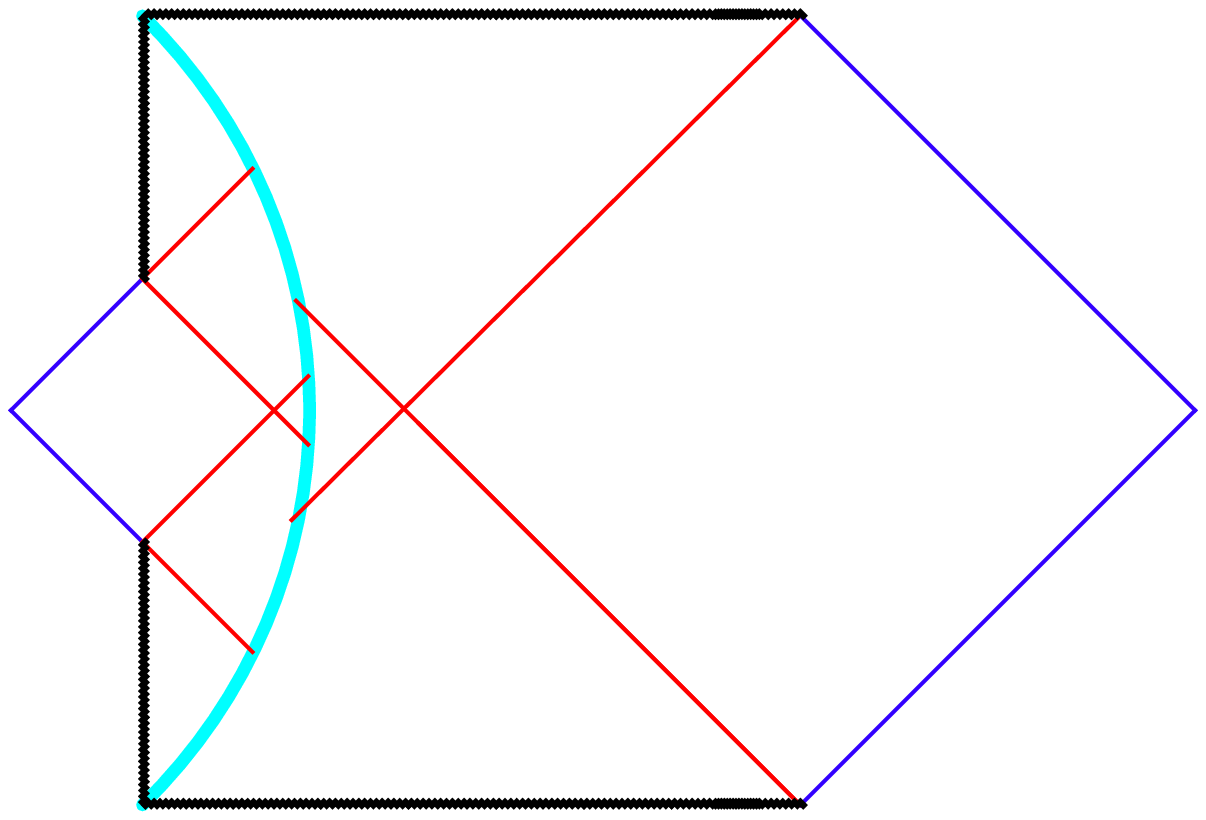}
\hskip0.1cm
\hfill
\includegraphics[angle=0,width=0.48\textwidth]{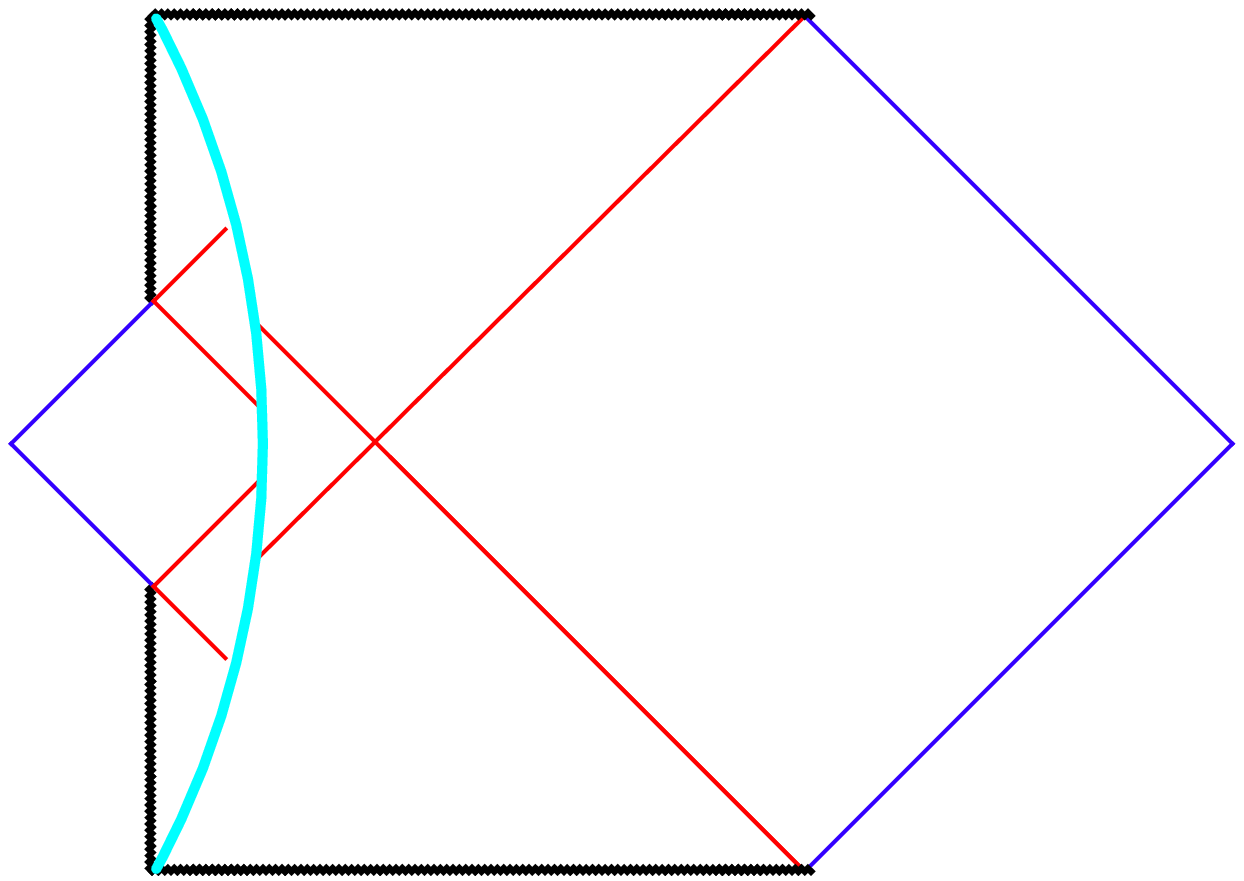}
\end{center}
\caption{Finite motion of the shell at $e^2=GM^2<Gm^2_{\rm in}$.
The left panel: at $ 0>\Delta m/M>
 \sqrt{(m_{\rm in}^2/M)^2-1} - (m_{\rm in}/M)$. The right panel: at $ -1<\Delta m/M<
 \sqrt{(m_{\rm in}^2/M)^2-1} - (m_{\rm in}/M)$.} \label{RN38}
\end{figure}

The conformal diagrams for finite motion at $e^2=GM^2$ and $\Delta
m<0$ are shown in Figs.~\ref{NS37}--\ref{RN38} for the cases $e^2=GM^2>GM^2m_{\rm in}$, $e^2=GM^2=GM^2m_{\rm in}$ and $e^2=GM^2<GM^2m_{\rm in}$ at
\begin{equation}
0>\frac{\Delta m}{M}>
 \sqrt{\frac{m_{\rm in}^2}{M^2}-1}-\frac{m_{\rm in}}{M}.
\end{equation}
and
\begin{equation}
-1<\frac{\Delta
 m}{M}< \sqrt{\frac{m_{\rm in}^2}{M^2}-1}-\frac{m_{\rm in}}{M}.
\end{equation}
The infinite motion in this case is impossible.

\subsection{Neutralizing shell at $\Delta m<0$}

We proceed now to the description of the possible alternative cases for the motion of the neutralizing shell at $\Delta m<0$. Mow the sign $\sigma_{\rm in}$ may be changed. For this reason the number of possible combinations of parameters is greatly increased. Now the internal game is going into the game and we need to consider the different types of the Reissner--Nordstr\"om metric: $e^2>Gm_{\rm in}$ --- naked singularity,
$e^2<Gm_{\rm in}^2$ --- black hole, and $e^2=Gm_{\rm in}^2$ --- extreme black hole.

We start our analysis from the simplest case from the point of view of the global Reissner--Nordstr\"om metric with naked singularity.

\subsubsection{Naked singularity at $e^2>Gm_{\rm in}^2$: infinite motion}

\begin{figure}[H]
\begin{center}
\includegraphics[angle=0,width=0.2\textwidth]{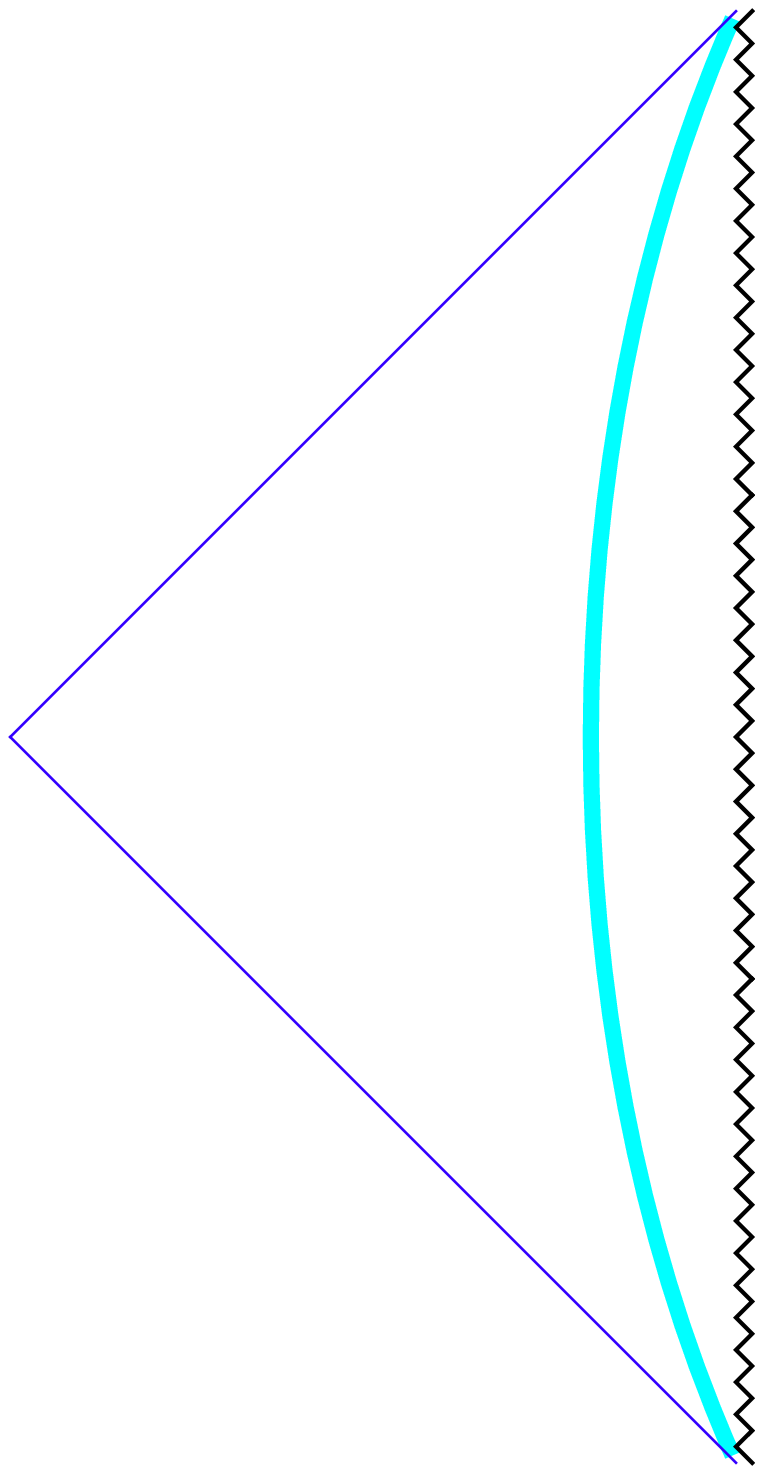}
\hfill
\includegraphics[angle=0,width=0.6\textwidth]{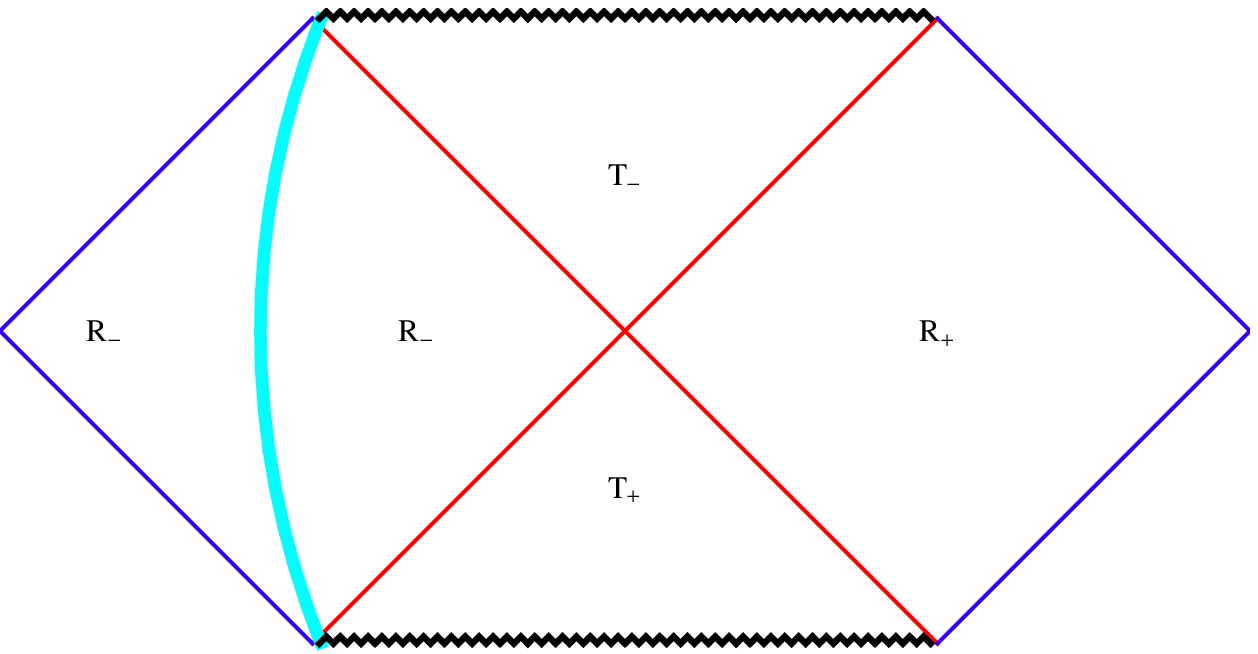}
\end{center}
\caption{Infinite motion of the self-repulsive shell at $\Delta m/M<-1$ and $e^2>G{m_{\rm in}}^2$. The left panel: internal metric. The right panel: the complete metric.} \label{inf47}
\end{figure}

As long as
\begin{equation}
 \sigma_{\rm in} = {\rm sign}[\Delta m+\frac{GM^2}{\rho}],
\end{equation}
then $\sigma_{\rm in}(0)=+1$ and $\sigma_{\rm in}(\infty)=-1$.
Therefore, at the infinite motion, starting at $\rho=\infty$, the trajectory can not reach the radial point $\rho=0$. In the case of the naked singularity the $T$-region is absent, and so it is absent the point, where $\sigma_{\rm in}$ may change the sign. For this reason it must inevitably the turning point in the $R_-$-region with a subsequent motion again to infinity. For this reason, the finite motion begins at $\rho=0$ in the $R_+$-region and must have the turning point at the same $R_+$-region (all from the point of you of the internal metric). In this case the conformal diagram for the internal metric has the form, shown in at the left panel in Fig.~\ref{inf47}. These qualitative conclusions must be still proved, i.\,e. it is requested to prove the real existence of these trajectories and define the necessary relations between the involved parameters: $M$, $\Delta m$ and $m_{\rm out}$ at the fixed $m_{\rm in}$ and $e$.

\begin{figure}
\begin{center}
\includegraphics[angle=0,width=0.9\textwidth]{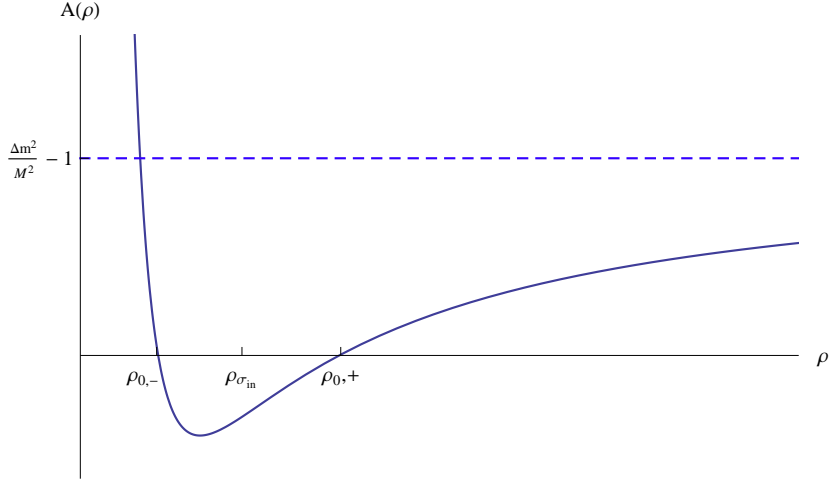}
\end{center}
\caption{The square trinomial $A(\rho)$ from equation (\ref{A}).} \label{Arho}
\end{figure}

At first consider relations for infinite motion of the shell:
\begin{eqnarray}
 \Delta m/M < -1\quad &\Rightarrow & \quad m_{\rm in}>M, \\
 e^2 > Gm_{\rm in}^2 \quad & \Rightarrow & \quad e^2 >GM^2.
\end{eqnarray}
This means, that in this case the shell is self-repulsive due to existence of the turning point. Further, let us see the equation for the turning point (\ref{bounce}). The both functions $B$ and $D$ may have, in principle, the different signs, since $\Delta m/M<-1$. We need to have $B<0$ and $D>0$. Only in this case there are two real roots and both of them are positive. We demonstrate that this is really the case. Strting from expression for $B$ in euation (\ref{BD}):
\begin{eqnarray}
 B &=& \frac{\Delta m}{M}(e^2\!+\!GM^2)\!+\!2GMm_{\rm in}, \\
  &=& \left(\frac{\Delta m}{M}+1\right)(e^2\!+\!GM^2)-
  (e^2\!-\!Gm_{\rm in}^2)-(Gm_{\rm in}^2-M)^2.
\end{eqnarray}
It is evident, that now $B < 0$.
Next, let us transform the expression (\ref{BD}) for $D$:
\begin{eqnarray}
 D &=& B^2-(GM^2-e^2)\left(\frac{{\Delta m}^2}{M^2}-1\right) \\
 &=& \!\!Ge^2\left[2\Delta m\!+\!\frac{m_{\rm in}}{e^2}(e^2\!+\!GM^2)\right]^2
 \!\!+\! (GM^2\!-\!e^2)^2\!\left(\!1\!-\!\frac{Gm_{\rm
 in}^2}{e^2}\!\right)\!>0,
\end{eqnarray}
since $e^2>Gm_{\rm in}^2$. We obtain for the used combination of parameters that equation for turning point really has two positive roots (formally there are two turning point: $\rho=\rho_{0,-}$ and $\rho=\rho_{0,+}\geq\rho_{0,-}$). It is clear, starting from infinity, the shell at first reaches the turning point, corresponding to the bigger root $\rho=\rho_{0,+}$, and after that again is moving toward the radial infinity. We need to clarify where it is placed the the point of the sign changing $\sigma_{\rm in}$, i.\,e., is it identically valid $\rho_{\sigma_{\rm in}}<\rho_{0,+}$, or there is additional constraint to parameters? Point is that $\rho_{\sigma_{\rm in}}$ always exists at $\Delta m<0$, but the shell cannot reach this radius because the necessary for sign changing $T$-region is absent at ll in the case of the naked singularity. Thus,
\begin{equation}
 \frac{1}{2\rho_{0,\pm}M}=\frac{-B\pm\sqrt{D}}{(e^2-GM^2)^2}, \quad
 \rho_{\sigma_{\rm in}}=-\frac{e^2+GM^2}{2\Delta m}
\end{equation}
To find the point $\rho_{\sigma_{\rm in}}$ we consider the square trinomial $A(\rho)$, With roots, corresponding to the turning points:
\begin{equation}
 A(\rho)=\frac{(e^2-GM^2)^2}{4\rho^2M^2}+\frac{2B}{2\rho M}+
 \frac{{\Delta m}^2}{M^2}-1.
 \label{A}
\end{equation}
The corresponding graphs is shown in Fig.~\ref{Arho}.
The value of $A(\rho)$ at the point $\rho_{\sigma_{\rm in}}$ is
\begin{equation}
 A(\rho_{\sigma_{\rm in}})=
 -\frac{G{m_{\rm in}}^2}{e^2}\left(\frac{\Delta m}{M}
 \frac{e^2}{{e^2}+GM^2}+1\right)^2-\left(1-\frac{G{m_{\rm
 in}}^2}{e^2}\right)<0.
\end{equation}
It is remarkable that now $\rho_{\sigma_{\rm in}}$ is just placed between bigger and smaller turning points.

In that way we see, that infinite motion at $\Delta m/M<-1$ and $e^2>G{m_{\rm in}}^2$ is possible only for the self-repulsive shell. In this case the conformal diagram for the internal metric has the form, shown in at the left panel in Fig.~\ref{inf47}.
At the same time, the point for sign changing of $\sigma_{\rm out}$ exists, in which connection $\rho_{\sigma_{\rm out}}<\rho_{\sigma_{\rm
in}}$. That is why the shell, moving from infinity does not reach the turning point $\rho_{\sigma_{\rm out}}$ and so $\sigma_{\rm out}=-1$. The corresponding complete conformal diagram is shown at he rigt panel in Fig.~\ref{inf47}. Left form the shell at this diagram there is not an electro-vacuum space-time with the infinity, but someone source with a total mass $m_{\rm in}$ and electric charge $e$. For example, it may be the thin shell, again with the Schwarzschild metric at left side from the shell.

A behavior of the parabolic trajectory at $\Delta m/M=-1$ is quite similar to the previously discussed the hyperbolic one. At this point we finish the description of the infinite motion.

\subsubsection{Naked singularity $e^2>Gm_{\rm in}^2$: finite motion}

Now we start a consideration of the finite motion at
\begin{equation}
 -1<\frac{\Delta m}{M}<0, \quad  e^2>G{m_{\rm in}}^2.
\end{equation}
Since $\sigma_{\rm in}(0)=+1$, and $T$-region is absent, then
$\sigma_{\rm in}=+1$ everywhere at the trajectory. The corresponding equation for the turning points has one positive and one negative root, i.\,e., there is only one turning point, which is physically acquitted.

\begin{figure}[h]
\begin{center}
\includegraphics[angle=0,width=0.95\textwidth]{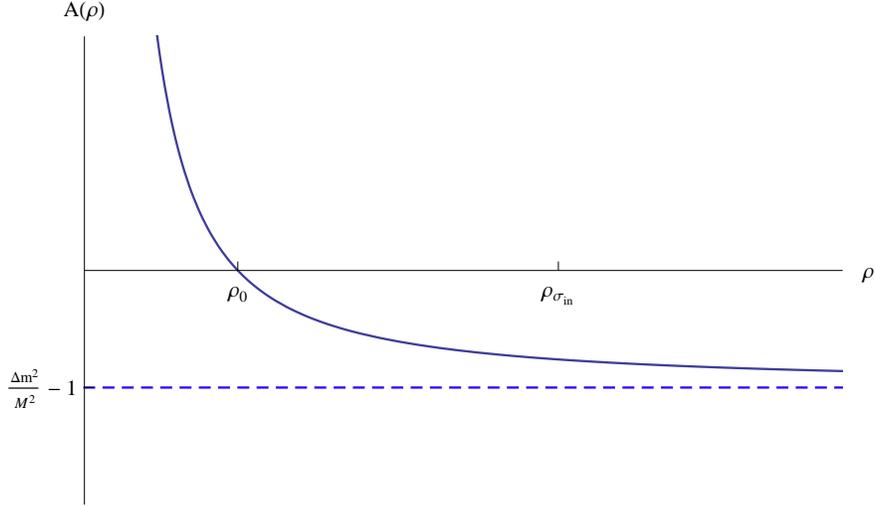}
\end{center}
\caption{Graph of the function $A(\rho)$ from (\ref{A}) for the case of the finite motion of the shell with one turning point.} \label{Arho49}
\end{figure}

The turning point $\rho_0$ in the equation (\ref{bounce}) corresponds to the positive value of the discriminant $D$, but $B$ may have any sign. Herewith there is point $\rho_{\sigma_{\rm in}}$ for sign changing in $\sigma_{\rm in}$, but we need to verify that $\rho_{\sigma_{\rm
in}}>\rho_0$. As previously, $A(\rho_{\sigma_{\rm in}})<0$, and graph of the function $A(\rho)$ from (\ref{A}) is shown in Fig.~\ref{Arho49}.
Note, that now there is possible as in the case $e^2>GM^2$ (self-repulsive shell) and $e^2<GM^2$ (self-attractive shell).

\begin{figure}[H]
\begin{center}
\includegraphics[angle=0,width=0.6\textwidth]{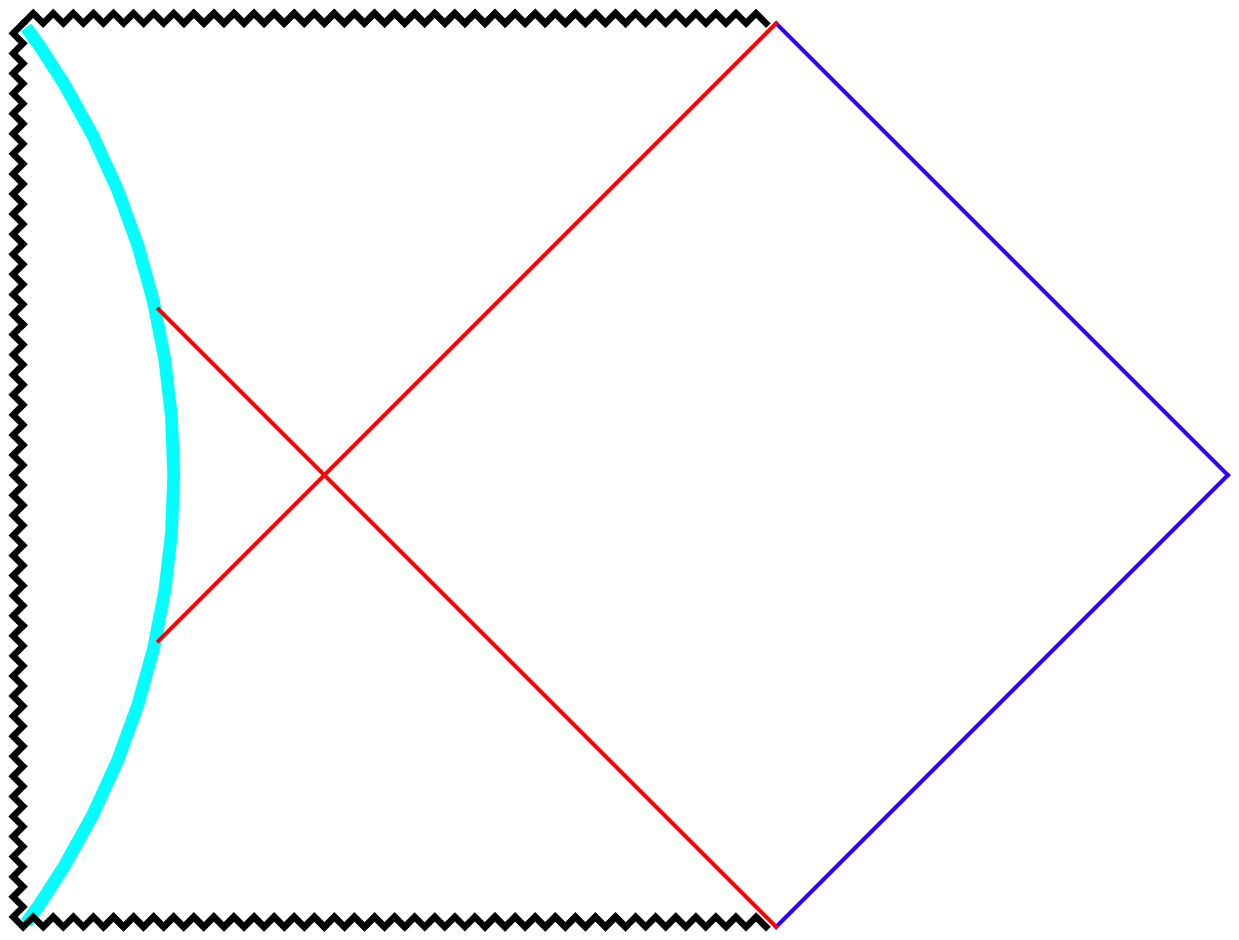}
\end{center}
\caption{Conformal diagram for the case $e^2<GM^2$, when $\rho_{\sigma_{\rm out}}$ does not exist and $\sigma_{\rm out}(0)=-1$.}
 \label{P50}
\end{figure}

We start from the case  $e^2<GM^2$, when $\rho_{\sigma_{\rm out}}$ does not exist and $\sigma_{\rm out}(0)=-1$. This mean, that the moving shell has turning point in the  $R_-$-region of the external metric. The complete conformal diagram shown in Fig.~\ref{P50}.

If  $e^2>GM^2$, then  $\rho_{\sigma_{\rm out}}$ exists because $\Delta m<0$ and $\sigma_{\rm out}(0)=1$. Now there are possible two cases:

(1) $\rho_{\sigma_{\rm out}}$ is placed in the $T_-$-region. The shell is starting at $\rho=0$, then  $\sigma$ changes its sign and the shell is coming to the $R_-$-region, where there is turning point for the trajectory;

(2) $\rho_{\sigma_{\rm out}}$ is placed outside of the event horizon, i.\,e., $\rho_{\sigma_{\rm out}}>2Gm_{\rm out=r_g}$. In this case the turning point is placed in the $R_+$-region. Besides, it is held true $\rho_{\sigma_{\rm out}}>\rho_0$.

Consider the inequality
\begin{equation}
 \rho_{\sigma_{\rm out}}>r_g \quad \Rightarrow \quad
 -M<\Delta m<-\frac{e^2-GM^2}{4Gm_{\rm out}}.
\end{equation}
In this case the turning point is placed in the $R_-$-region. If it is satisfied
\begin{equation}
 -\frac{e^2-GM^2}{4Gm_{\rm out}}<\Delta m,
\end{equation}
then the turning point is placed in the $R_+$-region.

Now we find the location of the turning point $\rho_0$. To do this we calculate
\begin{equation}
 A(\rho_{\sigma_{\rm out}})=
 1-\frac{4\Delta m\,m_{\rm out}}{e^2-GM^2}<0.
\end{equation}
Since $A(\rho_{\sigma_{\rm out}})<0$, then $\rho_{\sigma_{\rm
out}}>\rho_0$, and the shell is moving without changing the sign of $\sigma$.
The finite motion is realized at $-M<\Delta m<0$. This means that
$m_{\rm in}-M<m_{\rm out}<m_{\rm in}$ and, consequently,
\begin{equation}
 -M<\Delta m<-\frac{e^2-GM^2}{4Gm_{\rm out}}
\end{equation}
In this case there is a turning point in the $R_-$-region. At
\begin{equation}
0>\Delta m>-\frac{e^2-GM^2}{4Gm_{\rm out}}
\end{equation}
The shell is оболочка is passing through the $R_+$-region of the external metric. At last, at
\begin{equation}
m_{\rm out}<\frac{e^2-GM^2}{4GM},
\end{equation}
the shell trajectory always has the turning point in the $R_+$-region.

\subsubsection{Extreme black hole at $e^2=Gm_{\rm in}^2$: infinite motion}

In the case of the extreme Reissner--Nordstr\"om black hole with
$e^2=Gm_{\rm in}^2$ for internal metric we have
\begin{eqnarray}
 B &=& \left(\frac{\Delta m}{M}+1\right)(e^2\!+\!GM^2)-
  (e-\sqrt{G}M)^2 \\
  &=& \frac{\Delta m}{M}(e^2\!+\!GM^2)+2\sqrt{G}Me, \\
 D &=& \left(2e\Delta m +e^2+GM^2\right)^2>0.
\end{eqnarray}
For turning point we find
\begin{equation}
 \frac{1}{2\rho_0(\pm)M}=\frac{-\frac{\Delta m}{M}(e^2\!+\!GM^2)
 \!-\!
 2\sqrt{G}Me\pm|2e\sqrt{G}\Delta m\!+\!e^2\!+\!GM^2|}{(e^2-GM^2)^2}
\end{equation}
The difference of the case, considered previously for the naked singularity is the possibility for turning point to be placed as beyond and also inside of the double horizon of the internal metric $r_+=r_-=Gm_{\rm in}=\sqrt{G}e$. The discriminant $D$ now may be equal to zero at
\begin{equation}
\Delta m=-\frac{e^2+GM^2}{2e\sqrt{G}}.
\end{equation}
In this case we find that $\rho_0(+)=\rho_0(-)=r_\pm$, i.\,e., the double turning point is placed exactly at the double horizon. As before, in the case of infinite motion there are two positive roots, while at the finite motion one root is positive and the second one is negative.

\begin{figure}[H]
\begin{center}
\includegraphics[angle=0,width=0.32\textwidth]{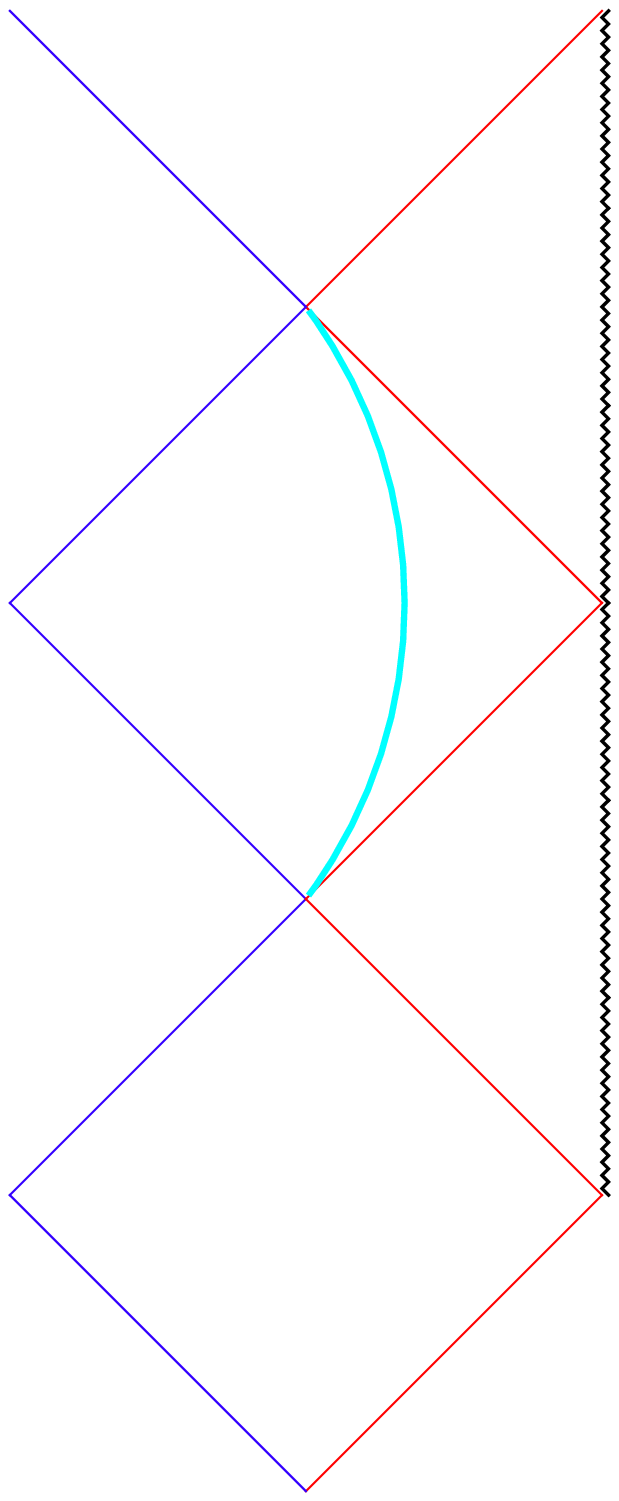}
\hskip1.5cm
\includegraphics[angle=0,width=0.32\textwidth]{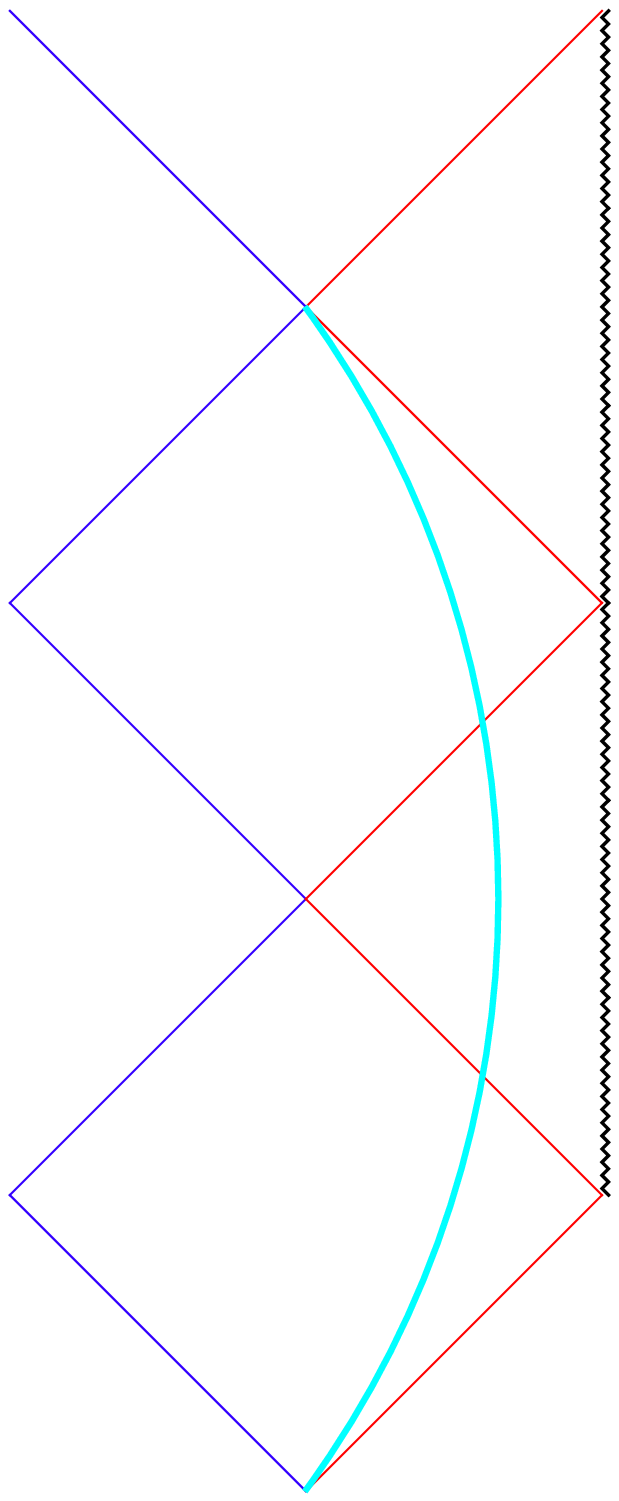}
\end{center}
\caption{Internal metric with extreme black hole. Infinite motion with the turning point.} \label{RNextremeP60}
\end{figure}

We start our analysis from the infinite motion at $\Delta m < -M$ when
$\sigma_{\rm in}(\infty)={\rm sign}(\Delta m)=-1$. The corresponding Carter--Penrose diagram for the internal metric is shown in Fig.~\ref{RNextremeP60} (at the left panel).

Let us initially
\begin{equation}
-\frac{e^2+GM^2}{2e\sqrt{G}}<\Delta m<-M.
\end{equation}
We verify now that these inequalities are compatible:
\begin{equation}
 -\frac{e^2+GM^2}{2e\sqrt{G}}<-M \quad \Rightarrow \quad
 -(e-\sqrt{G}M)^2\leq0.
\end{equation}
We find for the turning point
\begin{equation}
\rho_0(\pm)=\frac{(e\pm\sqrt{G}M)^2}{2M\left(-\frac{\Delta
m}{M}\pm1\right)}.
\end{equation}
Let us show that at
\begin{equation}
-\frac{e^2+GM^2}{2e\sqrt{G}}<\Delta m
\end{equation}
it is held the relation. Really, we have
\begin{equation}
 \frac{(e\!-\!\sqrt{G}M)^2}{2M\left(-1-\frac{\Delta
m}{M}\right)}>\frac{(e\!+\!\sqrt{G}M)^2}{2M\left(1-\frac{\Delta
m}{M}\right)}
 \quad \Rightarrow \quad
 \Delta m>-\frac{e^2\!+\!GM^2}{2\sqrt{G}e},
\end{equation}
as it must be. Similar to the case of the naked singularity, the point of the changing sign of $\sigma$ is placed between two roots. This is easily verified by the direct calculations:
\begin{equation}
 A(\rho_\sigma)<0 \quad \Rightarrow \quad
 \rho_0(+)<\rho_\sigma<\rho_0(-).
\end{equation}
For the point of the double horizon $r_\pm$ we find:
\begin{equation}
 A(r_\pm)=\left(\frac{e^2+GM^2}{2GMr_\pm}+\frac{\Delta
m}{M}\right)^2.
\end{equation}
By the direct calculations we verify that
\begin{equation}
 r_\pm<\rho_0(+) \quad \Rightarrow \quad
 \Delta m>-\frac{e^2\!+\!GM^2}{2\sqrt{G}e},
\end{equation}
i.\,e., at $\Delta m>-(e^2\!+\!GM^2/(2\sqrt{G}e)$ the smaller turning point $\rho_0(+)$ is placed outside the double horizon. The shell, starting at infinity, is reaching the turning point, which is the larger root, and further is moving to infinity without meeting the smaller turning point. The corresponding diagrams is shown in Fig.~\ref{RNextremeP60} (at the left panel). At $\Delta m/M\to\infty$ this turning point tends to infinity, and at $\Delta m\to-(e^2\!+\!GM^2/(2\sqrt{G}e)$ the turning point точка tends to the double horizon, where the both turning points are also doubling. At $\Delta  m=-(e^2\!+\!GM^2/(2\sqrt{G}e)$, as we already know, $\rho_0(+)=\rho_0(-)=\rho_\sigma=r_\pm$.
\begin{figure}[H]
\begin{center}
\includegraphics[angle=0,width=0.5\textwidth]{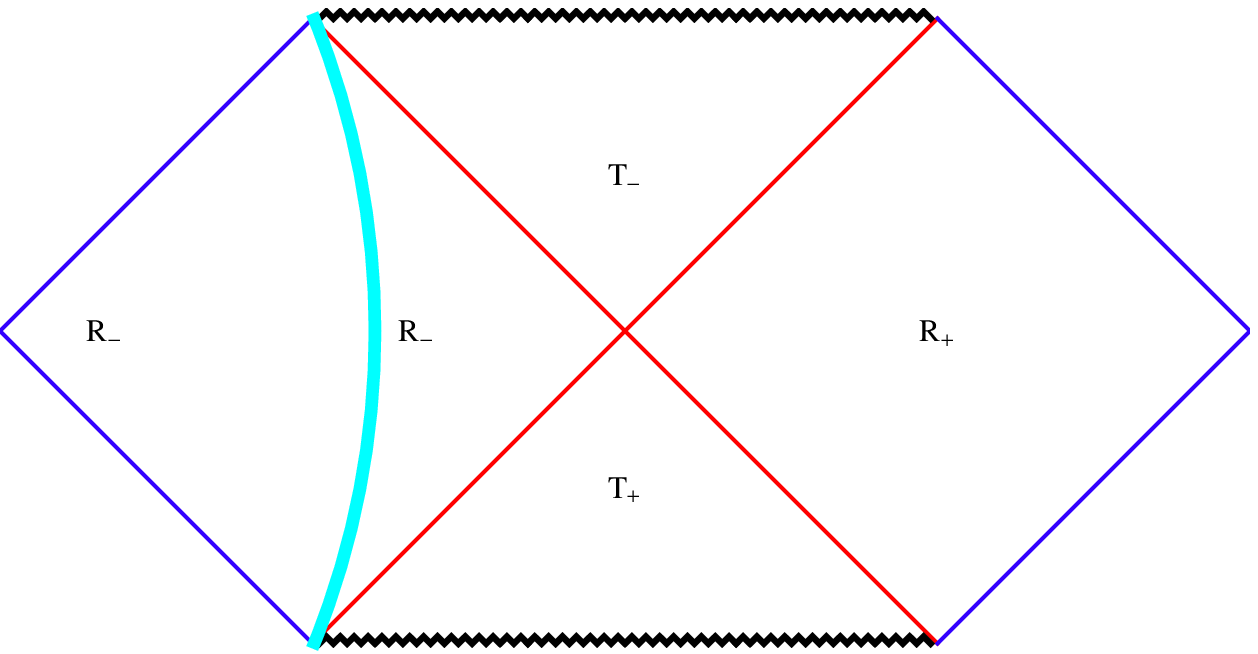}
\hfill
\includegraphics[angle=0,width=0.45\textwidth]{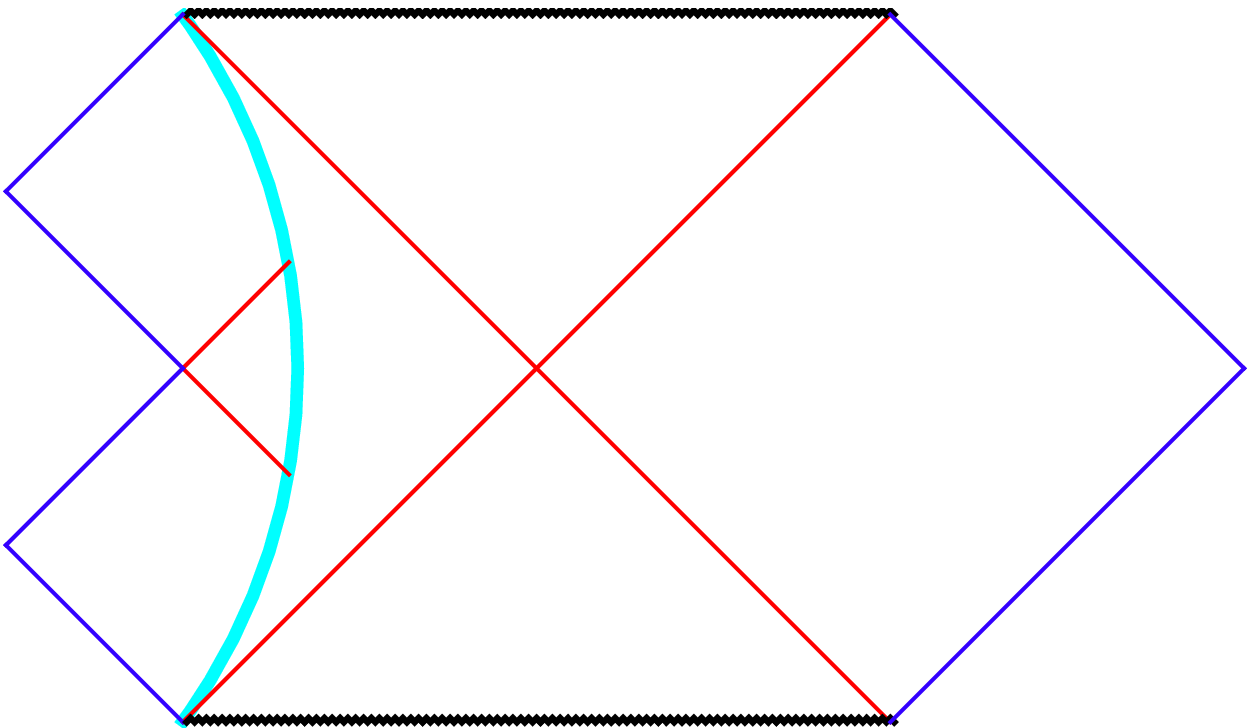}
\end{center}
\caption{The case of extreme black hole. Infinite motion with the turning point: at $-(e^2\!+\!GM^2/(2\sqrt{G}e)<\Delta m<-M$ (at the left panel) and at $\Delta m<-(e^2\!+\!GM^2/(2\sqrt{G}e)<-M$ (at the right panel). }
 \label{RNextremeP63}
\end{figure}

In the event $\Delta m<-(e^2\!+\!GM^2/(2\sqrt{G}e)$, the turning points
$\rho_0(+)$ and $\rho_0(-)$ switch their positions, i.\,e.,
\begin{equation}
\rho_0(\pm)=\frac{(e\mp\sqrt{G}M)^2}{2M\left(-\frac{\Delta
m}{M}\mp1\right)},
\end{equation}
As before, $\rho_0(+)<\rho_\sigma<\rho_0(-)$, however, $r_\pm>\rho_0(-)$, and, consequently, the both turning points are placed inside the double horizon. Remind, that from $\Delta m<-M$ it follows that
$M<m_{\rm in}=e/\sqrt{G}$. The corresponding diagrams is shown in Fig.~\ref{RNextremeP60}  (at the right panel).

For the external metric at $e^2>GM^2$ there exists the point $\rho_{\sigma_{\rm out}}$. It is evident, that $\rho_{\sigma_{\rm out}}<\rho_{\sigma_{\rm
\in}}<\rho_0$, where $\rho_0$ --- is the larger root, and, so $\sigma_{\rm out}(\rho_0)=-1$. The corresponding conformal diagrams ar show in Fig.~\ref{RNextremeP63}. At the left panel it is clear seen, that $r_g=2Gm_{\rm out}<r_\pm=Gm_{\rm in}=\sqrt{G}e$, as it must be at  $\Delta m<-(e^2\!+\!GM^2/(2\sqrt{G}e)$. It is impossible to draw differently, because $\sigma_{\rm out}(\rho_0)=-1$.

\subsubsection{Extreme black hole at $e^2=Gm_{\rm in}^2$: finite motion}

Let us investigate the finite motion, when $\Delta m > -M$,
$\sigma_{\rm in}(0)=+1$ and, therefore, $\sigma_{\rm in}=+1$ everywhere
(see Fig.~\ref{RNextremal}). Now we have only one turning point $\rho_0$, since the second root is negative. It is necessary to establish relations between three crucial radii of the shell: $\rho_0$, $\rho_{\sigma_{\rm in}}$, $\rho_{\sigma_{\rm out}}$ and also the radius of the double horizon $r_\pm$. For the turning point we find
\begin{equation}
\rho_0=\frac{1+\frac{\Delta m}{M}}{(e+\sqrt{G}M)^2},
\label{rho04}
\end{equation}
It is simply to verify that $r_\pm<\rho_0<\rho_{\sigma_{\rm in}}$.
Consequently, the shell, starting from $r=0$, come out beyond the double horizon and meet the turning point, before the reaching the point
$\rho_{\sigma_{\rm in}}$ (the point of changing sign of $\sigma$). This means, that $\sigma_{\rm in}=+1$ everywhere at the trajectory.

Now we investigate the point $\rho_{\sigma_{\rm out}}$. For the self-attractive shell with $e^2<GM^2$, the point of the changing sign of $\sigma_{\rm out}$ does not exist ($\Delta m<0$). Since $\sigma_{\rm
out}(0)=-1$, then $\sigma_{\rm out}=-1$ everywhere at the trajectory. Conformal diagram for the corresponding internal metric is shown at the left panel in Fig.~\ref{SchwFinP25}.

For the self-repulsive shell with $e^2>GM^2$ there is point of changing sign of $\sigma_{\rm out}$, and $\sigma_{\rm out}(0)=+1$. For this reason the shell, in principle, may come out to the  $R_-$-, and also to the $R_+$-region of the external metric. In the first case we have $\rho_{\sigma_{\rm out}}<r_g=2Gm_{\rm out}$, and, respectively, in the second case we have $\rho_{\sigma_{\rm out}}>\rho_0>r_g=2Gm_{\rm out}$.

Let us verify, at first, that $\rho_0\geq r_g$:
\begin{equation}
 \frac{(e\!+\!\sqrt{G}M)^2}{2(M\!-\!\Delta m)}>2Gm_{\rm out}
 \quad \Rightarrow \quad
 (e\!-\!\sqrt{G}M\!+\!\sqrt{G}\Delta m)^2\geq0.
\end{equation}
Consider, further, the case $\rho_{\sigma_{\rm out}}<r_g$, when turning point is placed in the  $R_-$-region of the external metric. Well, we have
\begin{equation}
 -\frac{e^2\!-\!GM^2}{2\Delta m}<2Gm_{\rm out}
 \quad \Rightarrow \quad
 \sqrt{G}(2m_{\rm out}-m_{\rm in})>-\frac{GM^2}{2}.
\end{equation}
The expression at the left part of the last inequality may be as positive and negative. For this reason, we well the both cases separately.

(1) If
\begin{equation}
 -\frac{GM^2}{2}<\sqrt{G}(2m_{\rm out}-m_{\rm in})<0,
\end{equation}
then
\begin{equation}
 -2m_{\rm out}+m_{\rm in})M \quad \Rightarrow \quad
 m_{\rm in}<0<2m_{\rm out} \quad \Rightarrow \quad
 \rho_{\sigma_{\rm out}}<r_g.
\end{equation}

\begin{figure}[H]
\begin{center}
\includegraphics[angle=0,width=0.39\textwidth]{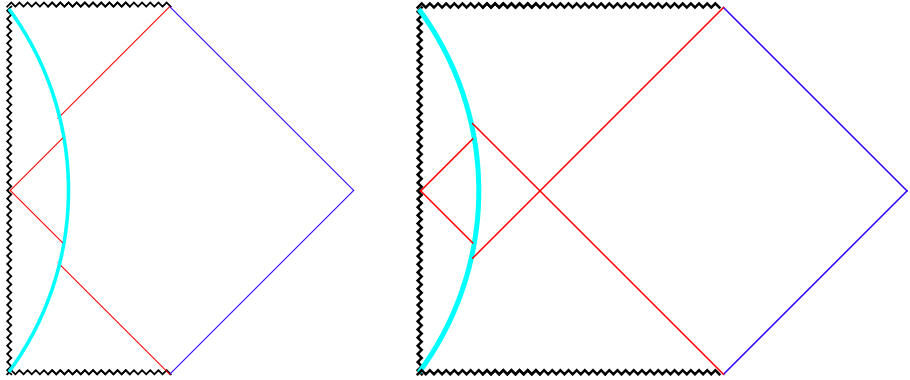}
\hskip0.2cm
\hfill
\includegraphics[angle=0,width=0.55\textwidth]{NSp37b.eps}
\end{center}
\caption{The case of extreme black hole. Finite motion at $2m_{\rm out}>M+m_{\rm in}$ and $\Delta m>(M-m_{\rm in})/2$.}
 \label{ExtrRNp71c}
\end{figure}

(2) If
\begin{equation}
 2\sqrt{G}\Delta m+e>0,
\end{equation}
then
\begin{equation}
 2\sqrt{G}\Delta m\!+\!e>0 \quad \Rightarrow \quad
 m_{\rm in}<2m_{\rm out}<M\!+\!m_{\rm in} \quad \Rightarrow \quad
 \rho_{\sigma_{\rm out}}<2r_g.
\end{equation}
In result, for implementation of the inequality $\rho_{\sigma_{\rm out}}<2r_g$, it is necessary
\begin{equation}
 m_{\rm out}<\frac{M+m_{\rm in}}{2}.
\end{equation}
The the moving shell is coming in the $R_+$-region, then $\rho_{\sigma_{\rm
out}}>2Gm_{\rm out}$ and $(m_{\rm in}2m_{\rm out})^2>M^2$. In the case
$m_{\rm in}>2m_{\rm out}$, the inequality is valid, $m_{\rm
in}-2m_{\rm out}>M$, and, therefore, $2m_{\rm out}<m_{\rm in}-M<0$.
However, the last inequality contradicts to condition $m_{\rm out}>0$.
Meantime, if $m_{\rm in}<2m_{\rm out}$, then
\begin{equation}
 2m_{\rm out}-m_{\rm in}>M \quad \Rightarrow \quad
 m_{\rm out}>\frac{M+m_{\rm in}}{2}.
\end{equation}
It is remained to verify that $\rho_0<\rho_{\sigma_{\rm
out}}$. Really, we obtain
\begin{equation}
 \frac{(e\!+\!\sqrt{G}M)^2}{2(M\!-\!\Delta m)}
 <-\frac{e^2-GM^2}{1\Delta m}
 \quad \Rightarrow \quad
2m_{\rm out}-m_{\rm in}>M.
\end{equation}
The total conformal diagram at $m_{\rm in}>2m_{\rm out}$
corresponds to the left panel in Fig.~\ref{NS27cd}, and at  $m_{\rm
in}<2m_{\rm out}<M+m_{\rm in}$ and $\Delta m<(M-m_{\rm in})/2$, respectively, to the right panel in Fig.~\ref{NS37}. At last, the total conformal diagram at $2m_{\rm out}>M+m_{\rm in}$ and  $\Delta m>(M-m_{\rm in})/2$ is shown in Fig.~\ref{ExtrRNp71c}.

\subsubsection{Black hole at $e^2<Gm_{\rm in}^2$: finite motion}

Now we initiate investigation of the most multivariant case with $\Delta m<0$, when inside the neutralizing shell there is the reissner--Nordstr\"om black hole with  $e^2<Gm_{\rm in}^2$. The Carter--Penrose diagram for this internal metric is shown in Fig.~\ref{RN}.

\begin{figure}
\begin{center}
\includegraphics[angle=0,width=0.48\textwidth]{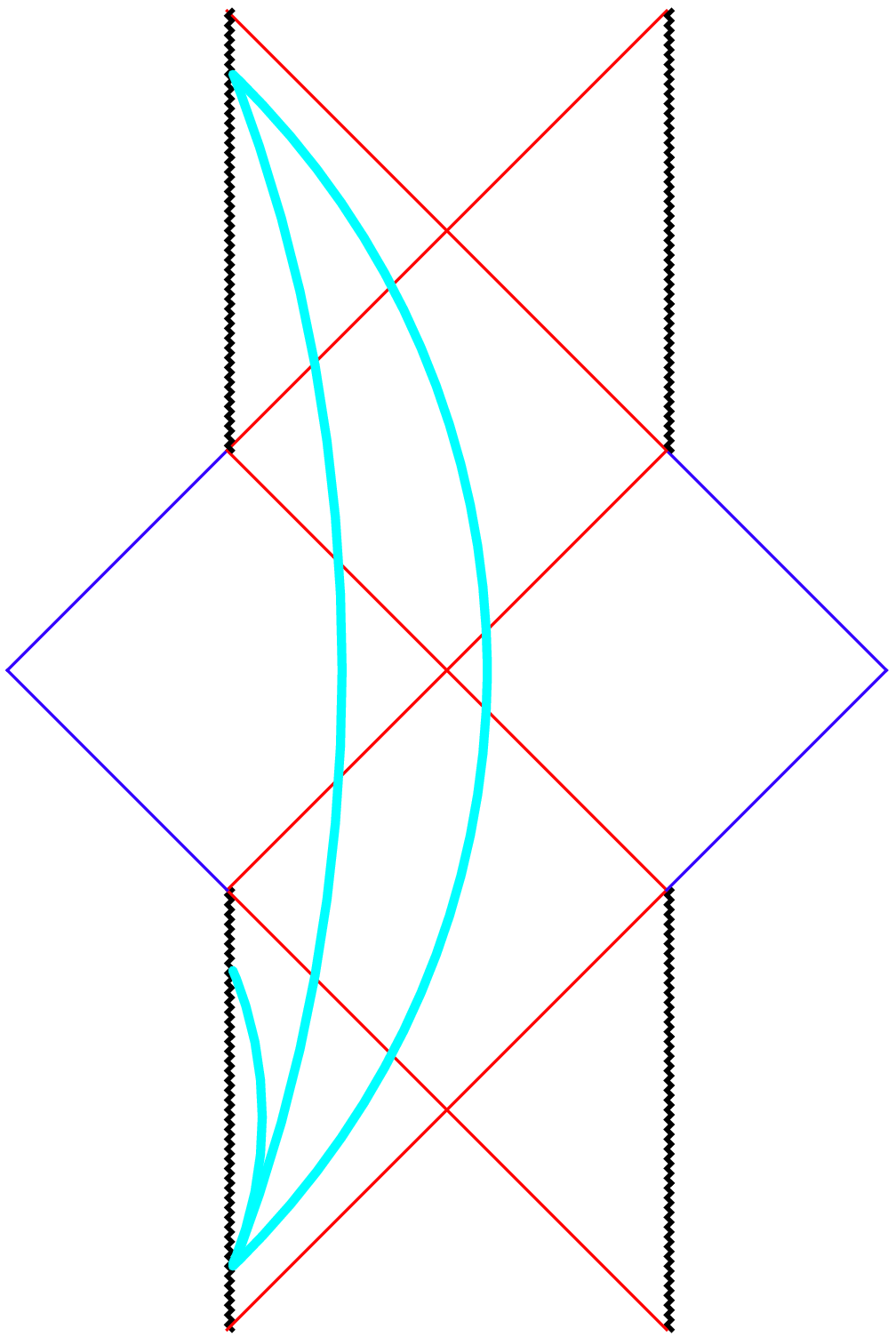}
\hfill
\includegraphics[angle=0,width=0.48\textwidth]{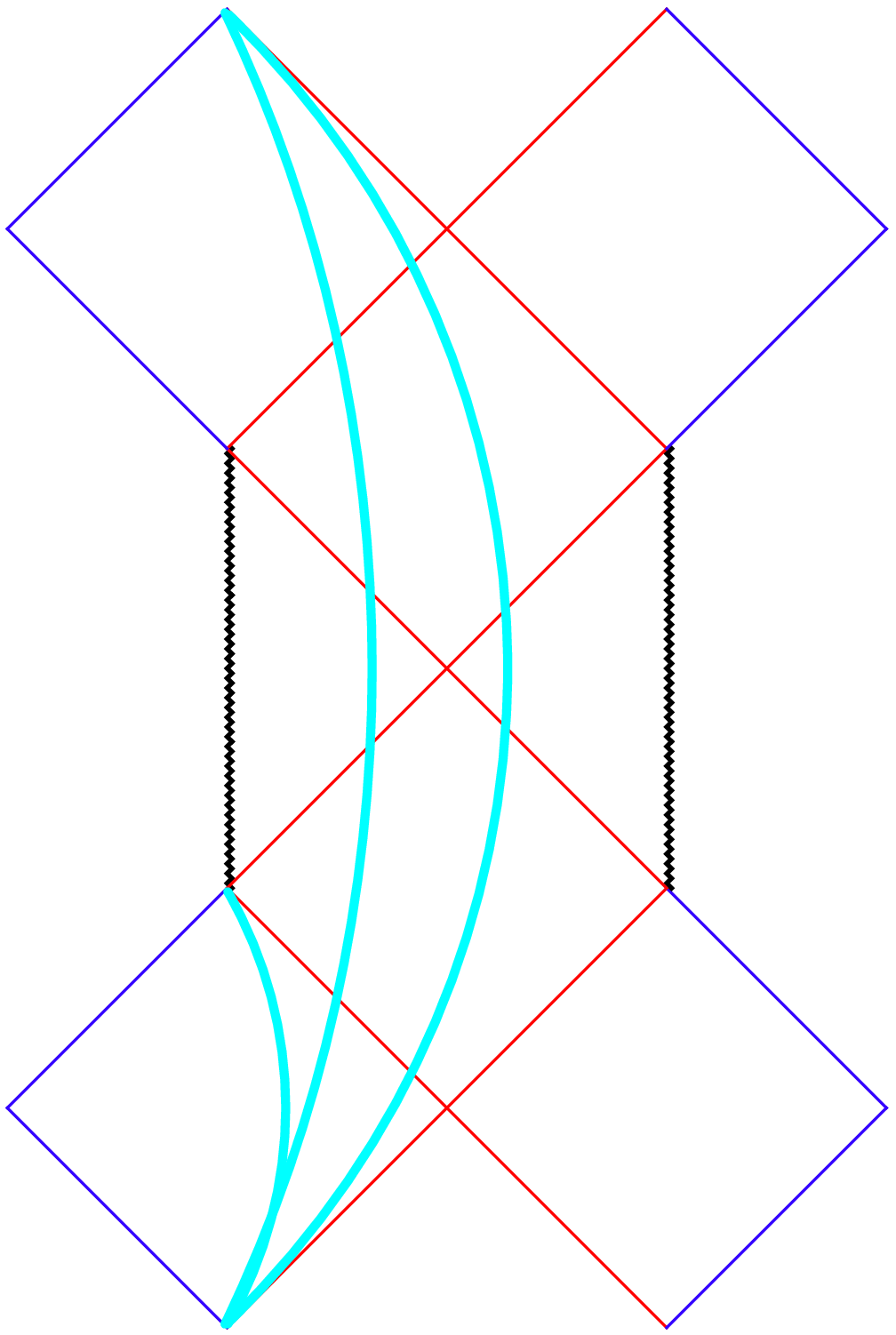}
\end{center}
\caption{All possible trajectories of the shell at $\Delta m<0$ and $e^2<Gm_{\rm in}^2$ for the internal metric.}
 \label{DeltaMinusInt}
\end{figure}

\begin{figure}
\begin{center}
\includegraphics[angle=0,width=0.9\textwidth]{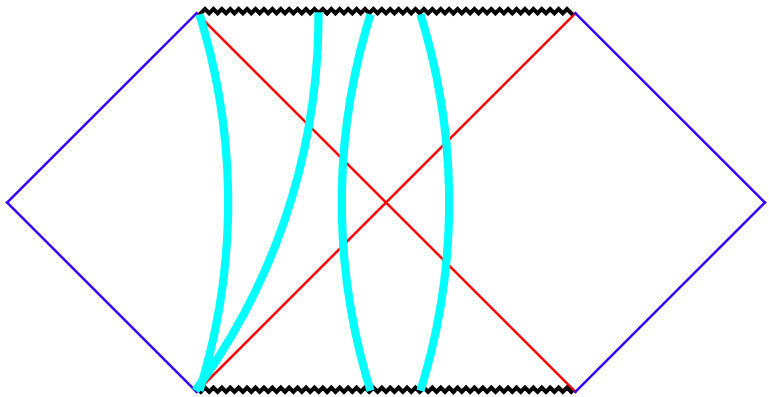}
\end{center}
\caption{All possible trajectories of the shell at $\Delta m<0$ and $e^2<Gm_{\rm in}^2$ for the external metric.}
 \label{DeltaMinusExt}
\end{figure}

Now there is the $T_\pm$-region, where the sign of $\sigma_{\rm in}$ may be changing, and two $R_\pm$-regions, where may be the turning points. All possibilities at $\Delta m<0$ and $e^2<Gm_{\rm in}^2$ for the internal metric are shown in Fig.~\ref{DeltaMinusInt}, and for the external metric, respectively, in Fig.~\ref{DeltaMinusExt}

\begin{figure}
\begin{center}
\includegraphics[angle=0,width=0.8\textwidth]{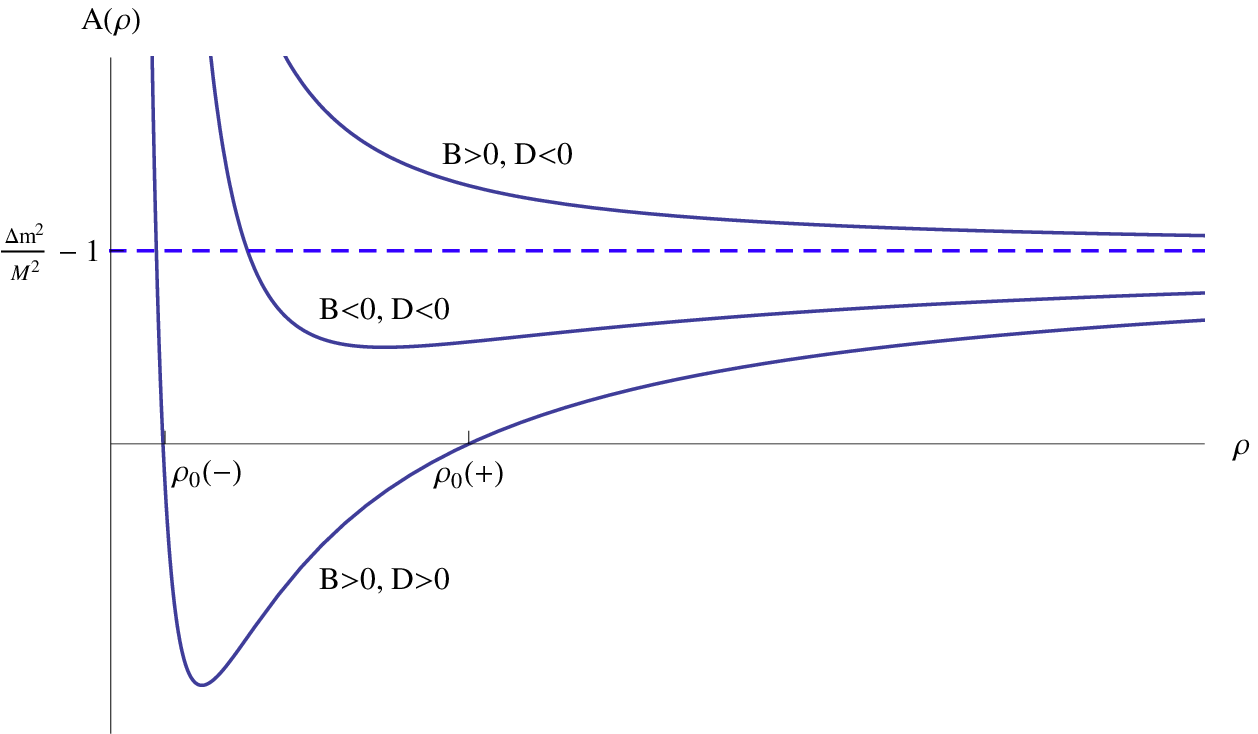}
\end{center}
\caption{The graphs for the trinomial $A(\rho)$ for infinite motion, when $\Delta
m/M<-1$. The allowed region for motion is at $A(\rho)\geq0$.}
 \label{Arho75a}
\end{figure}

\begin{figure}[H]
\begin{center}
\includegraphics[angle=0,width=0.8\textwidth]{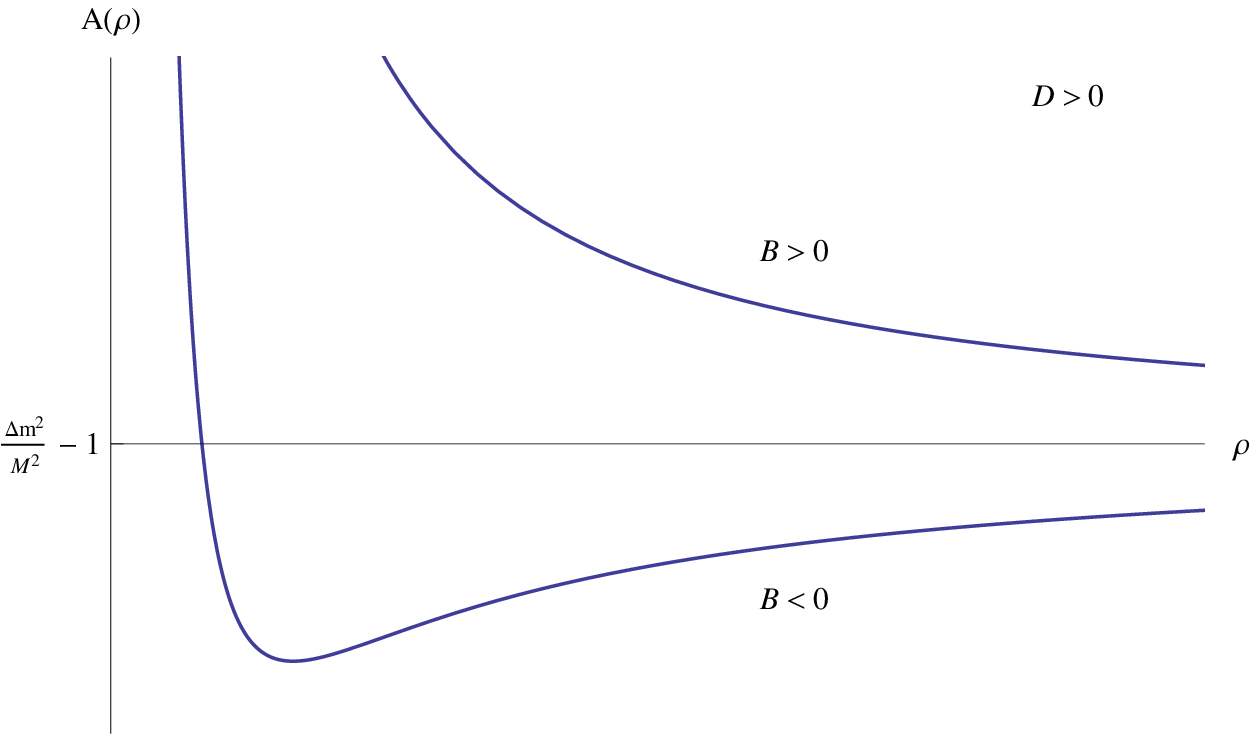}
\includegraphics[angle=0,width=0.8\textwidth]{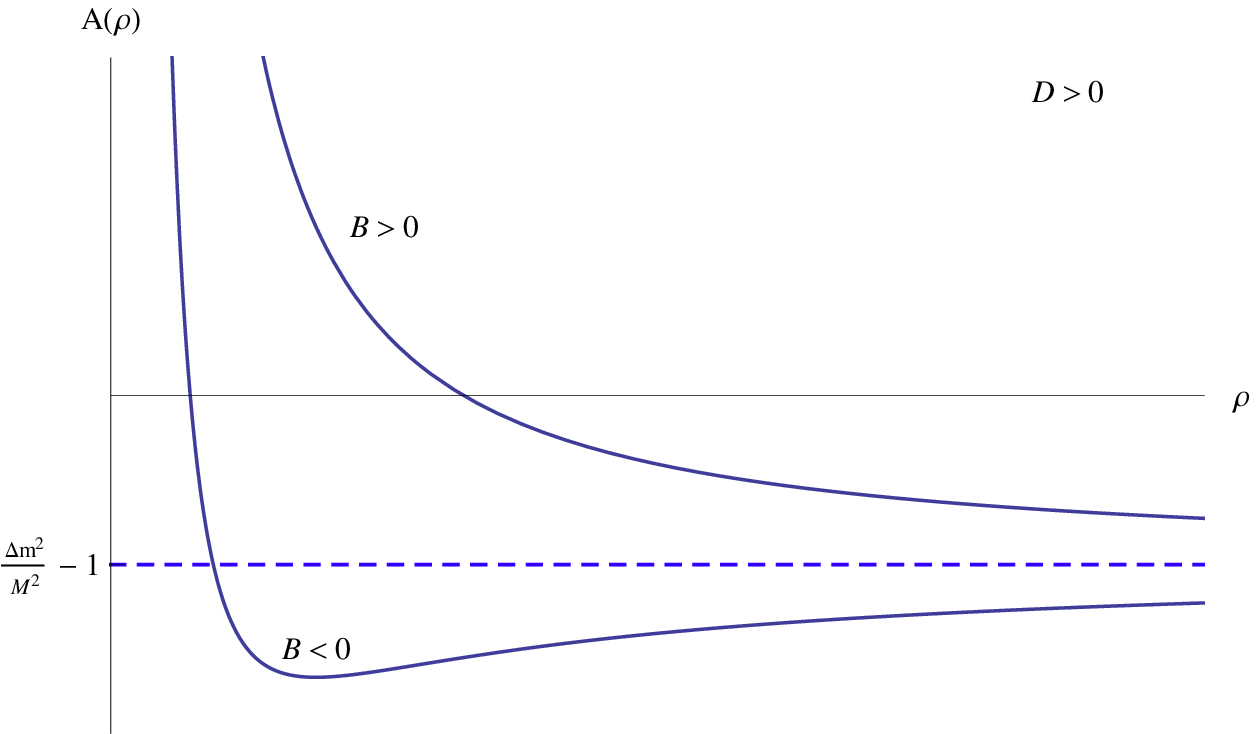}
\end{center}
\caption{The graphs for the trinomial $A(\rho)$ for parabolic motion (the upper panel),
when $\Delta m/M=-1$ and motion is possible only at $B>0$, and, respectively, for finite motion
(the lower panel), when $\Delta m/M>-1$ and there is only the one turning point.}
 \label{Arho75bc}
\end{figure}

Before to proceed further, let us see the graphs for the trinomial $A(\rho)$ in dependance of the relations between the specified parameters in Figs.~\ref{Arho75a}--\ref{Arho75bc}.
It is possible, in principle, the infinite motion at $\Delta
m/M<-1$, parabolic motion at $\Delta m/M<-1$ and finite motion at $0>\Delta
m/M>-1$ with the one turning point. We see, that for every type of motion,
the graphs in Figs.~\ref{Arho75a}--\ref{Arho75bc} differ in dependance of the sign of the coefficient $B$ and the sign of the discriminant $D$.

Let us consider the behavior of $B$ and $D$ in variation of the value $\Delta m<0$. Both $B$ and $D$ are the linear functions of $\Delta m<0$, growing in the case of the self-repulsive shell ($e^2>GM^2$), and decreasing for the self-attractive shell ($e^2<GM^2$). In the last case, since everywhere $B>0$, the turning points are absent for the infinite motion (the both roots of equation $A=0$ are negative).

\begin{figure}
\begin{center}
\includegraphics[angle=0,width=0.8\textwidth]{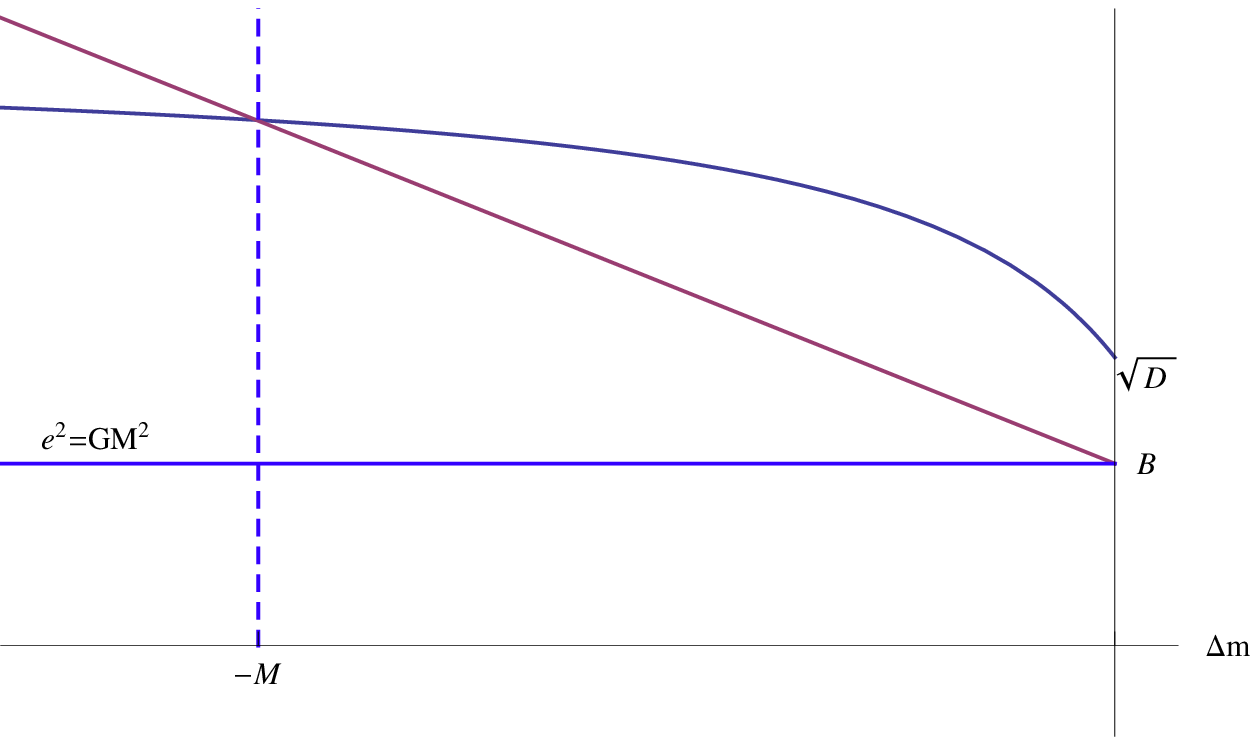}
\includegraphics[angle=0,width=0.8\textwidth]{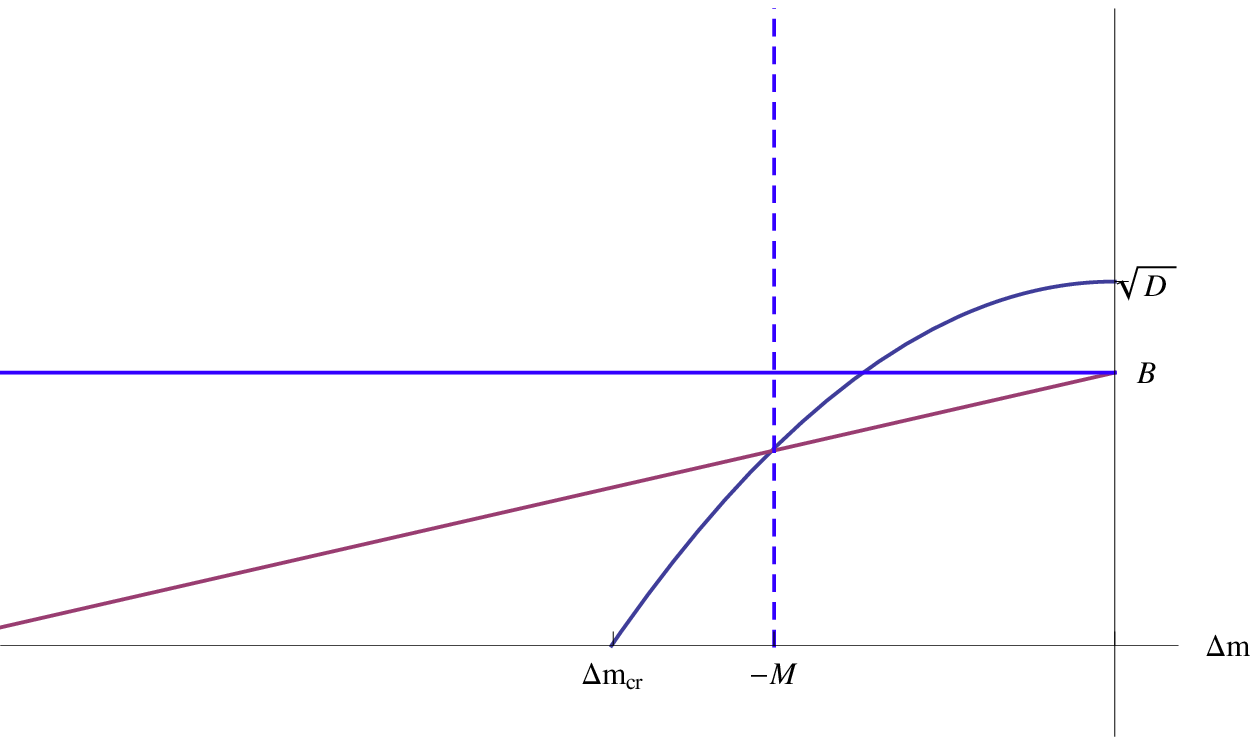}
\end{center}
\caption{The case $e^2=GM^2$, when $\sqrt{D}=B=2GMm_{\rm out}$
(a the upper panel). The case $B(-M)>0$ and, for comparison, the strict line $e^2=GM^2$,
when $\sqrt{D}=B=2GMm_{\rm out}$ (at the lower panel).}
 \label{P778}
\end{figure}

Meantime, at the finite motion one root of equation $A=0$ is positive and the other one is negative:
\begin{equation}
 \frac{1}{2\rho_0M}=\frac{-B+\sqrt{D}}{(e^2-GM^2)^2}.
\end{equation}
Note, that intersection of the curves on the graphs in Fig.~\ref{P778} takes place at
$\Delta m=-M$. For comparison on the graph is shown the case $e^2=GM^2$, when $\sqrt{D}=B=2GMm_{\rm
out}$. For the self-repulsive shell two of these possibilities are realized, depending on the positive or negative value of $B(\Delta m)$ at he point $\Delta m=-M$. On the graph in Fig.~\ref{P778} it is shown the case $B(-M)>0$ and, for comparison, the strict line $e^2=GM^2$, when $\sqrt{D}=B=2GMm_{\rm out}$.

\subsubsection{Black hole at $e^2<Gm_{\rm in}^2$: the case $GM^2<e^2<GM^2+2GMm_{\rm out}$}

Condition $B(-M)>0$, is held at if $GM^2<e^2<GM^2+2GMm_{\rm out}$.
From the point of view of the internal metric this case qualitatively is not different from the self-attractive shell: the infinite motion proceeds without the turning points ($B>0$, $D>0$ --- The both roots re negative or complex, if $D<0$), and in the case of finite motion there is one turning point, in which connection $\sqrt{D}>B$. However, there is difference with the external metric: $\sigma_{\rm in}(0)=+1$, but $\sigma_{\rm out}(0)=-1$ for the self-repulsive shell, and $\sigma_{\rm out}(0)=-1$ for the self-attractive shell. At the same time $\sigma_{\rm in}(\infty)=\sigma_{\rm out}(\infty)=-1$.

\subsubsection{Black hole at $e^2<Gm_{\rm in}^2$: the case $e^2>GM^2+2GMm_{\rm out}$}

\begin{figure}[H]
\begin{center}
\includegraphics[angle=0,width=0.65\textwidth]{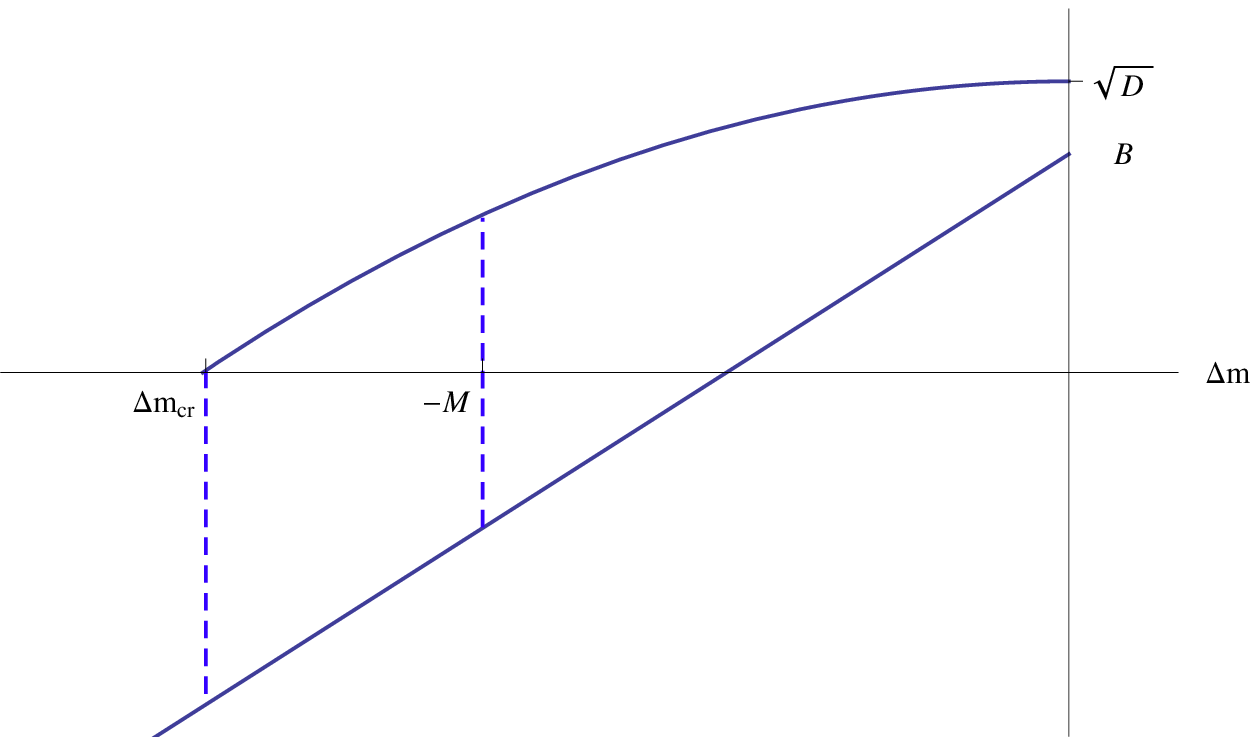}
\includegraphics[angle=0,width=0.65\textwidth]{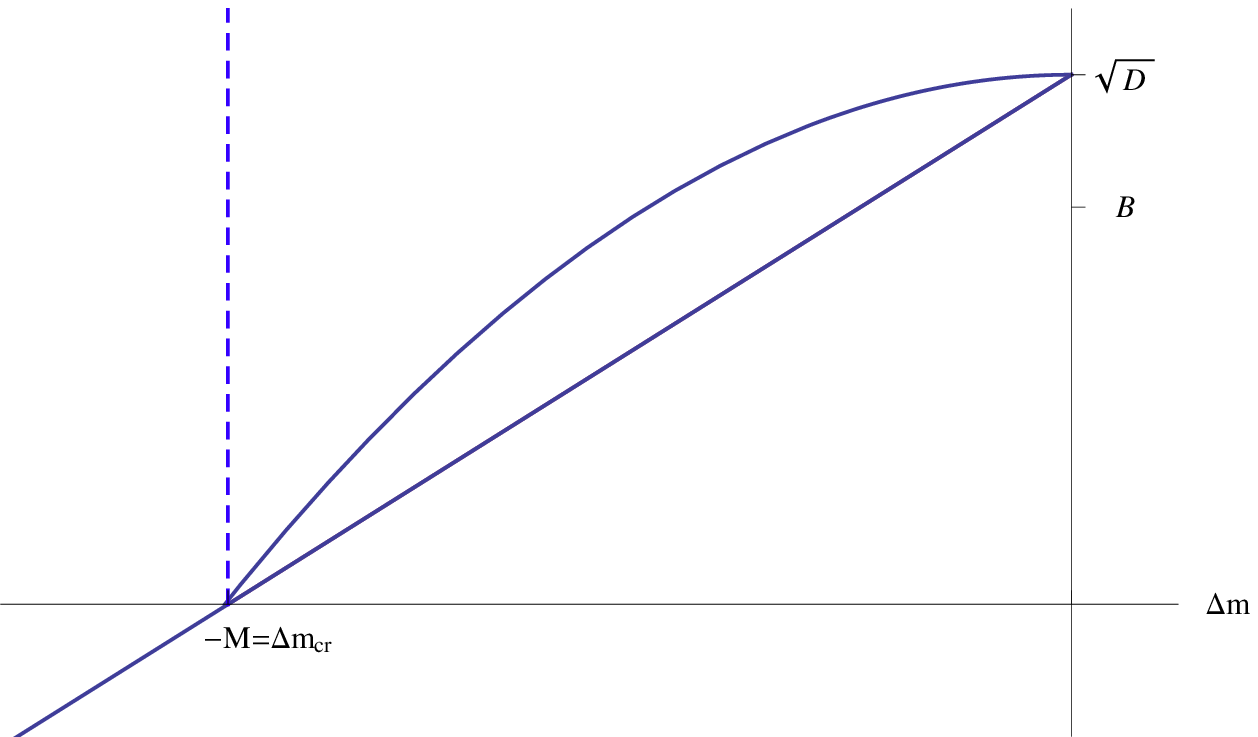}
\end{center}
\caption{The case $e^2>GM^2+2GMm_{\rm out}$. The intersection points of the graphs $\sqrt{D(\Delta m)}$ and $B(\Delta m)$ with the vertical line $\Delta m=-M$ are placed symmetrically with respect to the horizontal axes (the upper pairs of graphs). The lower  pairs of graphs corresponds to the transition case.}
 \label{P79}
\end{figure}

If $e^2>GM^2+2GMm_{\rm out}$, then the graph has the for, shown at he upper panel in Fig.~\ref{P79}. The intersection points of the graphs $\sqrt{D(\Delta m)}$ and $B(\Delta m)$ with the vertical line $\Delta m=-M$ are placed symmetrically with respect to the horizontal axes.

For finite motion, as previously, the change of the sign $B$ does not play any role, but situation now is quite the different for the infinite motion. If $\Delta m<\Delta m_{\rm cr}$, where
\begin{eqnarray}
 \Delta m_{\rm cr} &=& -\frac{(GM^2-e^2)^2+4G^2m^2_{\rm out}}
 {4Gm_{\rm out}(e^2-GM^2)} \nonumber \\
 &=& \frac{M}{2}\left(\frac{e^2-GM^2}{2GMm_{\rm out}}
 + \frac{2GMm_{\rm out}}{e^2-GM^2}\right)\leq M,
 \label{Deltacr}
\end{eqnarray}
(an equality is reached exactly at $e^2=GM^2+2GMm_{\rm out}$), then turning point are absent. However at  $\Delta m>\Delta m_{\rm cr}$ it appears the range of parameters, when $D>0$, and $B<0$, i.\,e., the equation $A=0$ has two positive roots, $B^2>D$. Since the shell is starting at infinity, then the turning point is the bigger root
\begin{equation}
 \frac{1}{2\rho_0M}=\frac{-B-\sqrt{D}}{(e^2-GM^2)^2}.
\end{equation}
At $\Delta m\to-M$ this turning point is going away to infinity. Afterwards the bigger turning point becomes the negative root for the finite motion ($\Delta m>M$), and the smaller root becomes the turning point
\begin{equation}
 \frac{1}{2\rho_0M}=\frac{-B+\sqrt{D}}{(e^2-GM^2)^2}.
\end{equation}

\subsubsection{Black hole at $e^2<Gm_{\rm in}^2$: the case $e^2=GM^2+2GMm_{\rm out}$}

At last, for comparison and clearness we demonstrate the graphs for the transition case, shown at he lower panel in Fig.~\ref{P79}.

\subsection{self-attractive shell}

For investigation the possible shell trajectories at $\Delta m<0$, it is requested to consider the relations between the turning points (when they exist), the point of changing sign of $\sigma_{\rm in}$ and $\sigma_{\rm out}$, the radii of horizons $r_\pm$ and the gravitational radius of the external metric $r_g$.

it appears, that it is useful to fix $\Delta m$, since in this choice the points of changing sign of  $\rho_{\rm in}$ and $\rho_{\rm out}$ are fixed. It is useful also to change the value of $m _{\rm out}$ (and simultaneously the value of $m_{\rm in}$), starting at $m_{\rm out}=0$. For the case we writhe the corresponding equation:
\begin{equation}
\sigma_{\rm in} \sqrt{\dot \rho^2 + F_{\rm in}} - \sigma_{\rm out} \sqrt{\dot \rho^2 + F_{\rm out}} = \frac{G  M }{\rho} \, ,
 \label{israel2}
\end{equation}
where
\begin{equation}
 F_{\rm in}=1-\frac{2Gm_{\rm in}}{\rho}+\frac{Ge^2}{\rho^2},
\quad  F_{\rm out}=1-\frac{2Gm_{\rm out}}{\rho}.
\label{Finout}
\end{equation}
The square of Israel equation gives
\begin{equation}
\frac{\Delta m}{M}\left(1 - \frac{\rho_{\sigma_{\rm in}}}{\rho}\right)=\sigma_{\rm in}
\sqrt{\dot \rho^2 + F_{\rm in}}
 \label{deltamout2}
\end{equation}
\begin{equation}
\frac{\Delta m}{M}\left(1 - \frac{\rho_{\sigma_{\rm out}}}{\rho}\right)=\sigma_{\rm out}
\sqrt{\dot \rho^2 + F_{\rm out}}
 \label{deltamout3}
\end{equation}
From this it follows, in particular, that if the turning point ($\dot \rho=0$) coincides with the one of horizons ($F=0$), then this point coincides with point of the changing sign of $\sigma$.

From equation for the turning point (\ref{A}) it follows the conditions
\begin{equation}
\dot\rho^2= A=\frac{(e^2-GM^2)^2}{4\rho^2M^2}+\frac{2B}{2\rho M}+
 \frac{{\Delta m}^2}{M^2}-1\geq0.
 \label{A2}
\end{equation}
For the turning point $\rho=\rho_0$, where $A=0$, we find with the help of (\ref{BD}):
\begin{equation}
 D = (e^2-GM^2)^2+4GM\Delta m(e^2-GM^2)m_{\rm out}+4G^2M^2m_{\rm out}^2.
 \label{D55}
\end{equation}
The permissible regions for the possible shell motions $A(\rho)\geq0$ are shown in Figs.~\ref{Arho75a} and \ref{Arho75bc}. We see, that behavior of the function $A(\rho)$, defining the turning point (or their absence), is qualitatively different for self-attractive, $e^2<GM^2$, and self-repulsive, $e^2>GM^2$, shells. Also, in the first case the function $B(m_{\rm out})$ is always positive (at $\Delta m<0$), but in the second case the function $B(m_{\rm out})$ is initially negative ($\Delta m=0$), and afterwards becomes the positive. Besides, the radius $\rho_{\sigma_{\rm out}}$ exists, only if $e^2>GM^2$. For this reason, all these cases must be analyzed separately.

\begin{figure}[H]
\begin{center}
\includegraphics[angle=0,width=0.9\textwidth]{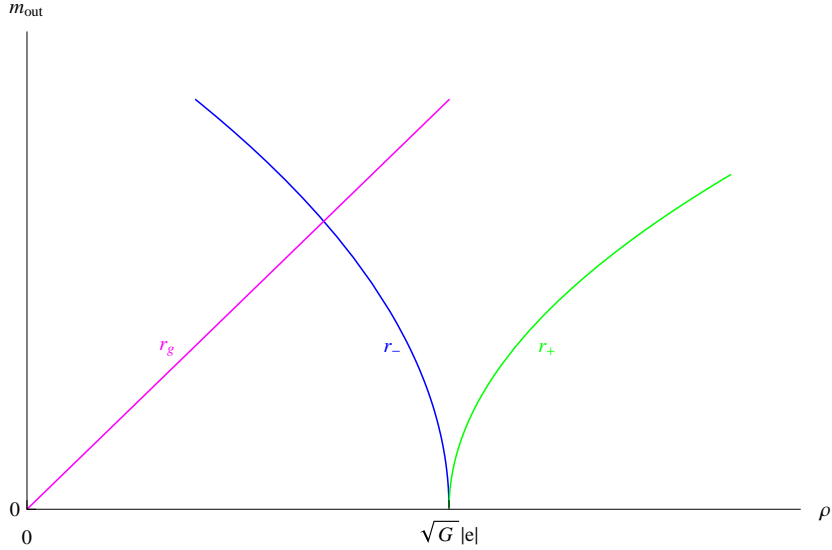}
\end{center}
\caption{The relative position of radii $r_\pm$ and $r_g$ in dependance of $m_{\rm out}$ in the case of the self-attractive shell.} \label{mout8}
\end{figure}

We start from the relatively simple of the self-attractive shell, $e^2<GM^2$, which is intuitively most easily understood, since there is  limit $e^2\to0$. In this case $0<\rho<\infty$, $\rho_{\sigma_{\rm
in}}(-\infty)=-1$, $0<\rho<\infty$ and $\rho_{\sigma_{\rm in}}(0)=+1$. Under these conditions the shell is inevitably comes through the point $\rho_{\sigma_{\rm in}}$, because everywhere at the trajectory $\sigma_{\rm out}=-1$. We need to verify the validity of inequalities $r_-<\rho_{\sigma_{\rm in}}<r_+$, i.\,e., the placement of the point $\rho_{\sigma_{\rm in}}$ in the $T$-region. To do this we consider the relations in the $T$-region of the internal metric:
\begin{equation}
 F_{\rm in}=1-\frac{2Gm_{\rm in}}{\rho}+\frac{Ge^2}{\rho^2}<0,
\label{Fin}
\end{equation}
It is enough to show, that $F_{\rm in}(\rho_{\sigma_{\rm
in}})<0$. At first we prove the validity of the required inequality for the limiting relation
$e^2=GM^2$. We have
\begin{equation}
 F_{\rm in}(\rho_{\sigma_{\rm in}}) =\frac{1}{M^2}
 [m^2_{\rm out}-(m_{\rm in}-M)((m_{\rm in}-M))]<0,
\end{equation}
since from the relation $\Delta m<-M$ it follows that $0<m_{\rm
out}<m_{\rm in}-M$, which was to be proved. Secondly, at the fixed  values of $M$, $m_{\rm in}$ and $\Delta m$, we will diminish the charge $e$. Herewith $\rho_{\sigma_{\rm in}}$ is diminishing, but $r_+$ is growing, i.\,e., the inequality $\rho_{\sigma_{\rm in}}<r_+$ is conserved. It remains to prove that the second inequality $r_-<\rho_{\sigma_{\rm in}}$ is also conserved, when both $\rho_{\sigma_{\rm in}}$ and $r_+$ are diminishing with the diminishing of the charge $e$. The function $F_{\rm in}(\rho)$ reaches the minimum$F_{\rm in}({\rm min})=1-Gm^2_{\rm in}/e^2<0$ at $\rho_{\rm min}=e^2/m_{\rm in}$. Prove now,
that always $\rho_{\sigma_{\rm in}}>\rho_{\rm min}$. It is really
\begin{eqnarray}
\rho_{\sigma_{\rm in}}>\rho_{\rm min}
 \: \Rightarrow \: e^2+GM^2>\frac{2\Delta m e^2}{m_{\rm in}}
\: \Rightarrow \: GM^2>e^2-2\frac{m_{\rm in}}{m_{\rm in}}e^2,
\end{eqnarray}
$GM^2>e^2$, wherefrom it follows the requested result. It is evident, in the case of the self-attractive shell always $D>0$, and $\sigma_{\rm out}=-1$ everywhere at trajectory, since $\rho_{\sigma_{\rm out}}$ does not exist ($\rho_{\sigma_{\rm out}}<0$), and $\sigma_{\rm out}(0)=\sigma_{\rm out}(\infty)=-1$. The following relations are valid:
\begin{equation}
\Delta m=m_{\rm out}-m_{\rm in}, \quad m_{\rm out}>0, \quad \Delta m>-m_{\rm in},
 \label{Delta42}
\end{equation}
\begin{equation}
 B_{\rm min} = -\frac{\Delta m}{M}(GM^2-e^2), \;
 \sqrt{D_{\rm min}} =GM^2-e^2, \;
 \rho_0=\frac{GM^2-e^2}{2(\pm M+\Delta m)}.
 \label{DBmin2}
\end{equation}
It is clear, that there is the turning point $\rho_0>0$ at $M>-\Delta m$ (by fixing the sign `$+$' in  \label{DBmin}), corresponding to the finite motion $0\leq\rho\leq\rho_0$ at $M<-\Delta m$ when is realized the infinite motion without the turning point, $0\leq\rho\leq\infty$. At infinite motion we have the following chain of inequalities:
\begin{equation}
e^2\leq GM^2\leq G\Delta m^2\leq Gm_{\rm in}^2, \quad m_{\rm out}\geq0.
 \label{infin43}
\end{equation}
At finite motion there are two separate cases. In the first case
\begin{equation}
e^2\leq G\Delta m^2\leq GM^2, \quad m_{\rm out}\geq0.
 \label{fin43}
\end{equation}
and $\Delta m^2\leq M^2$ at $m_{\rm out}=0$. Then, under the increasing of $m_{\rm out}$, the mass $m_{\rm out}$ happens to be greater than $M$. Respectively, in the second case $G\Delta m^2\leq e^2\leq  GM^2$. Therefore, we cannot start with $m_{\rm out}=0$, but only with
\begin{equation}
m_{\rm out,min}=\frac{|e|}{\sqrt{G}}+\Delta m>0.
 \label{fin43min}
\end{equation}
Remind, that $\rho_{\sigma_{\rm in}}=const$, $\rho_{\sigma_{\rm out}}$ do not exist, and the horizons $r_\pm$ of the internal metric with growing value of $m_{\rm in}(m_{\rm out})$ are changed in a following way:
\begin{equation}
r_\pm= Gm_{\rm in}\pm\sqrt{G^2m_{\rm in}^2-Ge^2},
 \label{rpm}
\end{equation}
\begin{equation}
\frac{\partial r_\pm}{\partial Gm_{\rm in}}=\frac{r_\pm}{\sqrt{G^2m_{\rm in}^2-Ge^2}}.
 \label{rpm2}
\end{equation}
The relative position of radii $r_\pm$ and $r_g$ in dependance of $m_{\rm out}$ is shown in Fig.~\ref{mout8}.

\begin{figure}[H]
\begin{center}
\includegraphics[angle=0,width=0.9\textwidth]{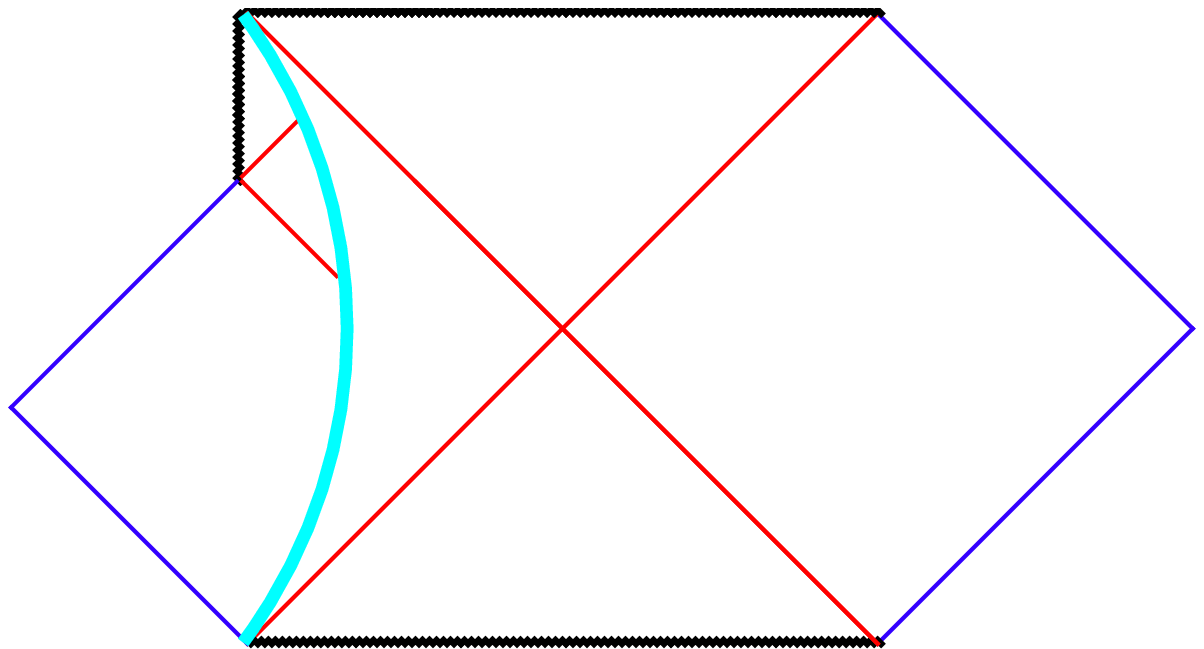}
\end{center}
\caption{The complete conformal Carter--Penrose for infinite motion of the self-gravitating shell.}
 \label{RNp84}
\end{figure}

Let us clarify now, where is placed the point $\rho_{\sigma_{\rm in}}$. To do this we verify the validity of inequality $\rho_{\sigma_{\rm in}}>r_\pm$ at the point, when $m_{\rm out}=0$ and $\Delta m=m_{\rm in}$:
\begin{equation}
GM>-G\Delta m\pm\sqrt{G^2\Delta m^2-Ge^2}=r_\pm,
 \label{rpm3}
\end{equation}
\begin{equation}
G(M+\Delta m)>\sqrt{G^2\Delta m^2-Ge^2}.
 \label{rpm4}
\end{equation}
It follows directly from here, that at the infinite motion, when $m_{\rm out}=0$ and $M+\Delta m<0$, the following relations are held
\begin{equation}
\rho_{\sigma_{\rm in}}<r_{+,min}<r_+,
 \label{rpmin}
\end{equation}
Respectively, at the finite motion, when $m_{\rm out}=0$ and $M+\Delta m>0$, the following relations are held
\begin{equation}
\rho_{\sigma_{\rm in}}>r_{-,\rm max}>r_-.
 \label{rmax}
\end{equation}
We demonstrate now that at the infinite motion $\rho_{\sigma_{\rm in}}>r_-$. It must be held the following inequality
\begin{equation}
\sqrt{G^2\Delta m^2-Ge^2}>G(M+\Delta m).
 \label{rpm5}
\end{equation}
Since $\Delta m<-M$, we have
\begin{equation}
GM^2+2GM\Delta m<-GM^2<-e^2.
 \label{rpm6}
\end{equation}
It is proved in a similar way, that at the finite motion $\rho_{\sigma_{\rm in}}<r_{+,min}$. At this step it is clear, that at the infinite motion there are no obstacles for shell to move from infinity to $r=0$, at $\sigma_{\rm in}(\infty)=-1$ and $\sigma_{\rm in}(0)=+1$, since $\rho_{\sigma_{\rm in}}$, it is proved, is placed in the $T$-region between the horizons. In other words, for the self-gravitating shell with $e^2<GM^2$ the only condition for the infinite motion is $\Delta m<-M$. The Carter--Penrose diagram for infinite motion of the self-gravitating shell at $\Delta m<0$, $e^2<Gm_{\rm in}^2$ coincides with diagram in Fig.~\ref{RNp84}.

\begin{figure}
\begin{center}
\includegraphics[angle=0,width=0.7\textwidth]{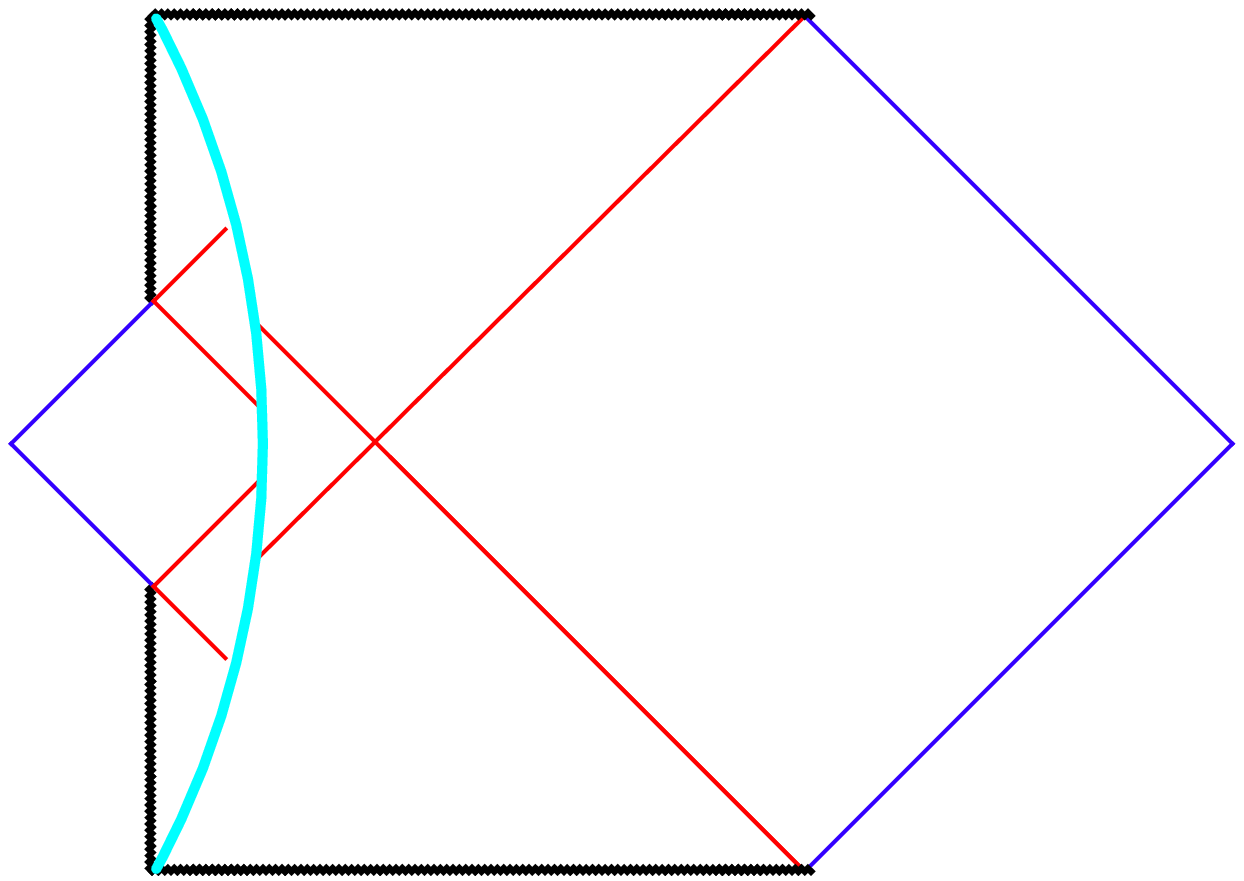}
\end{center}
\caption{The Carter--Penrose diagram for the case of finite motion of the self-attractive shell at $\Delta m<0$, $e^2<Gm_{\rm in}^2$.} \label{RN38b}
\end{figure}

Let us see now, what is happened at the finite motion. Now we have the turning point at $\rho_0$. The question is, where this point is placed with respect to $r_\pm$, and what is the sign of $\sigma(\rho_0)$. We already know, that there are two different cases. In the first case, when $e/\sqrt{G}<-\Delta m<M$, we have
\begin{equation}
m_{\rm out,min}=0, \quad \rho_{0,\rm min}=\frac{GM^2-e^2}{2(M-m_{\rm in})}.
 \label{mrho}
\end{equation}
Let us demonstrate, that $\rho_{0,\rm min}\geq r_{+,\rm min}$. Really, the inequality
\begin{equation}
\rho_{0,\rm min}=\frac{GM^2-e^2}{2(M-m_{\rm in})}>Gm_{\rm in}+\sqrt{G^2m_{\rm in}^2-Ge^2}
 \label{rho2}
\end{equation}
is transformed to the evident inequality
\begin{equation}
r_+^2-2GMr_++G^2M^2=(r_++GM)^2\geq0.
 \label{rho3}
\end{equation}
This means, that $\sigma_{\rm in}(\rho_{0,\rm min})=-1$, and the shell, starting from the zero radius, moves through the $T$-region into the $R$-region of the internal netric. With increasing of $m_{\rm out}$, there are also increased both the $\rho_0$ and $r_+$, buy, however, $\rho_{\sigma_{\rm in}}$ is not changed. Therefore the sign of  $\sigma$ is inevitably changing at the trajectory. This is confirmed also by the fact, that the coinciding of $\rho_0$ and $r_+$ means simultaneously the intersection at this point also with $\rho_{\sigma_{\rm in}}$, but, however, $\rho_{\sigma_{\rm in}}<r_+$. The Carter--Penrose diagram for the case of finite motion of the self-attractive shell at $\Delta m<0$, $e^2<Gm_{\rm in}^2$ is shown in Fig.~\ref{RN38b} and coinside with the right diagram in Fig.~\ref{RN38}.

\begin{figure}
\begin{center}
\includegraphics[angle=0,width=0.95\textwidth]{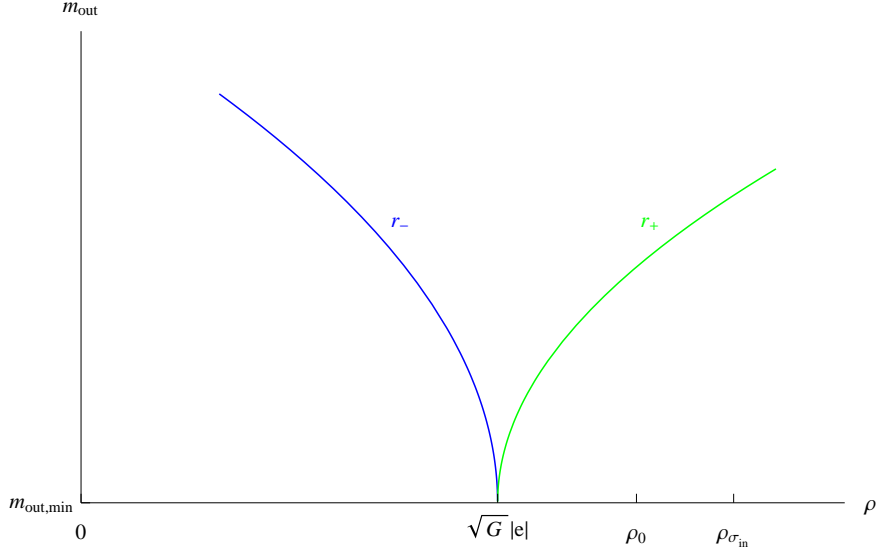}
\end{center}
\caption{The mutual positional relationship of the characteristic radii $0<\sqrt{G}e<r_0<\rho_{\sigma_{\rm in}}$ and $r_\pm$.} \label{mout14}
\end{figure}

It is remained to consider the second case, when $-\Delta m<e/\sqrt{G}<M$ and the minimal value of $m_{\rm out}$ is already nonzero:
\begin{equation}
m_{\rm out,min}=\frac{e}{\sqrt{G}}+\Delta m>0, \quad
m_{\rm in,min}=\frac{e}{\sqrt{G}}.
 \label{mout}
\end{equation}
It is not difficult to show, that что $\rho_{\sigma_{\rm in}}>e/\sqrt{G}=r_{-,\rm max}=r_{+,\rm min}$. Really, the inequality
\begin{equation}
\frac{GM^2-e^2}{2\Delta m}>e\sqrt{G}
 \label{rho4}
\end{equation}
transforms to the evident one
\begin{equation}
(e+\sqrt{G}\Delta m)^2+G(M^2-\Delta m^2)>0.
 \label{rho5}
\end{equation}
Let us calculate $\rho_{0,\rm min}$ in this case. We have:
\begin{eqnarray}
 \label{Bmin3}
B_{\rm min} &=& \frac{\Delta m}{M}(e^2+GM^2)+2\sqrt{G}Me, \\
 \label{Dmin3}
\sqrt{D_{\rm min}} &=& e^2+GM^2+2\sqrt{G}\Delta me>0, \\
 \rho_{0,\rm min}&=&\frac{(e+\sqrt{G}M)^2}{2(M-\Delta m)}.
 \label{rhomin3}
\end{eqnarray}
By direct calculations it is easily to verify, that $\rho_{0,\rm min}<\rho_{\sigma_{\rm in}}$ (see Fig.~\ref{mout14}). Therefore, $0<\sqrt{G}e<r_0<\rho_{\sigma_{\rm in}}$ at $m_{\rm out}=m_{\rm out,min}$.

\begin{figure}
\begin{center}
\includegraphics[angle=0,width=0.7\textwidth]{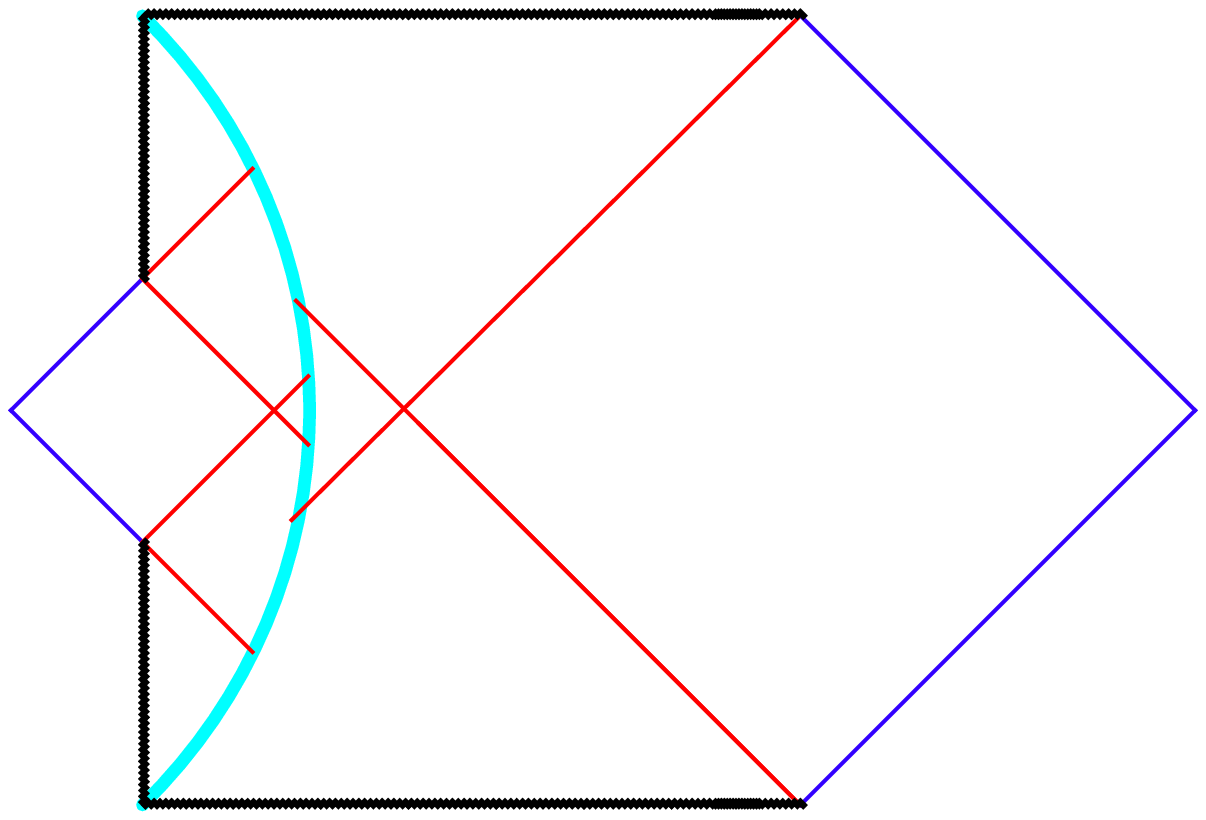}
\end{center}
\caption{The Carter--Penrose diagram for the case of finite motion of the self-attractive shell at $-\Delta m<\sqrt{G}e<M$ and $m_{\rm in}>m_{\rm in,cr}$.} \label{RN38c}
\end{figure}

With the growing of $m_{\rm out}$ (ad $m_{\rm in}$) the radius $r_+$ inevitably intersects   $\rho_{\sigma_{\rm in}}$, i.\,e., $\rho_{\sigma_{\rm in}}$ is inevitably coming into the $T$-region. At the same point $\rho_0$ intersects $\rho_{\sigma_{\rm in}}$ and touches with $r_+$. This means, that $\rho_0$ from $R_+$-region beyond the $r_+$ goes into $R_-$-region beyond the $r_+$. This event proceeds at the critical value of $m_{\rm in}=m_{\rm in,cr}$, where
\begin{equation}
m_{\rm in,cr}=-\frac{e}{2\sqrt{G}}\left(\frac{e^2+GM^2}{2\sqrt{G}e\Delta m}+
\frac{2\sqrt{G}e\Delta m}{e^2+GM^2} \right) > \frac{e}{\sqrt{G}}.
 \label{mcr}
\end{equation}
The corresponding Carter--Penrose diagram for the case of finite motion of the self-attractive shell at $-\Delta m<\sqrt{G}e<M$ and $m_{\rm in}>m_{\rm in,cr}$ is shown in Fig.~\ref{RN38c} and coincides with the left diagram in Fig.~\ref{RN38}.

\subsection{self-repulsive shell}

In the case of the self-repulsive shell, since $GM^2<e^2<Gm_{\rm in}^2$, then $M<m_{\rm in}$, and form condition  $Gm_{\rm out}=m_{\rm in}+\Delta m\geq0$ it follows, that $\Delta m\geq-m_{\rm in}$. At $m_{\rm out}=0$ we obtain $\Delta m\geq-m_{\rm in}<-M$. At the same time, this means that at $m_{\rm out}=0$ we start our analysis from the condition $|\Delta m|\geq M$, when
\begin{equation}
 B_{\rm min} = \frac{\Delta m}{M}(e^2-GM^2)<0, \quad
 \sqrt{D_{\rm min}} = e^2-GM^2>0,
 \label{DBmin4}
\end{equation}
We already know that in this case there are two turning points:
\begin{equation}
\rho_0,\pm=\frac{B\pm\sqrt{D}}{2M\left(1-\frac{\Delta m^2}{M^2}\right)}
=\frac{e^2-GM^2}{2(-\Delta m\pm M)},
\label{rho02}
\end{equation}
in which connection $\rho_{0,+}<\rho_{0,-}$. It is evident, that $\rho_{0,+}<\rho_{\sigma_{\rm out}}<\rho_{0,-}$. We demonstrate below, that $\rho_{0,-}\leq r_-$. Really, this inequality easily transforms to the the evident one, $(GM-r_-)^2\geq0$.

As regards the relative placement of $\rho_{\sigma_{\rm in}}$, it only may say, that $\rho_{\sigma_{\rm out}}<\rho_{\sigma_{\rm in}}<r_+$. The first inequality is evident, and let us prove the second one:
\begin{eqnarray}
\frac{e^2+GM^2}{2m_{\rm in}}&<&Gm_{\rm in}+\sqrt{G^2m_{\rm in}^2-Ge^2} \,
\Rightarrow \\
e^2+GM^2&<&2Gm_{\rm in}^2+2m_{\rm in}\sqrt{G^2m_{\rm in}^2-Ge^2} \, \Rightarrow \\
G^2M^2&<&r_+^2 \, \Rightarrow \, GM<r_+.
\label{rhoin}
\end{eqnarray}
Since $-\Delta m=m_{\rm in}>M$ at $m_{\rm out}=0$, than it is equitable the same chain of inequalities, but with the change from $r_+$ to $r_-$. In result, we obtain, that $\rho_{\sigma_{\rm in}}<r_-$, if $GM<\rho_{\sigma_{\rm in}}$, and, consequently, $\rho_{\sigma_{\rm in}}>r_-$, if $GM>\rho_{\sigma_{\rm in}}$. Note, that inequality $GM<\rho_{\sigma_{\rm in}}$ is equivalent to the inequality $\rho_{\sigma_{\rm in}}<\rho_{0,-}$. In this case we have the following relation between the characteristic parameters: $GM<\rho_{\sigma_{\rm in}}<\rho_{0,-}<r_-$.

\begin{figure}
\begin{center}
\includegraphics[angle=0,width=0.95\textwidth]{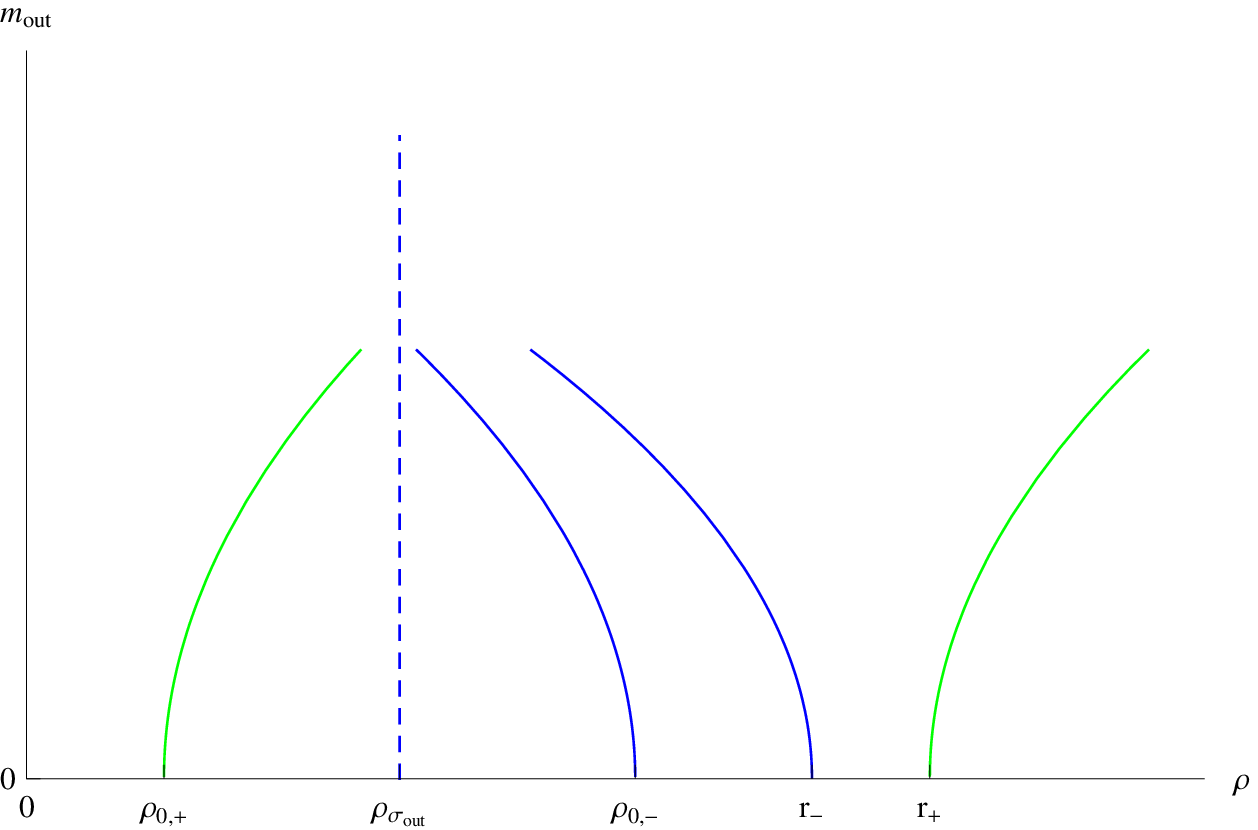}
\end{center}
\caption{The relative positions of the characteristic radii in dependance of $m_{\rm out}$.} \label{mout19}
\end{figure}

\begin{figure}
\begin{center}
\includegraphics[angle=0,width=0.95\textwidth]{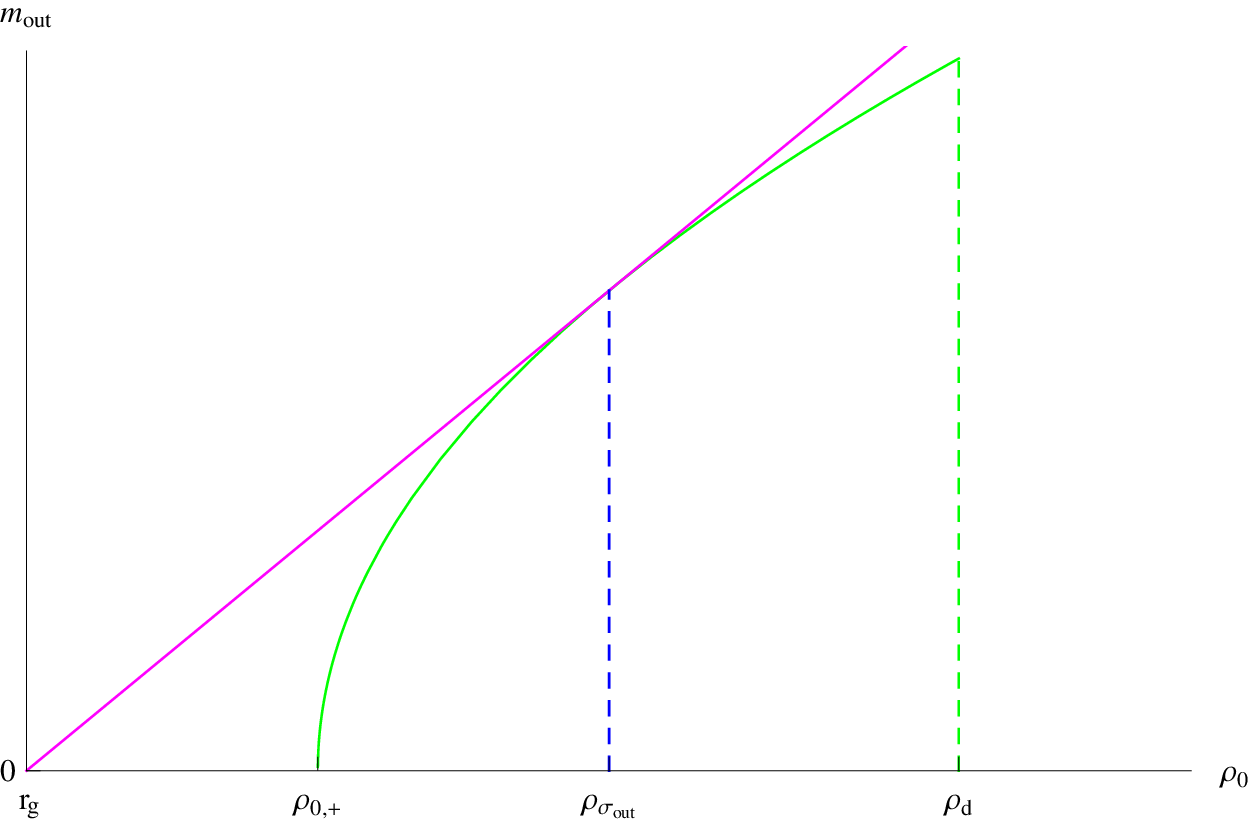}
\end{center}
\caption{The relative positions of the characteristic radii in dependance of $m_{\rm out}$.} \label{mout22}
\end{figure}

What happens during the growing of $m_{\rm out}$? We fix the value of $\rho_{\sigma_{\rm out}}$ for subject to agreement. As we know, the radius of the inner horizon $r_-$ is decreasing up to zero, and radius of the outer horizon  $r_+$ is growing to infinity. Further, we have
\begin{equation}
\frac{\partial\rho_{0,-}}{\partial(2Gm_{\rm out})} =
\frac{1\pm\frac{B}{\sqrt{D}}}{2M\left(1-\frac{\Delta m^2}{M^2}\right)}
\frac{\partial B}{\partial(2Gm_{\rm out})}=\pm\frac{\rho_{0,\pm}}{\sqrt{D}},
\label{partial}
\end{equation}
i.\,e., $\rho_{0,+}$ is growing, and $\rho_{0,-}$ is diminishing. Ultimately, they coalesce, whenthe discriminant $D=0$, but $B$ is still negative. It happens at
\begin{equation}
m_{\rm out}=m_{\rm out,1}=\frac{e^2-GM^2}{2GM}\left(-\frac{\Delta m}{M}-\sqrt{\frac{\Delta m^2}{M^2}-1}\right).
\label{mout1}
\end{equation}
The double root equals
\begin{equation}
\rho_d=\frac{e^2-GM^2}{2\sqrt{\delta m^2-M^2}}>\rho_{\rm out}.
\label{rhod}
\end{equation}
At the further growing of $m_{\rm out}$ the real roots of equation $A=0$ are both disappeared, and discriminant is diminished to the minimal value, when $B=0$, at
\begin{equation}
m_{\rm out}=-\frac{\Delta m}{M}\frac{e^2-GM^2}{2GM}.
\label{mout10}
\end{equation}
Then, the discriminant is growing again up to the zero at
\begin{equation}
m_{\rm out,2}=\frac{e^2-GM^2}{2GM}\left(-\frac{\Delta m}{M}+\sqrt{\frac{\Delta m^2}{M^2}-1}\right).
\label{mout2}
\end{equation}
At the further growing of the $m_{\rm out}$, the two real roots of equation $A=0$ are appearing again, but now both of them are negative. This means that the motion of the shell occurs without the turning points.  The relative positions of the characteristic radii is shown in Figs.~\ref{mout19} and \ref{mout22}.

We start investigation from the region $0<m_{\rm out}<m_{\rm out,1}$, when there are two turning points, separately for the finite motion at $0<\rho_0\leq\rho_{0,+}\leq r_g$, and for the infinite motion at $\rho_d\leq\rho_0\leq\rho_{0,-}$ and  $\rho_0\leq\rho<\infty$.

\subsubsection{Finite motion with turning point}
\begin{figure}
\begin{center}
\includegraphics[angle=0,width=0.9\textwidth]{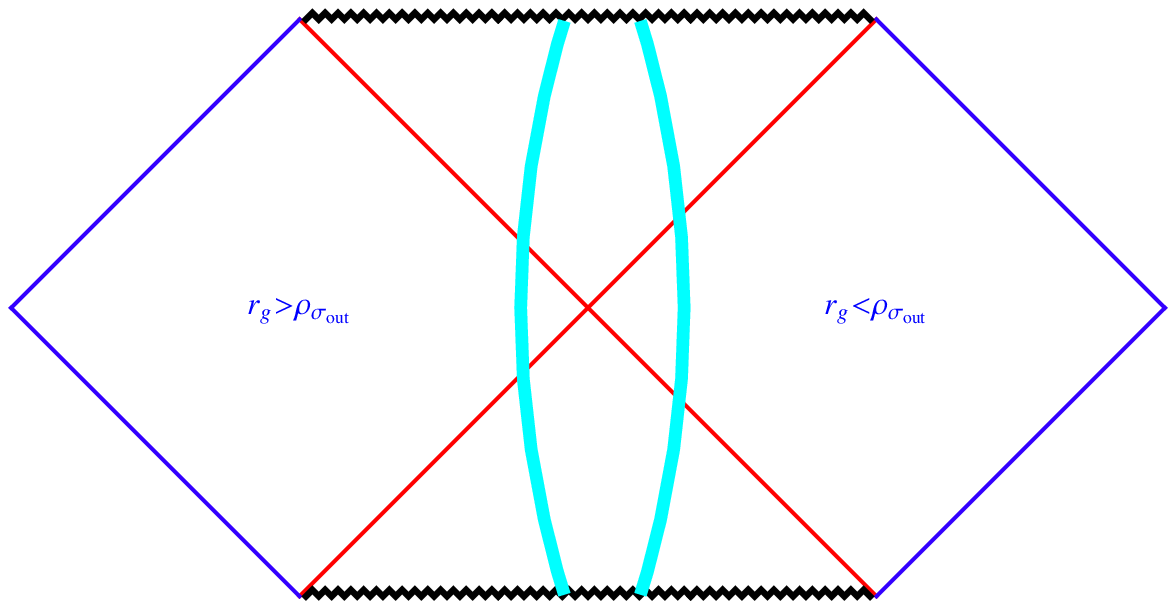}
\end{center}
\caption{The possible conformal diagrams for the external Schwarzschild metric at $r_g\gtrless\rho_{\sigma_{\rm out}}$.}
 \label{DeltaMinusExt2}
\end{figure}

Let us describe the finite motion. We know, that at $m_{\rm out}=0$ the following relations are held: $0=r_g<\rho_{0,+}\leq\rho_{\sigma_{\rm out}}$. It is clear, that now $\rho_d>\rho_{\sigma_{\rm out}}$. this mean, that with the growing of $m_{\rm out}$, the curve $\rho_0$ intersects the line $\rho_{\sigma_{\rm out}}$, i.\,e., at first $\sigma_{\rm out}(\rho_0)=+1$, and then $\sigma_{\rm out}(\rho_0)=-1$. It is also $\rho_0=r_g$ at the intersection point, and at the further growing of $m_{\rm out}$, it must be  $\rho_0>r_g$ (the trajectory turns out in the $R$-region of the external metric). Verify, that this is really takes place for $\rho_d$:
\begin{eqnarray}
\rho_d>r_g &\quad& \Rightarrow \quad \frac{e^2-GM^2}{2\sqrt{\Delta m^2-M^2}}>2Gm_{\rm out}
\quad \Rightarrow \\
&&\left(\frac{\Delta m}{M}+\sqrt{\frac{\Delta m^2}{M^2}-1}\right)^2>0.
\label{rd}
\end{eqnarray}
The possible conformal diagrams for the external Schwarzschild metric are shown in Fig.~\ref{DeltaMinusExt2}.

To look into the internal metric, it is requested to find the place of $\rho_{\sigma_{\rm in}}$. it is evident, that $\rho_{\sigma_{\rm in}}>\rho_{\sigma_{\rm out}}$. Find at first the condition, when $\rho_{\sigma_{\rm in}}<r_-$ at $m_{\rm out}$, i.\,e., at $m_{\rm in}=\Delta m$:
\begin{eqnarray}
\rho_{\sigma_{\rm in}}&<&r_- \, \Rightarrow \,
\frac{e^2+GM^2}{2m_{\rm in}}<Gm_{\rm in}-\sqrt{G^2m_{\rm in}^2-Ge^2} \, \Rightarrow \\
GM&<&r_- \, \Rightarrow \, GM<\rho_{\sigma_{\rm in}}.
\label{rhoinm}
\end{eqnarray}
It appears,  that at $GM<\rho_{\sigma_{\rm in}}$ simultaneously $\rho_{\sigma_{\rm in}}<r_-$ at $m_{\rm out}=0$. With the growing of $m_{\rm out}$ the radius $\rho_{\sigma_{\rm in}}$ is standing at the same place, and, consequently, $\rho_0$ is growing, but $r_-$ is diminishing. What are their relative values at the point $\rho_d$? We prove, that $\rho_d<\rho_{\sigma_{\rm in}}$:
\begin{figure}
\begin{center}
\includegraphics[angle=0,width=0.4\textwidth]{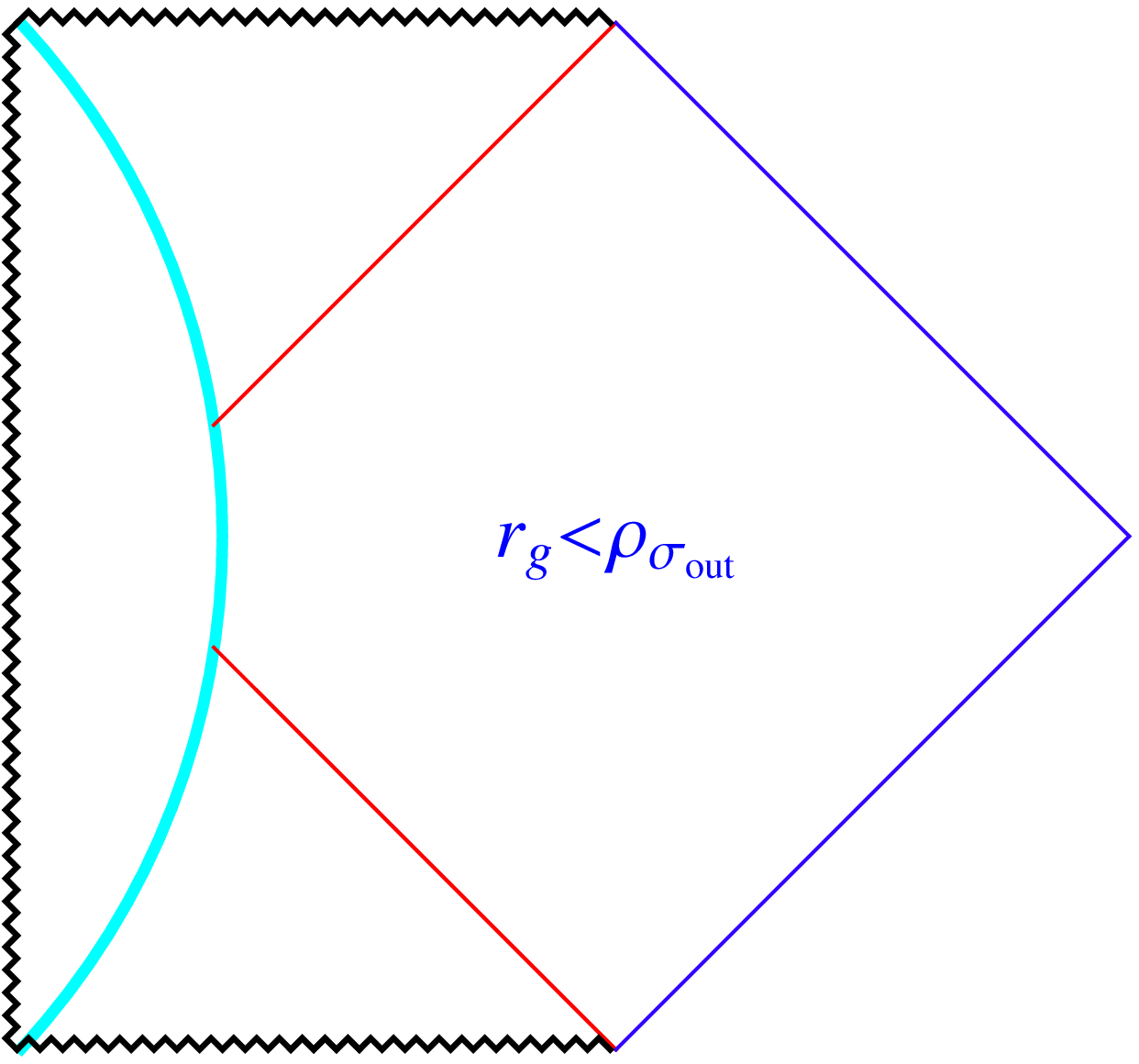}
\hskip0.2cm
\hfill
\includegraphics[angle=0,width=0.45\textwidth]{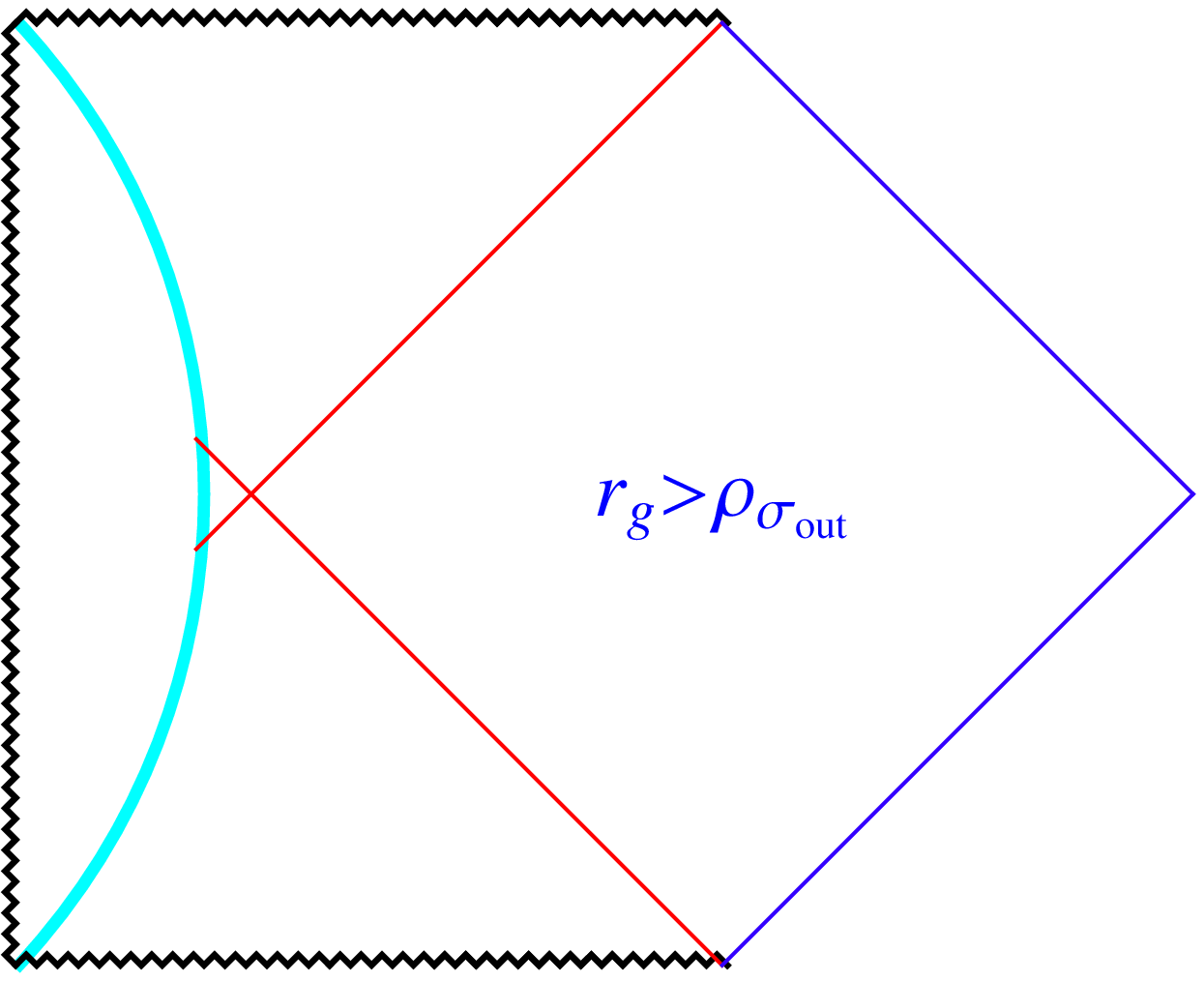}
\end{center}
\caption{The Carter--Penrose diagrams for finite motion of the shell at $r_g\gtrless \rho_{\rm out}$.} \label{p25}
\end{figure}
\begin{eqnarray}
\frac{e^2-GM^2}{2\sqrt{\Delta m^2-M^2}}<-\frac{e^2+GM^2}{2\Delta m} \, \Rightarrow
\rho_{\sigma_{\rm in}}<\sqrt{G}e.
\label{rdin}
\end{eqnarray}
However, in the described case this is valid from the outset (at $m_{\rm out}=0$), if only $GM<\rho_{\sigma_{\rm in}}$. Meantime, if $GM>\rho_{\sigma_{\rm in}}$, then $\rho_{\sigma_{\rm in}}>r_-$ at $m_{\rm out}=0$, and this inequality is held with the growing $m_{\rm out}$ (since $r_-$ is diminishing). Since the turning point $\rho_{0,-}<r_-$ at $m_{\rm out}=0$, then, $\rho_{0,-}<\rho_{\sigma_{\rm in}}$ at $GM>\rho_{\sigma_{\rm in}}$. Additionally, since $\rho_{0,-}$ is diminishing up to $\rho_0=\rho_d$, then  $\rho_d<\rho_{\sigma_{\rm in}}$. In result, our self-gravitating is expanded under the  finite motion at $\Delta m <-M$, starting from $\rho=0$, up to the turning point, which is smaller of the inner horizon $r_-$ and without the intersection of $\rho_{\sigma_{\rm in}}$, i.\,e., $\sigma_{\rm in}(\rho_0)=+1$. The corresponding Carter--Penrose diagrams for finite motion of the shell at $r_g\gtrless \rho_{\rm out}$ are shown in Fig.~\ref{p25}.

\subsubsection{Infinite motion with turning point}

Now we move to the infinite motion with the turning point $\rho_{0,-}$. The motion of the shell proceeds in the radial interval $\rho_{0,-}\leq\rho<\infty$. At $m_{\rm out}=0$ the relation $\rho_{\sigma_{\rm out}}<\rho_{0,-}$ is held, and afterwards $\rho_{0,-}$ is diminishing up to $\rho_d$. Since, as it was before, $\rho_d>r_g$ and $\rho_d>\rho_{\sigma_{\rm out}}$, then in the external metric the motion takes place in the $R_-$-region and $\sigma_{\rm out}(\rho_0)=-1$. In the internal metric the the of the shell trajectory depends on the relation between $r_\pm$, $\rho_{\sigma_{\rm in}}$ and $\rho_{0,-}$. We already know, that at $m_{\rm out}=0$ we have $\rho_{0,-}<r_-$. Respectively, for $\rho_{\sigma_{\rm in}}$ we have, $\rho_{0,-}<r_-$ at $GM<\rho_{\sigma_{\rm in}}$ and $\rho_{0,-}<r_-<\rho_{\sigma_{\rm in}}<r_+$ at $GM>\rho_{\sigma_{\rm in}}$.

\begin{figure}[H]
\begin{center}
\includegraphics[angle=0,width=0.47\textwidth]{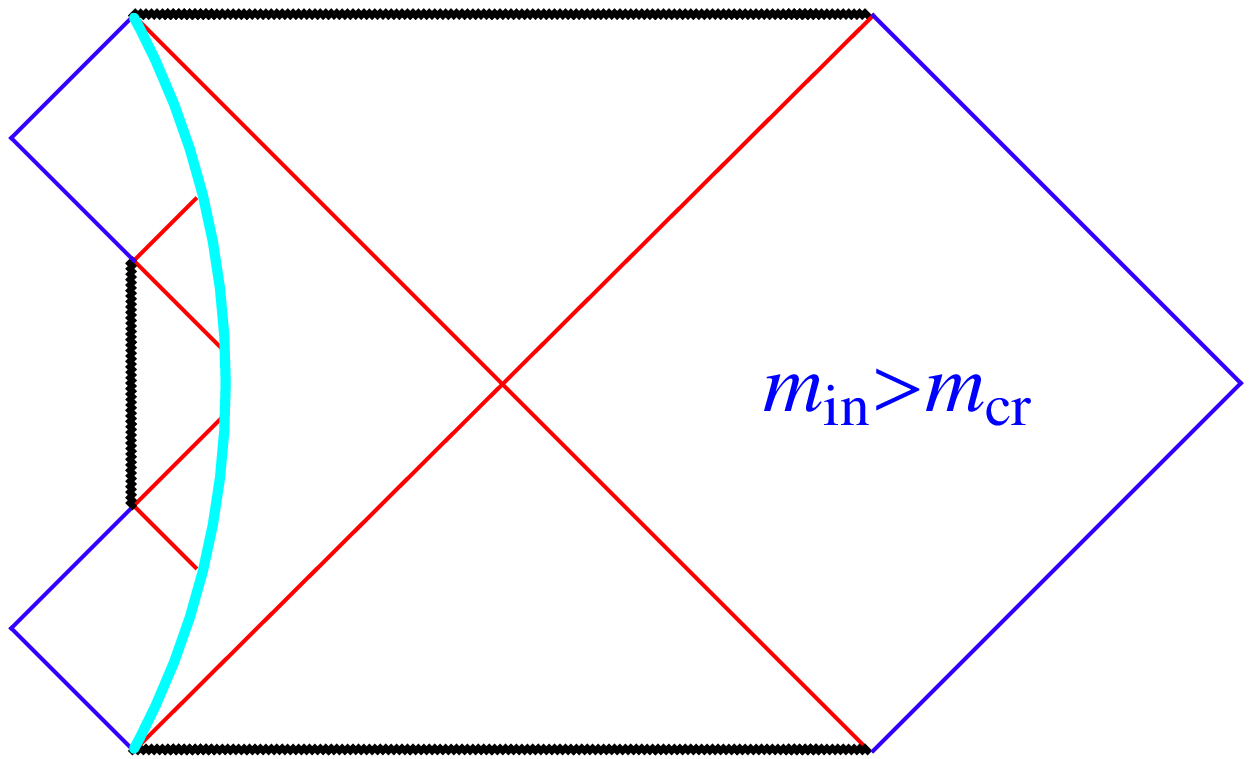}
\hskip0.2cm
\hfill
\includegraphics[angle=0,width=0.47\textwidth]{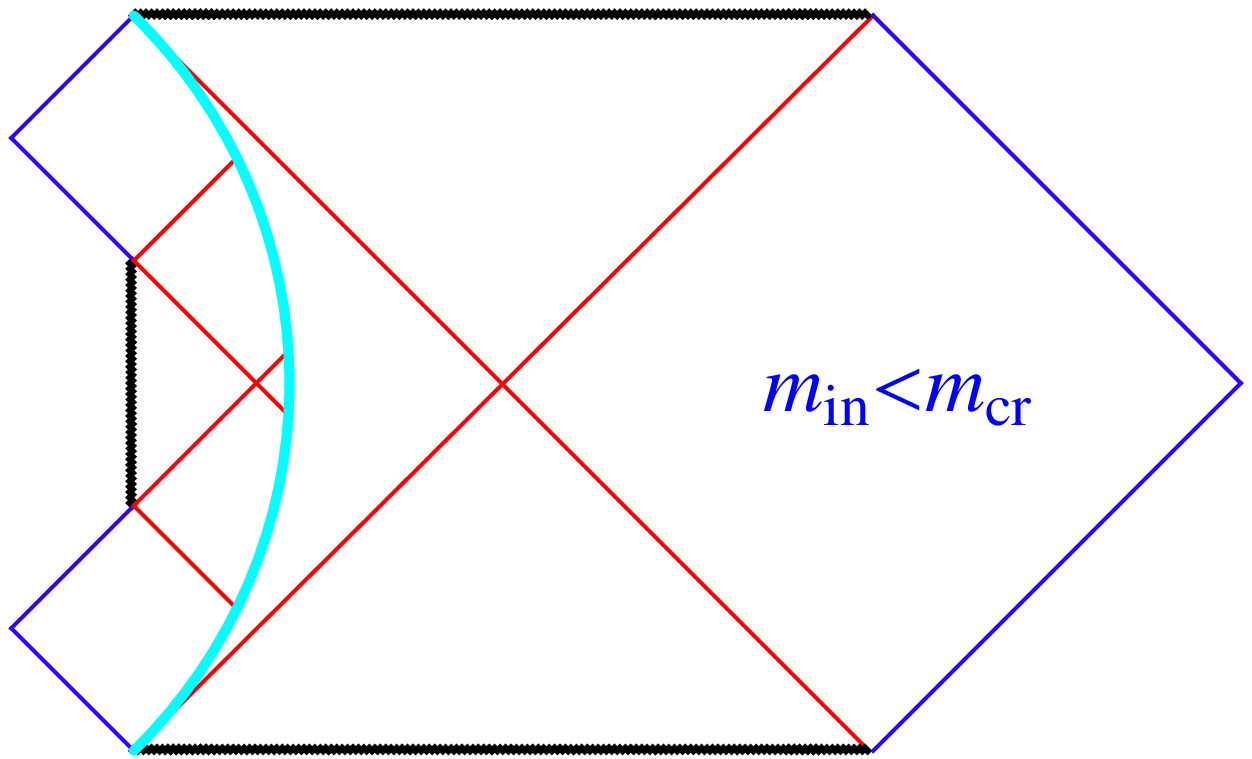}
\end{center}
\caption{The Carter--Penrose diagrams for infinite motion of the self-attractive shell at $m_{\rm in}\gtrless m_{\rm in,cr}$.} \label{p27}
\end{figure}

Let us consider at first the last case, which is more simpler. It is clear, that the shell, contracting from the infinity, where $\sigma_{\rm in}(\infty)$, comes through the $T_+$-region between the horizons of the internal metric $r_\pm$, where the sign of $\sigma_{\rm in}$ is changing, and reach afterward theturning point $\rho_{0,-}$ in $R_+$-region near the central singularity. It happens at $0\leq m_{\rm out}<m_{\rm out,1}$, since $\rho_{0,-}$ is diminishing with the growth of $ m_{\rm out}$, while $\rho_{\sigma_{\rm in}}$ remains at the same place. The corresponding Carter--Penrose diagram is shown at the left panel in Fig.~\ref{p27}.
Meanwhile, if $G M<\rho_{\sigma_{\rm in}}$, then at $m_{\rm out}=0$ the moving shell does not meet at the trajectory the turning point $\rho_{\sigma_{\rm in}}$, i.\,e., $\sigma_{\rm in}(\rho_0)=-1$, and the turning point is placed in $R_-$-region near the central singularity. Since $\rho_d<\rho_{\sigma_{\rm in}}$ (see the text above), then with growing of $m_{\rm out}$ inevitably takes place the meeting: $\rho_{\sigma_{\rm in}}=r_-=\rho_{0,-}$.
It happens at
\begin{equation}
m_{\rm in} = m_{\rm in,cr} =
\frac{e}{2\sqrt{G}}\left( \frac{\rho_{\sigma_{\rm in}}}{\sqrt{G}e} +
\frac{\sqrt{G}e}{\rho_{\sigma_{\rm in}}} \right)>M.
\label{mincr}
\end{equation}
Afterwards, $\rho_{0,-}$ becomes smaller than $\rho_{\sigma_{\rm in}}$, i.\,e., $\rho_{0,-}<\rho_{\sigma_{\rm in}}<r_-$ and therefore $\sigma_{\rm in}(\rho_0)=+1$, as in the case of $M>\rho_{\sigma_{\rm in}}$. The corresponding Carter--Penrose diagram is shown at the right panel in Fig.~\ref{p27}.

\begin{figure}[H]
\begin{center}
\includegraphics[angle=0,width=0.8\textwidth]{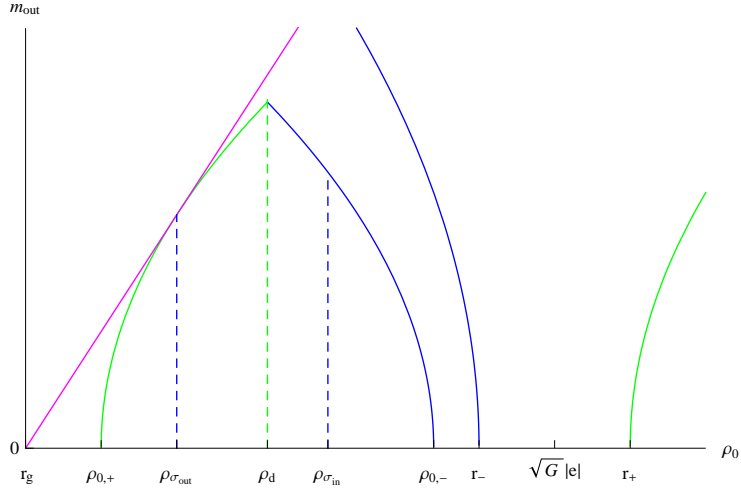}
\end{center}
\caption{The relative displacement of $r_g$, $r_\pm$, $\rho_0$, $\rho_{\sigma_{\rm in}}$, $\rho_{\sigma_{\rm out}}$ and $\rho_0$ in dependance of $m_{\rm out}$ for infinite motion of the self-attractive shell.} \label{mout25a}
\end{figure}

\begin{figure}[H]
\begin{center}
\includegraphics[angle=0,width=0.9\textwidth]{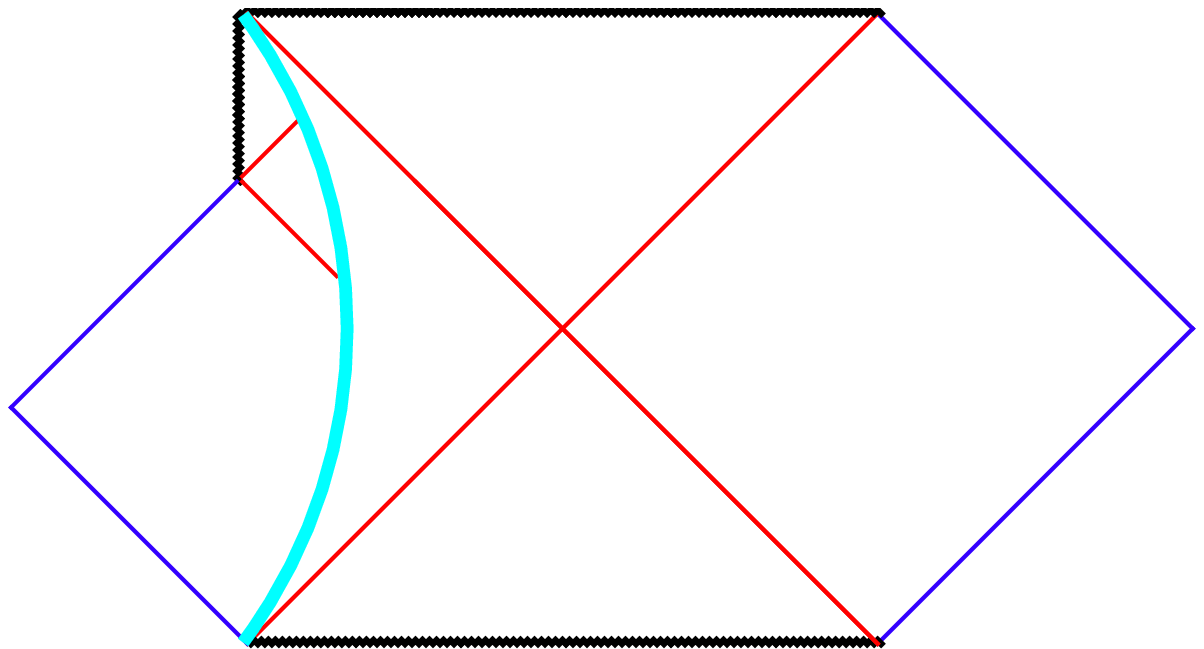}
\end{center}
\caption{The Carter--Penrose diagram for infinite motion of the self-attractive shell in the case $m_{\rm out}>m_{\rm out,1}$.}
 \label{RNp84f}
\end{figure}

At a further growing of $m_{\rm out}>m_{\rm out,1}$ we transfer to the region, where there are no turning points for the case of $\Delta m<-M$. Consequently, the shell, contracting from infinity to the zero radius, should intersect the points of the sign changing of $\sigma_{\rm in}$ and $\sigma_{\rm out}$, which now should be placed in the corresponding $T$-regions. At the same time, it should be $\rho_{\sigma_{\rm out}}<r_g$ and $r_-<\rho_{\sigma_{\rm in}}<r_+$, which is evident from the graph in Fig.~\ref{mout25a} for the behavior of $r_g$, $r_\pm$, $\rho_0$, $\rho_{\sigma_{\rm in}}$, $\rho_{\sigma_{\rm out}}$ and $\rho_0$ in dependance of $m_{\rm out}$.
The Carter--Penrose diagram for infinite motion of the self-attractive shell in the case $m_{\rm out}>m_{\rm out,1}$ is shown in Fig.~\ref{RNp84f} and coincides with similar one in in Fig.~\ref{RNp84}.

\subsubsection{Finite motion at $\Delta m>-M$ and $e^2>GM^2$}

\begin{figure}
\begin{center}
\includegraphics[angle=0,width=0.85\textwidth]{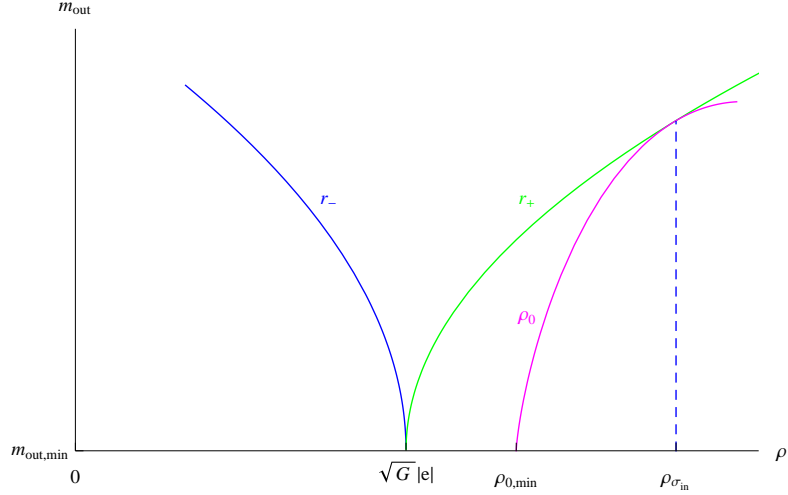}
\end{center}
\caption{The relative positions of the characteristic radii $\rho_0$, $e\sqrt{G}<\rho_{0,\rm min}<\rho_{\sigma_{\rm in}}$ and $r_\pm$.} \label{mout33}
\end{figure}

It is remained to consider the finite motion at $\Delta m>-M$ and $e^2>GM^2$. We already know, that in this case there is only one turning point
\begin{equation}
\rho_0=\frac{B+\sqrt{D}}{2M\left(1-\frac{\Delta m^2}{M^2}\right)},
\quad \frac{\partial\rho_0}{\partial(2Gm_{\rm out})} =
\frac{\rho_0}{\sqrt{D}}>0, \quad D\geq B^2\geq0.
\label{rho0f}
\end{equation}
Now we look on the inequalities between parameters $e$, $M$, $m_{\rm in}$ and $m_{\rm out}$:
\begin{equation}
Gm_{\rm in}^2>e^2>GM^2 \; \Rightarrow \; m_{\rm in}>M>-\Delta m,
\label{min2}
\end{equation}
\begin{equation}
\Delta m=m_{\rm out}-m_{\rm in}>-M \; \Rightarrow \;
m_{\rm in}>m_{\rm out}>m_{\rm in}-M>0.
\label{min3}
\end{equation}
This means, that we cannot start our consideration from $m_{\rm out}=0$ as previously. Now we need to to start from the nonzero minimal value $m_{\rm out,min}>0$ from (\ref{mout}). This minimal value $m_{\rm out,min}$ in (\ref{mout}) at the fixed bare mass $M$ depends on the corresponding minimal value of $m_{\rm in,min}$ at the fixed charge $e$. At the same time, the expression for $\rho_{0,\rm min}$ coincides with (\ref{rhomin3}). It is easily to verify, that relation $\sqrt{D_{\rm min}}= e^2+GM^2+2\sqrt{G}\Delta me>0$ in (\ref{Dmin3}) is equivalent to inequalities $e\sqrt{G}<\rho_{0,\rm min}<\rho_{\sigma_{\rm in}}$. Graphically these inequalities are shown in Fig.~\ref{mout33}.

From the point of view of the internal metric at $m_{\rm out,min}<m_{\rm out}<m_{\rm cr}+\Delta m$, where $m_{\rm cr}$ is defined by expression (\ref{mcr}), the shell, starting from the zero radius expands up to the turning point, by passing through the $T_+$-region between the both horizons without changing the sign of $\sigma_{\rm in}(\rho_0)=+1$. Therefore, $\rho_0$ is placed in the (right) $R_+$-region beyond the horizon. At the same time, if $m_{\rm out}>m_{\rm cr}+\Delta m$, then the turning point is placed beyond the horizon $r_+$, but in the (left) $R_-$-region.

How all this looking from the point of view of the external metric? Now it is needed to compare the relative values of $\rho_{0,\rm min}$, $\rho_{\sigma_{\rm out}}$ and $r_{g,\rm min}$. It appears, that always $r_{g,\rm min}\leq\rho_{0,\rm min}$:
\begin{eqnarray}
r_{g,\rm min} = 2\sqrt{G}e+2G\Delta m&<&\rho_{0,\rm min}=\frac{GM^2-e^2}{2(M-m_{\rm in})} \; \Rightarrow \\
(e-\sqrt{G}M+2\sqrt{G}\Delta m)^2&\geq&0.
\label{min4}
\end{eqnarray}
If the condition $e-\sqrt{G}M+2\sqrt{G}\Delta m<0$ is held in (\ref{min4}), it is verified by direct substitution, that $\rho_{\sigma_{\rm out}}<\rho_{0,\rm min}$. The last relation provides also the inequality $\rho_{\sigma_{\rm out}}<r_{g,\rm min}$:
\begin{eqnarray}
\frac{e^2-GM^2}{-2\Delta m} < 2G\left(\frac{e}{\sqrt{G}}+\Delta m\right) \; \Rightarrow \\
 (e-\sqrt{G}M+2\sqrt{G}\Delta m)(e+\sqrt{G}M+2\sqrt{G}\Delta m)<0.
\label{min5}
\end{eqnarray}
As we already know, the function in the second parenthesis is always positive under the considered conditions. This means, that the shell, starting from the zero radius, passes through the $T_+$-region of the external metric, the sign of $\sigma_{\rm out}$ is changing, and come out to the $R_-$-region, where $\sigma_{\rm out}(\rho_0)=-1$.

\begin{figure}[H]
\begin{center}
\includegraphics[angle=0,width=0.48\textwidth]{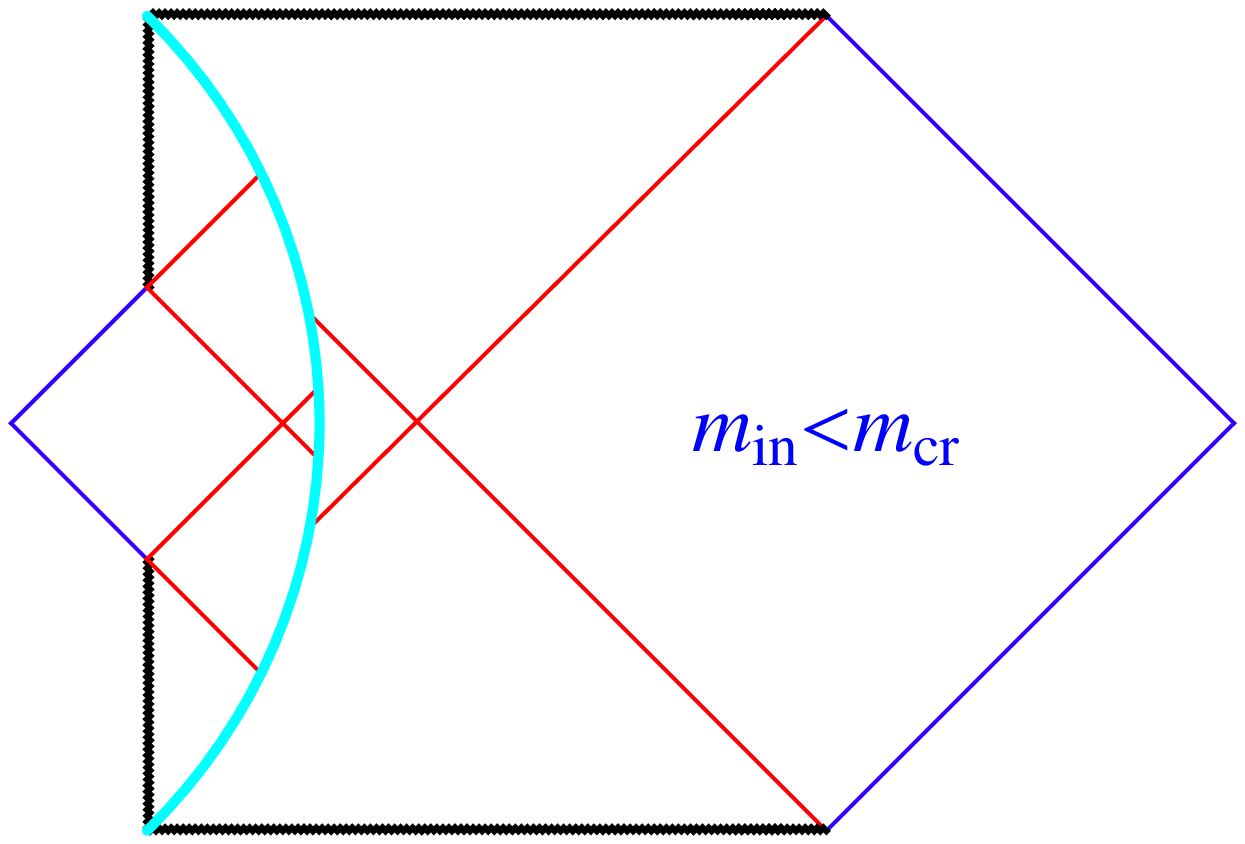}
\hfill
\includegraphics[angle=0,width=0.46\textwidth]{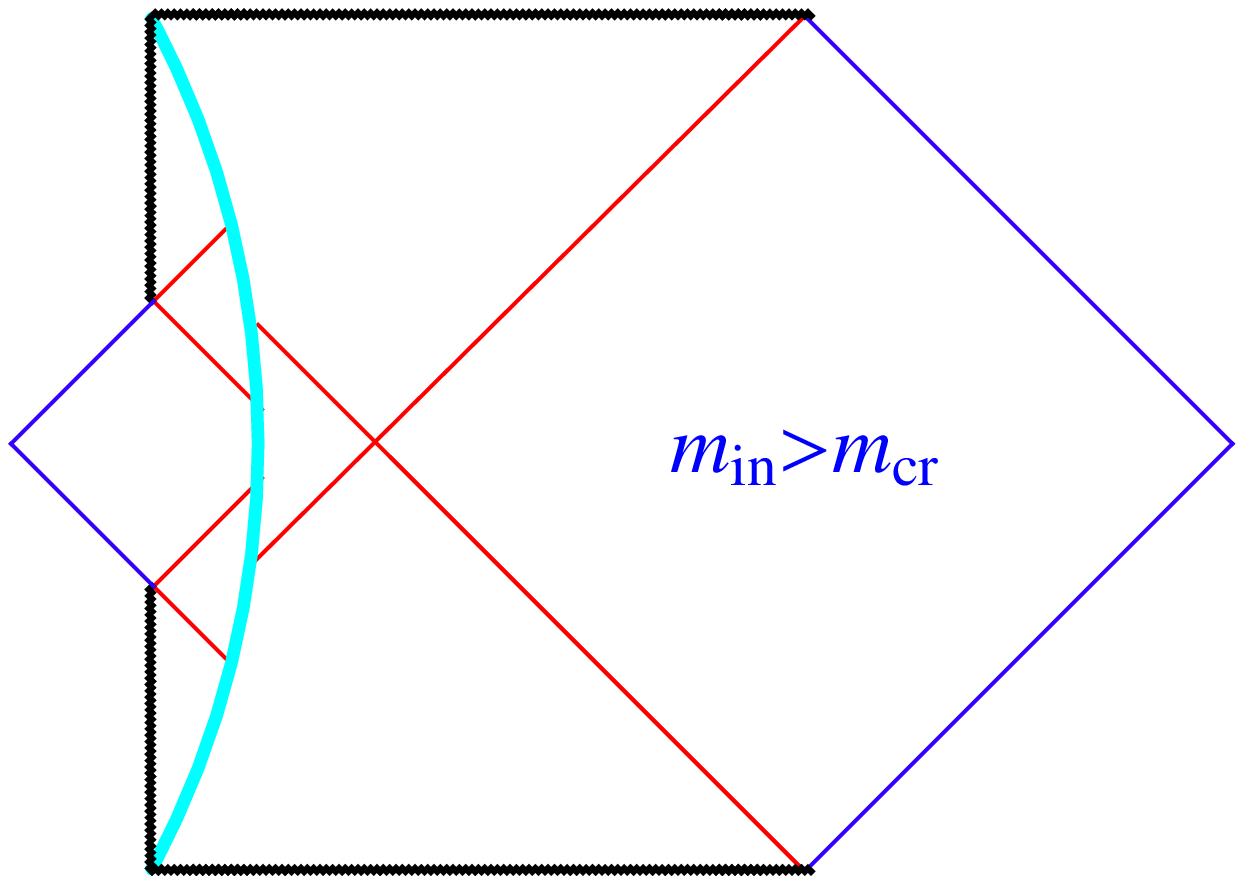}
\end{center}
\caption{The Carter--Penrose diagrams for finite motion of the self-gravitating shell in the case $\Delta m<-M$ and $e^2>GM^2$, $e-\sqrt{G}M+2\sqrt{G}\Delta m<0$ and $m_{\rm in}\gtrless m_{\rm cr}$.}
  \label{p36}
\end{figure}

The corresponding Carter--Penrose diagrams for finite motion of the self-gravitating shell in the case $\Delta m<-M$ and $e^2>GM^2$, $e-\sqrt{G}M+2\sqrt{G}\Delta m<0$ and $m_{\rm in}\gtrless m_{\rm cr}$ are shown in Figs.~\ref{p36}.

\begin{figure}[H]
\begin{center}
\includegraphics[angle=0,width=0.95\textwidth]{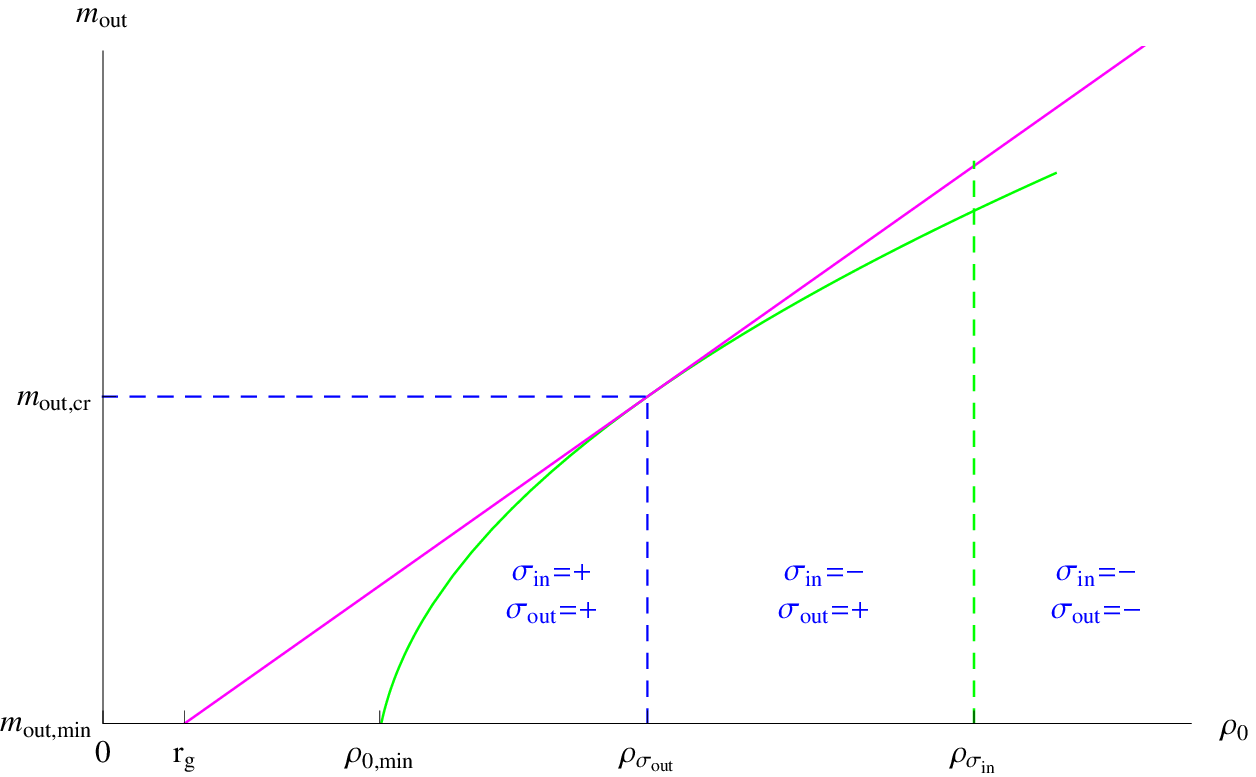}
\end{center}
\caption{The relative positions of the characteristic radii $r_g$, $\rho_{0,\rm min}$, $\rho_{\sigma_{\rm in}}$ and $\rho_{\sigma_{\rm out}}$.} \label{mout37}
\end{figure}

\begin{figure}[H]
\begin{center}
\includegraphics[angle=0,width=0.44\textwidth]{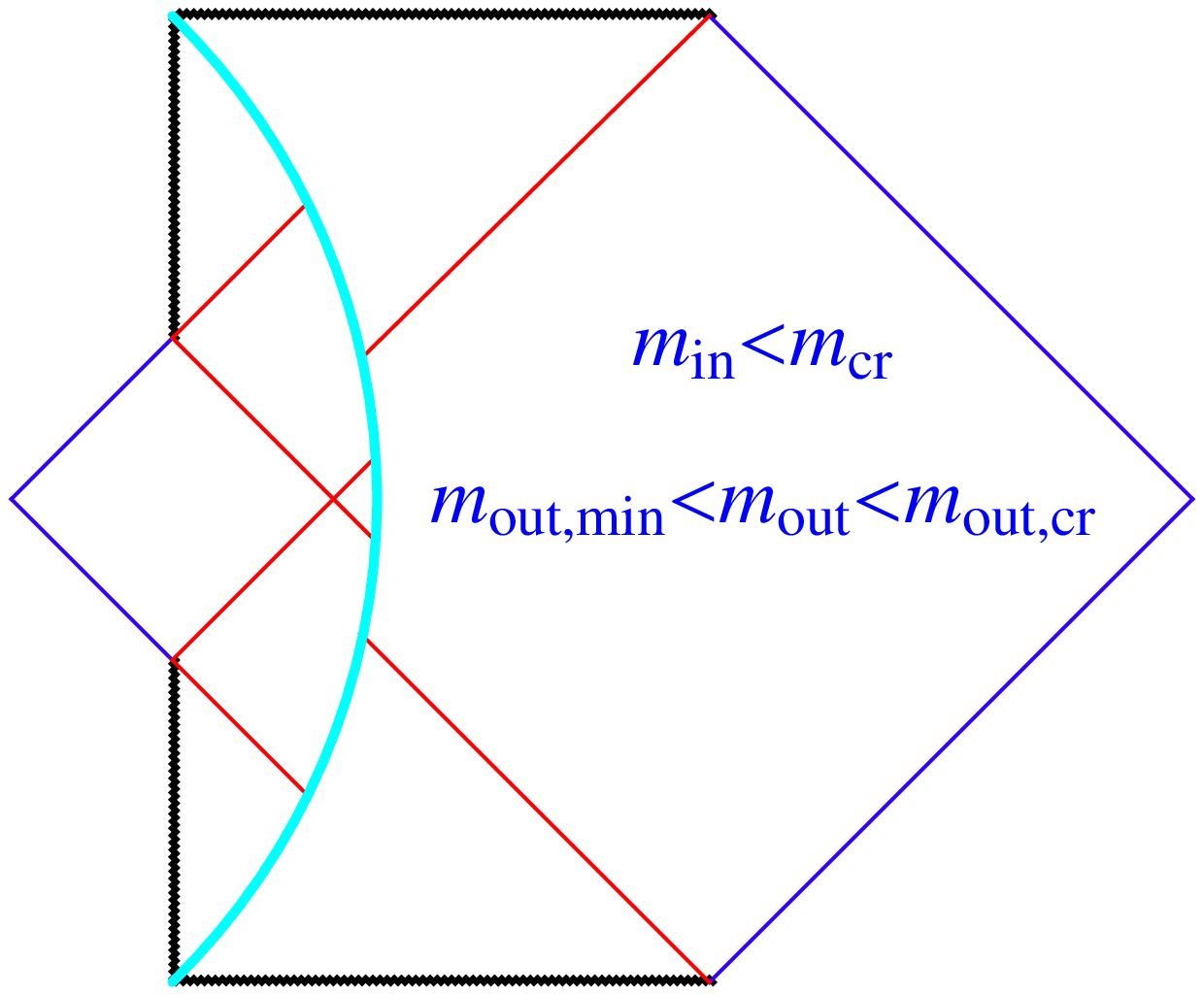}
\hfill
\includegraphics[angle=0,width=0.53\textwidth]{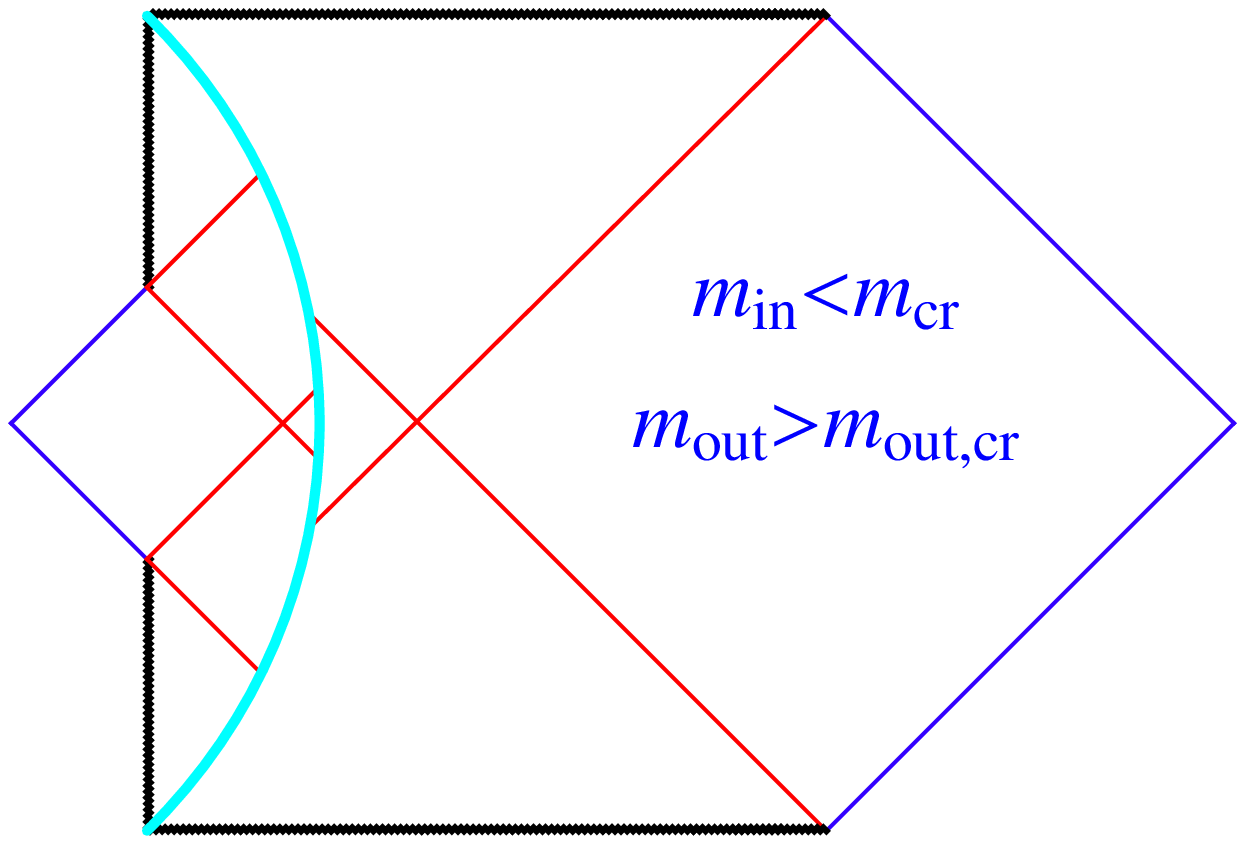}
\end{center}
\caption{The Carter--Penrose diagrams for finite motion of the self-gravitating shell in the case
$m_{\rm in}<m_{\rm in,cr}$ and $m_{\rm out,min}<m_{\rm out}<m_{\rm out,cr}$ (the left panel), and, respectively, in the case of $m_{\rm in}<m_{\rm in,cr}$, $m_{\rm out}>m_{\rm out,cr}$ (the right panel).}
  \label{p38}
\end{figure}

Now, at last, we consider the case of $e-\sqrt{G}M+2\sqrt{G}\Delta m>0$, when
$r_g<\rho_{0,\rm min}<\rho_{\sigma_{\rm out}}<\rho_{\sigma_{\rm in}}$.
Graphically these inequalities are shown in Fig.~\ref{mout37}.
We see, that there are three regions with the different combinations of $\sigma_{\rm in}$ and $\sigma_{\rm out}$. In the first case, at $m_{\rm in}<m_{\rm in,cr}$ and $m_{\rm out,min}<m_{\rm out}<m_{\rm out,cr}$, where
\begin{equation}
m_{\rm out,cr}=-\frac{e^2-GM^2}{2G\Delta m},
 \label{min6}
\end{equation}
there are realized $\sigma_{\rm in}(\rho_0)=+1$ and $\sigma_{\rm out}(\rho_0)=+1$.
The corresponding Carter--Penrose diagram is shown at the left panel in Fig.~\ref{p38}.

In the second case, at $m_{\rm in}<m_{\rm in,cr}$, but $m_{\rm out}>m_{\rm out,cr}$, there are realized $\sigma_{\rm in}(\rho_0)=+1$ and $\sigma_{\rm out}(\rho_0)=-1$. This second case corresponds the the left panel in Fig.~\ref{p38}.

\begin{figure}[H]
\begin{center}
\includegraphics[angle=0,width=0.6\textwidth]{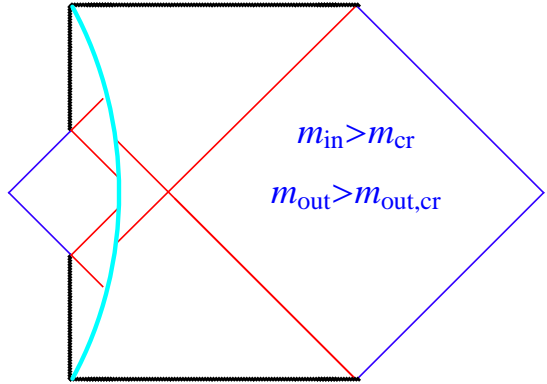}
\end{center}
\caption{The Carter--Penrose diagrams for finite motion of the self-gravitating shell in the case $m_{\rm in}>m_{\rm in,cr}$, $m_{\rm out}>m_{\rm out,cr}$.}
  \label{p39}
\end{figure}

At last, in the third case, at $m_{\rm in}>m_{\rm in,cr}$ and $m_{\rm out}>m_{\rm out,cr}$, it is realized $\sigma_{\rm in}(\rho_0)=-1$ and $\sigma_{\rm out}(\rho_0)=-1$. The corresponding Carter--Penrose diagram for this third case is shown in Fig.~\ref{p39}.

\section{Conclusion}
It is elaborated the complete classification of the possible types of the spherically symmetric global geometries for two types of electrically charged shells:

(1) The charged shell as a single source of the gravitational field, when internal space-time is flat, and external space-time is the Reissner--Nordstr\"om metric;

(2) The neutralizing shell with an electric charge opposite to the charge of the internal source with the Reissner--Nordstr\"om metric and with the Schwarzschild metric outside the shell.

\vskip 1cm
Authors acknowledges the Russian Fund of the Fundamental Research for financial support through grant 13-02-00257. On of us, V.A.B., acknowledges also the Ministry of Education of the Russian Federation, grant № 8412.


\begin{thebibliography}{99}

\bibitem{Israel} W. Israel, Phys. Rev, 153, 1388 (1967).

\bibitem{BerKuzTkachPRD87} V.\,A. Berezin, V.\,A.Kuzmin, I.\,I.
Tkachev, Phys. Rev. D 36, 2919 (1987).

\bibitem{BerKuzTkach4} V.\,A. Berezin, V.\,A.Kuzmin, I.\,I. Tkachev,
Phys. Lett. B 120, 91 (1983).

\bibitem{BerKuzTkach6} V.\,A. Berezin, V.\,A.Kuzmin, I.\,I. Tkachev,
Phys. Lett. B 124, 479 (1983).

\bibitem{BerKuzTkach5} V.\,A. Berezin, V.\,A.Kuzmin, I.\,I. Tkachev,
Phys. Lett. B 130, 23 (1983).

\bibitem{BerKuzTkach2} V.\,A. Berezin, V.\,A.Kuzmin, I.\,I. Tkachev,
Pis'ma ZhETF 41, 446 (1985).

\bibitem{BerKuzTkachJETP} V.\,A. Berezin, V.\,A.Kuzmin, I.\,I. Tkachev,
ZhETF, 86, 785 (1984).

\bibitem{BerDokErosSmir} V. Berezin, V. Dokuchaev, Yu. Eroshenko,
A. Smirnov, Class. Quant. Grav. 22, 4443 (2005).

\bibitem{BerOkh} V. Berezin, M. Okhrimenko, Class. Quant. Grav, 18, 2195 (2001).

\bibitem{BerKuzTkach3} V.\,A. Berezin, V.\,A.Kuzmin, I.\,I. Tkachev,
ZhETF 93, 1159 (1987).

\end{thebibliography}
\end{document}